\documentclass[camera,letterpaper,nomarginnotes,nonarrowgutter]{jpaper}
\usepackage[sort,compress]{cite}
\usepackage{amsmath,amssymb,amsfonts}

\usepackage{graphicx}
\usepackage{subcaption}
\usepackage{textcomp}
\usepackage{xcolor}
\usepackage{fancyhdr}
\usepackage{adjustbox}
\usepackage{multirow}
\usepackage{siunitx}
\usepackage{pifont}
\usepackage{balance}
\usepackage{stfloats}  
\usepackage{listings}
\usepackage{datetime}
\usepackage{setspace}

\usepackage[bookmarks=true,breaklinks=true,letterpaper=true,colorlinks,citecolor=blue,linkcolor=blue,urlcolor=blue]{hyperref}

\usepackage[colorinlistoftodos,prependcaption,textsize=small]{todonotes}

\newcommand{\figref}[1]{Fig.~\ref{#1}}
\newcommand{\secref}[1]{§\ref{#1}}
\newcommand{\obslbl}[0]{Finding}

\newcommand{\powerSavingWorksAllCitations}[0]{\cite{%
bhati2013coordinated,liu2012raidr,seshadri2013rowclone,patel2017reaper,
ozturk2006data,malladi2012rethinking,song2011flikker,lee2016simultaneous,
khan2017detecting,khan2016parbor,khan2014efficacy,fan2001memory,
diniz2007limiting,deng2011memscale,delaluz2002scheduler,delaluz2001dram,
ha2016improving,cooper2010fine,oconnor2017fine,zhang2014half,lee2017partial,
olgun2023sectored,chang2017understanding,chang2017thesis,bhati2015flexible,
lee2021greendimm,koppula2019eden,nguyen2021zem,alawneh2023high,
zheng2008minirank,aggarwal2008power,agrawal2014mosaic,david2011memory,
luz2002automatic,deng2012active,kandemir2001improving,lebeck2000power,
lin2012secret,mukundan2013understanding,ohsawa1998optimizing,
qureshi2015avatar,venkatesan2006rapid,udipi2010rethinking,ware2006improving,
lee2013tiered,lee2015adaptive,hassan2016chargecache,chang2016lisa,
hassan2019crow,luo2020clrdram,seshadri2017ambit,chang2014improving,
baek2014refresh,yoon2011adaptive,chatterjee2017architecting,lee2018reducing,
kim2012case}}
\newcommand{\powerModelsAllCitations}[0]{\cite{%
steiner2025drampower5,drampowergithub,vogelsang2010understanding,
ghose2018what,li2009mcpat,fan2001memory,micron2007power,stow2019power,
chen2012cacti,poddar2017mesap,park20183ddate,weis2016dramspec,
thoziyoor2008comprehensive,wang2005dramsim,rosenfeld2011dramsim2,
li2020dramsim3,dramsim3github,mathew2017bank,jung2016new,naji2015high,
shih2014dart,oconnor2017fine,chatterjee2017architecting,adiletta2026rpu}}

\newlength{\HeightReference}
\newlength{\DepthReference}
\newlength{\Width}%

\newcommand{\MyColorBox}[2][red]%
{%
    \settowidth{\Width}{#2}%
    \colorbox{#1}%
    {%
        \raisebox{-\DepthReference}%
        {%
                \parbox[b][\HeightReference+\DepthReference][c]{\Width}{\centering#2}%
        }%
    }%
}
\newcounter{obs}
\newcommand\observation[1]{%
   \refstepcounter{obs}
   \noindent
   \MyColorBox[gray!20]{\textbf{\obslbl{} \theobs.}} \emph{#1}}
   
\newcounter{take}
\newcommand\take[1]{%
   \refstepcounter{take}
   \vspace{1mm}
  \noindent
  \begin{tabular}{|p{0.95\linewidth}|}
       \hline
       \textbf{{Takeaway \thetake}.} {{#1}}\\
       \hline 
  \end{tabular}
  \vspace{1mm}
}

\definecolor{iy}{rgb}{0.0, 0.5, 0.2}
\definecolor{nbc}{rgb}{0.88, 0.07, 0.37}
\definecolor{moegi}{rgb}{0.357, 0.537, 0.188}
\definecolor{amethyst}{rgb}{0.6, 0.4, 0.8}

\newcommand{\dingOne}{\ding{172}}
\newcommand{\dingTwo}{\ding{173}}
\newcommand{\dingThree}{\ding{174}}
\newcommand{\dingFour}{\ding{175}}
\newcommand{\dingFive}{\ding{176}}
\newcommand{\dingSix}{\ding{177}}
\newcommand{\dingSeven}{\ding{178}}

\newif\ifdraft
\newif\ifsubmission
\newif\ifrevision
\newif\ifiterations
\draftfalse
\submissiontrue
\revisionfalse
\iterationsfalse

\newcommand{\extended}[1]{#1}

\ifdraft
    \paperwidth=\dimexpr \paperwidth + 5cm\relax
    \oddsidemargin=\dimexpr\oddsidemargin + 2cm\relax
    \evensidemargin=\dimexpr\evensidemargin + 2cm\relax
    \marginparwidth=\dimexpr \marginparwidth + 2cm\relax

    \newcommand{\atbcomment}[1]{\todo[size=\scriptsize, linecolor=blue, bordercolor=blue, backgroundcolor=white]{\textcolor{blue}{\textbf{@atb:} #1}}}

    \newcommand{\ieycomment}[1]{\todo[size=\scriptsize, linecolor=iy, bordercolor=iy, backgroundcolor=white]{\textcolor{iy}{\textbf{@iey:} #1}}}
    \newcommand{\nbcomment}[1]{\todo[size=\scriptsize, linecolor=nbc, bordercolor=nbc, backgroundcolor=nbc!5]{\textcolor{nbc}{\textbf{@nb:} #1}}}
    \newcommand{\agycomment}[1]{\todo[size=\scriptsize, linecolor=nbc, bordercolor=nbc, backgroundcolor=white]{\textcolor{orange}{\textbf{@agy:} #1}}}
    \newcommand{\agyurgentcomment}[1]{\todo[size=\scriptsize, linecolor=nbc, bordercolor=nbc, backgroundcolor=white]{\textcolor{red}{\textbf{[!ALARM!]@agy:} #1}}}
    \newcommand{\gfcomment}[1]{\todo[size=\scriptsize, linecolor=blue, bordercolor=blue, backgroundcolor=white]{\textcolor{blue}{\textbf{@gf:} #1}}}
      
    \newcommand{\param}[1]{\textcolor{red}{#1}}
    \newcommand{\copied}[1]{\textcolor{gray}{#1}}
    \newcommand{\outline}[1]{\noindent\textcolor{orange}{\textbf{Outline:} \emph{#1}}}
    \newcommand{\TODO}[1]{\textcolor{red}{\textbf{#1}}}

    \newcommand{\nb}[1]{\textcolor{nbc}{#1}}

    \newcommand{\microrevcommon}[1]{#1}
    \newcommand{\microreva}[1]{#1}
    \newcommand{\microrevb}[1]{#1}
    \newcommand{\microrevc}[1]{#1}
    
    \newcommand{\microreve}[1]{#1}

    \newcommand{\microlabel}[1]{}

    \newcommand{\atbcr}[2]{#2}
    \newcommand{\modelcr}[2]{#2}
    \newcommand{\omcr}[2]{#2}
    \newcommand{\atbcrcomment}[2]{}
    \newcommand{\omcrcomment}[2]{}
\else
\ifrevision

    \paperwidth=\dimexpr \paperwidth + 3cm\relax
    \oddsidemargin=\dimexpr\oddsidemargin + 1cm\relax
    \evensidemargin=\dimexpr\evensidemargin + 1cm\relax
    \marginparwidth=\dimexpr \marginparwidth + 1cm\relax

    \newcommand{\atbcomment}[1]{}
    \newcommand{\ieycomment}[1]{}
    \newcommand{\nbcomment}[1]{}
    \newcommand{\agycomment}[1]{}
    \newcommand{\agyurgentcomment}[1]{}
    \newcommand{\gfcomment}[1]{}

    \newcommand{\param}[1]{#1}
    \newcommand{\copied}[1]{#1}
    \newcommand{\outline}[1]{}
    \newcommand{\TODO}[1]{}

    \newcommand{\nb}[1]{#1}

    \newcommand{\microrevcommon}[1]{\textcolor{blue}{#1}}
    \newcommand{\microreva}[1]{\textcolor{orange}{#1}}
    \newcommand{\microrevb}[1]{\textcolor{brown}{#1}}
    \newcommand{\microrevc}[1]{\textcolor{teal}{#1}}
    
    \newcommand{\microreve}[1]{\textcolor{red}{#1}}

    \newcommand{\microlabel}[1]{\todo[size=\scriptsize, linecolor=blue, bordercolor=blue, backgroundcolor=white]{\textcolor{blue}{#1}}}

    \newcommand{\atbcr}[2]{#2}
    \newcommand{\modelcr}[2]{#2}
    \newcommand{\omcr}[2]{#2}
    \newcommand{\atbcrcomment}[2]{}
    \newcommand{\omcrcomment}[2]{}
\else
\ifiterations
    \paperwidth=\dimexpr \paperwidth + 5cm\relax
    \oddsidemargin=\dimexpr\oddsidemargin + 2cm\relax
    \evensidemargin=\dimexpr\evensidemargin + 2cm\relax
    \marginparwidth=\dimexpr \marginparwidth + 2cm\relax

    \newcommand{\atbcomment}[1]{}
    \newcommand{\ieycomment}[1]{}
    \newcommand{\nbcomment}[1]{}
    \newcommand{\agycomment}[1]{}
    \newcommand{\agyurgentcomment}[1]{}
    \newcommand{\gfcomment}[1]{}

    \newcommand{\param}[1]{#1}
    \newcommand{\copied}[1]{#1}
    \newcommand{\outline}[1]{}
    \newcommand{\TODO}[1]{}

    \newcommand{\nb}[1]{#1}

    \newcommand{\microrevcommon}[1]{#1}
    \newcommand{\microreva}[1]{#1}
    \newcommand{\microrevb}[1]{#1}
    \newcommand{\microrevc}[1]{#1}
    
    \newcommand{\microreve}[1]{#1}

    \newcommand{\microlabel}[1]{}

    \newcommand{\atbcr}[2]{\ifnum#1=\value{version}\textcolor{red}{#2}\else{#2}\fi}
    \newcommand{\modelcr}[2]{\ifnum#1=\value{version}\textcolor{purple}{#2}\else{#2}\fi}
    \newcommand{\omcr}[2]{\ifnum#1=\value{version}\textcolor{blue}{#2}\else{#2}\fi}
    \newcommand{\atbcrcomment}[2]{\ifnum#1=\value{version}\todo[size=\scriptsize, linecolor=orange, bordercolor=orange, backgroundcolor=white]{\textcolor{red}{Atb:~#2}}\else{}\fi}
    \newcommand{\omcrcomment}[2]{\ifnum#1=\value{version}\todo[size=\scriptsize, linecolor=orange, bordercolor=orange, backgroundcolor=white]{\textcolor{blue}{Onur:~#2}}\else{}\fi}
\else
\ifsubmission
    \newcommand{\atbcomment}[1]{}
    \newcommand{\ieycomment}[1]{}
    \newcommand{\nbcomment}[1]{}
    \newcommand{\agycomment}[1]{}
    \newcommand{\agyurgentcomment}[1]{}
    \newcommand{\gfcomment}[1]{}

    \newcommand{\param}[1]{#1}
    \newcommand{\copied}[1]{#1}
    \newcommand{\outline}[1]{}
    \newcommand{\TODO}[1]{}

    \newcommand{\nb}[1]{#1}

    \newcommand{\microrevcommon}[1]{#1}
    \newcommand{\microreva}[1]{#1}
    \newcommand{\microrevb}[1]{#1}
    \newcommand{\microrevc}[1]{#1}
    
    \newcommand{\microreve}[1]{#1}

    \newcommand{\microlabel}[1]{}

    \newcommand{\atbcr}[2]{#2}
    \newcommand{\modelcr}[2]{#2}
    \newcommand{\omcr}[2]{#2}
    \newcommand{\atbcrcomment}[2]{}
    \newcommand{\omcrcomment}[2]{}
\fi
\fi
\fi
\fi

\def\UrlBreaks{\do\/\do-\/\do.\/\do:}

\expandafter\def\expandafter\UrlBreaks\expandafter{\UrlBreaks
  \do\a\do\b\do\c\do\d\do\e\do\f\do\g\do\h\do\i\do\j
  \do\k\do\l\do\m\do\n\do\o\do\p\do\q\do\r\do\s\do\t
  \do\u\do\v\do\w\do\x\do\y\do\z\do\A\do\B\do\C\do\D
  \do\E\do\F\do\G\do\H\do\I\do\J\do\K\do\L\do\M\do\N
  \do\O\do\P\do\Q\do\R\do\S\do\T\do\U\do\V\do\W\do\X
  \do\Y\do\Z}

\newcommand{\numStacks}[0]{\param{36}}
\newcommand{\numFPGAs}[0]{\param{18}}

\newcommand{\versionnum}[1]{3.0}

\newcounter{version}
\ifiterations
\else
\fi

\AtBeginDocument{%
  }

\makeatletter
\g@addto@macro{\normalsize}{%
  \setlength{\abovedisplayskip}{3pt plus 1pt minus 1pt}
  \setlength{\belowdisplayskip}{2pt plus 1pt minus 1pt}
  \setlength{\abovedisplayshortskip}{0pt}
  \setlength{\belowdisplayshortskip}{0pt}
  \setlength{\intextsep}{2pt plus 1pt minus 1pt}
  \setlength{\textfloatsep}{0pt plus 1pt minus 1pt}
  \setlength{\skip\footins}{2pt plus 1pt minus 1pt}
  \setlength{\abovecaptionskip}{2pt plus 1pt minus 1pt}}
\makeatother

\usepackage{titlesec}
\titlespacing\section{0pt}{4pt plus 1pt minus 1pt}{2pt plus 1pt minus 1pt}
\titlespacing\subsection{0pt}{4pt plus 1pt minus 1pt}{2pt plus 1pt minus 1pt}
\titlespacing\subsubsection{0pt}{4pt plus 1pt minus 1pt}{2pt plus 1pt minus 1pt}

\title{\Large{\omcr{1}{Understanding the Power Consumption of Modern High Bandwidth Memory:\\Experimental Analysis and Modeling Using Real HBM2 DRAM Chips}}}

\author{Ataberk Olgun$^{\ddagger{}}$\quad
Spiros Galanopoulos$^{\ddagger{}}$\quad 
Andreas Kosmas Kakolyris$^{\ddagger{}}$\quad 
\\
\.{I}smail Emir Y\"{u}ksel$^{\ddagger{}}$\quad 
Haocong Luo$^{\ddagger{}}$\quad 
F. Nisa Bostanc{\i}$^{\ddagger{}}$\quad 
Onur Mutlu$^{\ddagger{}}$$^{\dagger{}}$\vspace{-3mm}\\\\
$^{\ddagger{}}$\emph{ETH Zurich} \quad $^{\dagger{}}$\emph{New York University}}

\fancypagestyle{iterationsfirstpage}
{
    \fancyhead{}
    \fancyhead[C]{
    \textcolor{red}{CONFIDENTIAL DRAFT -- DO NOT DISTRIBUTE -- TO APPEAR IN HPCA'25} \\ 
    \emph{\textcolor{blue}{Version 5.0~---~\today, \ampmtime}}
    }
}

\fancypagestyle{firstpage}
{
    \fancyhead{}
    \begin{tikzpicture}[remember picture,overlay]
    \node [xshift=150mm,yshift=-10.5mm]
    at (current page.north west) {\href{https://www.acm.org/publications/policies/artifact-review-and-badging-current}{\includegraphics[width=1.7cm]{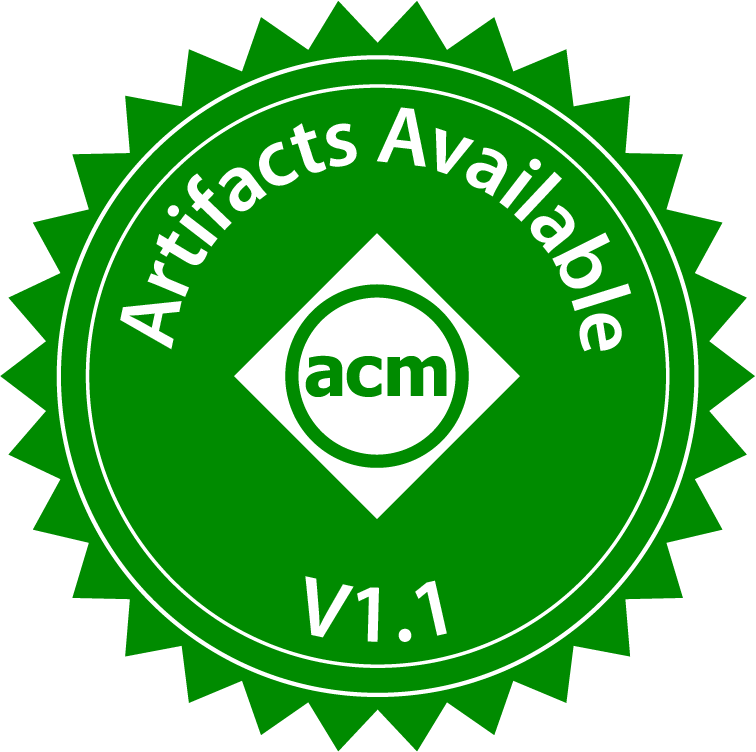}}} ;
    \node [xshift=168mm,yshift=-10.5mm]
    at (current page.north west) {\href{https://www.acm.org/publications/policies/artifact-review-and-badging-current}{\includegraphics[width=1.7cm]{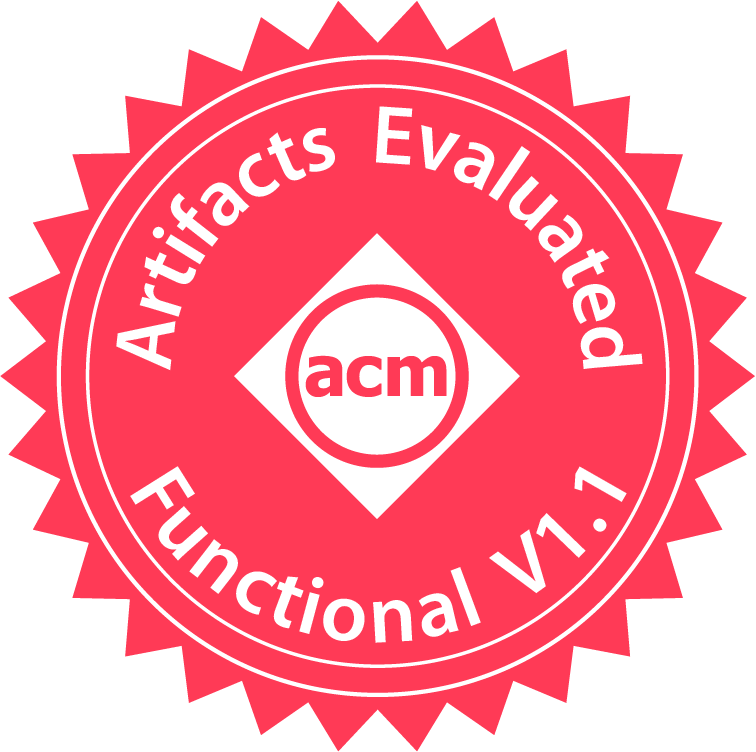}}} ;
    \node [xshift=186mm,yshift=-10.5mm]
    at (current page.north west) {\href{https://www.acm.org/publications/policies/artifact-review-and-badging-current}{\includegraphics[width=1.7cm]{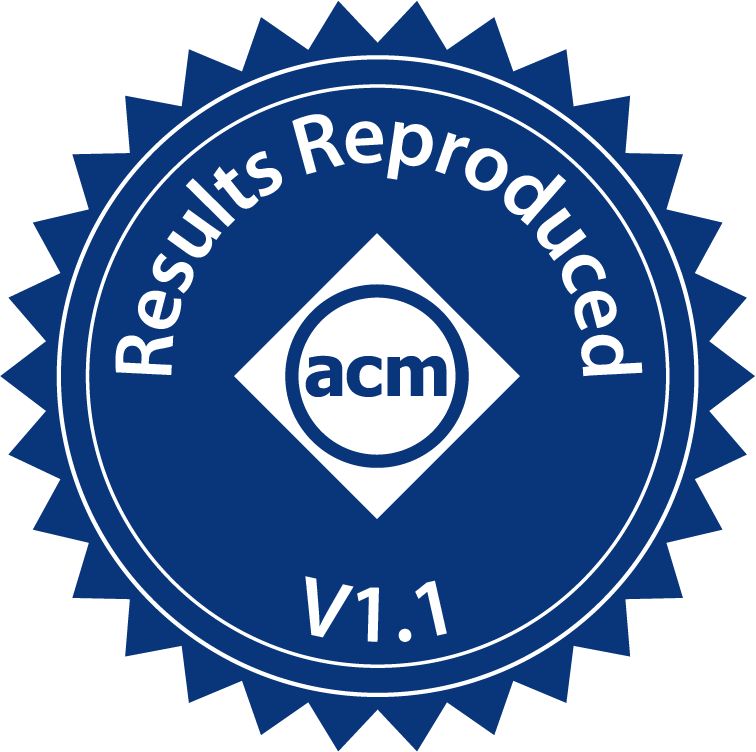}}} ;
    \end{tikzpicture}
  
  \pagenumbering{arabic}
  \vspace{1em}
  \fancyfoot[C]{\large\thepage}
}

\begin{document}
\bstctlcite{IEEEexample:BSTcontrol}
\maketitle


\ifiterations
\fancyhead{}
\fancyhead[C]{
\textcolor{red}{CONFIDENTIAL DRAFT -- DO NOT DISTRIBUTE -- TO APPEAR IN MICRO'26} \\
\textcolor{MidnightBlue}{\emph{Version \versionnum~---~\today, \ampmtime}}
}
\else
\pagestyle{plain}
\thispagestyle{firstpage}
\fi

\sloppy
\begin{abstract}

We perform the first comprehensive experimental characterization of power
consumed by real HBM2 \omcr{1}{DRAM chips}. Our extensive characterization of
\numStacks{} HBM2 \omcr{1}{chips} yields 1)~the empirical basis for developing
new,\nbcomment{and?} accurate HBM power models or validating existing power
models and 2)~key insights into structural and data-pattern-dependent variation
of power in HBM2 \omcr{1}{chips}. Out of our \microrevc{\param{13}} new findings and
\microrevc{\param{6}} key takeaway messages, we highlight: 1)~\atbcr{4}{a} read
operation consumes higher current than any other tested operation including
refresh (by \param{$3.1\times{}\!$}\!),
2)~HBM2 read
current significantly varies across bank groups (\omcr{1}{by} \param{8.1\%}) and banks
(\omcr{1}{by} \param{5.3}\%), and 3)~the 
toggled data bus signals in a data transfer burst \omcr{1}{significantly}
affect current \omcr{1}{draw}. For example, toggling all HBM2-internal bank
group bus signals causes \param{29.4}\% higher \omcr{3}{power} consumption than keeping
them constant. Leveraging our empirical insights into HBM2 power
\omcr{1}{consumption}, we carefully engineer an access pattern that yields
\param{23.6}\% (\param{35.4}\%) higher power consumption on average (maximum)
across all tested HBM2 \omcr{1}{chips} than the highest-power-consuming
JEDEC-specified current measurement loop. 

Our results have important implications for HBM-based system design and
validation. For example, relying solely on JEDEC- or vendor-specified test loops
to bound HBM power could cause errors or system failures, thermal throttling,
and thermal shutdowns in real-world high-power, high-performance computing or
AI/ML workloads. To improve 3D-stacked DRAM power modeling fidelity in open
source architectural simulation infrastructure\atbcr{3}{s},
\microlabel{CC1}\microrevcommon{we develop a new HBM power model, guided by our
characterization and built on DRAMPower, that captures structural and
data-pattern-dependent power variation. Our model predicts the power of
\numStacks{} real HBM2 \omcr{1}{chips} across diverse workloads (including a
real LLM decode workload) with \param{\modelcr{2}{7.6--8.1}}\% mean absolute
percentage error, substantially more accurately than state-of-the-art HBM power
models \omcr{1}{(\omcr{2}{which have} error between \modelcr{2}{20--42}\%)}. We
validate our model against real HBM3E GPU memory power \atbcr{1}{consumption.
The power consumption predicted by our model for a power-intensive
microbenchmark closely matches (\omcr{2}{with \atbcr{3}{2.4\% to 5.3\%}
error} \modelcr{2}{across HBM3E's specified supply voltage range}) the average
memory power consumed by three tested NVIDIA H200 GPUs.} \atbcr{1}{We} use the
power model to project the read energy of future HBM4 \omcr{1}{chips}.} We
publicly release our power model at
\omcr{1}{\url{https://github.com/CMU-SAFARI/HBM-Power}}.

\end{abstract}

\section{Introduction}

High Bandwidth Memory (HBM) is heavily \microrevb{and increasingly} used as a
fundamental building block for memory systems \omcr{1}{especially} in modern
\omcr{1}{AI/ML and HPC systems, \atbcr{3}{and}} machine learning accelerators.
\omcr{1}{This trend is} driven by the rapid increase in modern
application\omcr{1}{s' capacity and bandwidth demands~\cite{he2025papi,
gu2025cent, boroumand2021google, mutlu2025memory, ahn2015tesseract,
mutlu2013memory, mutlu2025modern}}. In modern \omcr{1}{AI/ML} training and
\omcr{1}{serving} systems, HBM can account for a substantial portion of the
total system power budget. \microlabel{CC5}\microrevcommon{For example, we
measure that GPU memory (HBM3E) power is as high as \param{\SI{248}{\watt}} on
average across three real NVIDIA H200 GPUs (\secref{sec:hbm3e-validation}),
approximately \param{35}\% of the GPU's \SI{700}{\watt} power
budget~\cite{nvidia-h200-datasheet}.} Consequently, understanding and accurately
predicting HBM power consumption is necessary to develop \omcr{1}{efficient} and
high-performance HBM\omcr{1}{-based computing} systems. \nbcomment{this is
better than the sentence in the abstract for motivation}

Computer architects heavily rely on architectural power models
(e.g.,~\cite{chen2012cacti, vogelsang2010understanding, naji2015high,
shih2014dart, park20183ddate, thoziyoor2008comprehensive}) to make informed
design decisions during \nb{the} early stages of 3D-stacked memory system
development, and to design and evaluate techniques that reduce DRAM power
\atbcr{1}{\powerSavingWorksAllCitations{}}. Empirical power measurements are
critical to \omcr{1}{validate} and improve the fidelity of such power models.
For example, the VAMPIRE paper~\cite{ghose2018what} demonstrates that relying
solely on vendor \atbcr{3}{datasheets} for DDR3L DRAM leads to \nb{a} significant
discrepancy in predicted vs measured DRAM power because real-world power
consumption deviates from \nb{the} vendor-specified \omcr{1}{datasheet} values. 
Inaccurate HBM power models could lead to misguided power budgeting in
the design of an HBM-based computing system and consequently incur important
robustness\omcr{1}{, efficiency,} and performance implications, such as job
failures~\cite{kokolis2025revisiting}, thermal throttling, and performance
degradation.

\begin{figure}[!ht]
\centering
\includegraphics[width=1.0\linewidth]{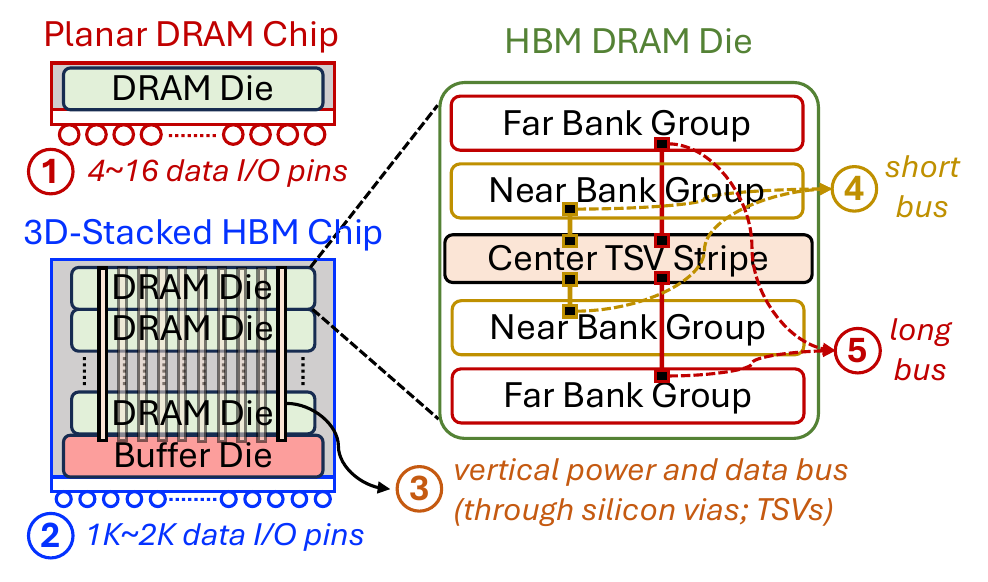}
\caption{\atbcr{2}{Planar DRAM chip vs 3D-stacked HBM DRAM chip design, and HBM-die-internal data buses.}}
\label{fig:intro_figure}
\end{figure}

Unfortunately, despite HBM's prevalence and importance, \omcr{1}{measured} power
\omcr{1}{consumption data \atbcr{1}{of real HBM chips is
scarce~\cite{stow2019power, larimi2021understanding} and limited to
coarse-grained access patterns (e.g., streaming and random)}}. Moreover, unlike
planar DDRx DRAM, HBM has a 3D-stacked architecture \nb{in which}
multiple DRAM dies are layered on top of each other.
\atbcr{2}{\figref{fig:intro_figure} shows the key structural differences between
planar DDRx DRAM and 3D-stacked HBM chips.}
\microlabel{B1}\microrevb{\atbcr{2}{An} HBM \omcr{1}{chip}
\atbcr{2}{qualitatively} differs from a planar DDRx DRAM chip in two key ways
that affect its power consumption. First, \atbcr{2}{a planar DRAM chip typically
has 4 -- 16 data I/O pins (\textcolor{red}{\ding{172}}) whereas an HBM chip has
thousands (\textcolor{blue}{\ding{173}}).}} {Second, data is transferred over a
larger variety of internal wiring\atbcr{2}{, such as through-silicon vias
(}TSVs; \textcolor{orange}{\ding{174}}) and bank group buses
\atbcr{2}{(\textcolor{olive}{\ding{175}}--\textcolor{red}{\ding{176}})}
\atbcr{2}{with different lengths, widths, and electrical characteristics} in an
HBM \omcr{1}{chip} than in a planar DRAM chip. As we show in this work, both
differences have concrete quantitative consequences for HBM power
\omcr{1}{consumption}.} 
\atbcr{2}{Beyond these qualitative differences,} th\atbcr{2}{e} 3D structure of
HBM chips introduces new variables, e.g., vertical power delivery
constraints~\atbcr{1}{\cite{shevgoor2013quantifying, healy2010power,
bose2021characterization}}, that could affect HBM power
consumption\nbcomment{consumption? management?} in currently unknown ways 
because there \nb{are}
\emph{no} publicly available empirical DRAM command current measurements for
HBM.\nbcomment{too long. I'd split into two? "publicly available empirical
low-level DRAM command current measurements for HBM" is too long} The lack of
such measurements prevent\nb{s} researchers from accurately and quantitatively
reasoning about the power impact of fine-grained structural variations (e.g.,
power variation across bank groups in an HBM2 channel).

In this work, we present the first rigorous experimental characterization of
HBM2 power consumption using \numStacks{} HBM2 \omcr{1}{chip}s in \numFPGAs{}
AMD Alveo U55C FPGA boards~\atbcr{1}{\cite{alveo-u55-c, alveo-u55c-manual}}.
\omcr{1}{Building on DRAM Bender~\cite{safari-drambender, olgun2023drambender,
olgun2026laboratory, hassan2017softmc} \atbcr{1}{and prior HBM2 testing
infrastructure~\cite{olgun2023hbm, olgun2024read}},} we develop a
\microrevb{new} FPGA-based HBM2 testing platform that can execute arbitrary
low-level DRAM command sequences targeting fine-grained DRAM structures across
any combination of HBM2 channels. Using this platform, {we perform a four-step
characterization study}. First, we measure \emph{IDD values}
\microrevb{(consumed current)} using standard current measurement loops that are
meant to represent common\nbcomment{common across what? or are these called
common in general?} high-level DRAM operations (e.g., reading cache lines in a
continuous data transfer burst) in the HBM2 specification~\cite{jedechbm2}. The
IDD values form the empirical basis for calibrating existing DRAM power
models~\atbcr{1}{\powerModelsAllCitations{}}. Second, we study HBM2 current's
sensitivity to temperature and HBM2 bandwidth \omcr{1}{consumption}. Third,
using the standard \microrevb{current measurement} loops, we investigate if and
how power varies across HBM2 structures (e.g., channels, pseudo-channels, bank
groups, banks) and with data patterns transferred across various
HBM2-\omcr{1}{chip}-internal data buses (e.g., through-silicon vias (TSVs) and
the bank group bus~\atbcr{1}{\cite{lee2016hbm, lee2020hbm2e, cho2018hbm2,
ryu2023hbm3, park2022hbm3, lee2014high, oh2020hbm2e, chun2020hbm2e, kim2022low,
jun2017hbm, jedechbm2}}). Fourth, building on our characterization results, we
carefully engineer data and access patterns designed to \omcr{1}{induce} higher
power consumption than the standard \microrevb{current measurement} loops
specified in the HBM2 standard.

Our characterization yields \microrevc{\param{13}} new findings and
\microrevc{\param{6}} key takeaway messages on HBM2 power consumption that could
substantially improve High Bandwidth Memory power modeling
methodologies~\atbcr{1}{\powerModelsAllCitations{}}. We highlight \param{3}
takeaways from our experimental characterization study of structural variation
and data pattern dependence of HBM2 power. First, we observe significant
variation in power \omcr{1}{consumption} across HBM2 structures:
\omcr{1}{drawn} current varies by \param{8.1}\% between bank groups and
\param{5.3}\% across banks on average across all tested HBM2 \omcr{1}{chip}s.
\microrevb{Thus, state-of-the-art power models}
\omcr{1}{(e.g.,~\cite{steiner2025drampower5, li2020dramsim3, oconnor2017fine,
chen2012cacti})} that attribute equal power consumption to all exercised
\atbcr{2}{bank groups or banks} in \microrevb{an HBM2 channel
\omcr{1}{\emph{cannot} accurately}} capture the real power distribution across
3D-stacked HBM2 channels. Second, HBM2 current is highly sensitive to data
patterns transferred over burst read operations~\cite{jedechbm2}. We find that
\omcr{1}{drawn} current increases with the number of toggled \atbcr{1}{data bus
signals}
\microrevb{(\secref{sec:dpd-current})}. Third, 
we find that IDD current measurement loops \atbcr{1}{specified in the
HBM2 standard} do \emph{not} yield the highest power consumption for the tested HBM2
\omcr{1}{chip}s. We show that a
new and carefully engineered \microrevb{current measurement} loop consumes an
average of \param{23.6}\% (and up to \param{35.4}\%) more power than the most
power\nb{-}consuming current measurement loop specified in the \omcr{1}{HBM2}
standard (IDD4R~\cite{jedechbm2}). These findings have important implications
for HBM2-based system design and HBM power modeling. For example, an HBM2-based
system designed using JEDEC-provided IDD current measurement loops may be
vulnerable to thermal shutdowns or \atbcr{1}{thermal
throttling}~\atbcr{1}{\cite{lin2007thermal, liu2011hardware, eckert2014thermal,
son2023thermal, paul2015harmonia, li2020dramsim3}} or power delivery
failures~\atbcr{1}{\cite{shevgoor2013quantifying, fan2007power, wu2016dynamo,
kokolis2025revisiting}} when real workloads \omcr{1}{(such as memory-intensive
AI/ML training or serving)} trigger high-power
states~\atbcr{1}{\cite{ganesan2010sympo, fan2007power, wu2016dynamo,
hajyahya2022agilewatts, hajyahya2021ichannels, antoniou2022agilepkgc}}.

To improve 3D-stacked DRAM power modeling fidelity in open source architectural
simulation infrastructure\omcr{1}{s}, \microlabel{CC1}\microrevcommon{we develop
a new HBM power model, guided by our characterization results and built on
DRAMPower~\cite{steiner2025drampower5}, that captures both structural and
data-pattern-dependent power variation. \atbcr{1}{We make three key observations
from our evaluation of the power model. First,} across \numStacks{} HBM2
\omcr{1}{chip}s and a diverse set of workloads (\param{6} microbenchmarks and a
real Llama3.1-8B~\atbcr{1}{\cite{grattafiori2024llama}} decode workload), our
model predicts power \omcr{1}{consumption} with \param{\modelcr{2}{7.6--8.1}}\%
mean absolute percentage error (\omcr{2}{see~\secref{sec:hbm2-power-model}}),
\omcr{1}{which is substantially lower than the error rates of} state-of-the-art
power models (e.g., \param{\modelcr{2}{21.6}}\% for DRAMSim3 and
\param{\modelcr{2}{22.6}}\% for FGDRAM \modelcr{2}{with the all-0s data
pattern}). \atbcr{1}{Second, we measure power consumed by a power-intensive
microbenchmark using various data patterns
\omcr{2}{(\secref{sec:data-dependent-power})}.\atbcrcomment{3}{fgdram has
slightly higher error than dramsim for all-0s} We carefully construct each data
pattern to induce unique toggling activity across the HBM data bus hierarchy. A
state-of-the-art power model~\cite{oconnor2017fine} that describes all data bus
(inside the HBM chip) toggling activity with a single parameter yields a
prediction error of \modelcr{3}{11.4\%} \modelcr{3}{(in read energy per bit)}
even after we calibrate it to our measurements. Our model separately accounts
for toggling on the vertical connections between 3D-stacked HBM dies and on the
data bus on an HBM die, and reduces the prediction error to \modelcr{3}{2.3\%}
\omcr{2}{(\secref{sec:data-dependent-power})}.}
\atbcr{1}{Third, we validate our model against real HBM3E GPU memory power
consumption \omcr{2}{(\secref{sec:hbm3e-validation})}. The power consumption
predicted by our model for a power-intensive microbenchmark closely matches
(\omcr{2}{with \atbcr{3}{2.4\% to 5.3\%} error} \modelcr{2}{across HBM3E's
specified supply voltage range}) the average memory power consumed by three
tested NVIDIA H200 GPUs. We use the power model to project the read energy of
future HBM4 chips \omcr{2}{(\secref{sec:hbm4-estimation})}.} \atbcr{1}{We
open source our power model as part of Ramulator2.1~\cite{luo2026ramulator21,
luo2024ramulator, ramulator2github, kim2016ramulator, bostanci2026cleaning} and
independently~\cite{hbmpower-github} (\omcr{2}{as well as} characterization data
and infrastructure).}}

\microlabel{B1}\microrevb{Our quantitative results reflect the two
\omcr{2}{major} qualitative differences between HBM and planar DRAM. First,
\atbcr{1}{an HBM chip transfers data over more pins and channels than a planar
DRAM chip. Thus, data transfer is a larger fraction of HBM chip power
consumption. In our tested HBM2 chips, a read operation draws 3.1$\times{}$
higher current than refresh (\secref{sec:standard-idd}). In planar DDR3L DRAM,
refresh draws more current than any other operation~\cite{ghose2018what}.}
Second, \atbcr{1}{data read from an HBM2 die crosses several internal buses
before reaching the HBM chip's data pins. Each bus has a different width and
operates at a different clock frequency. Thus, a data pattern that frequently
toggles one bus might \emph{not} toggle another bus. This makes the
highest-power data pattern difficult to identify \omcr{2}{in HBM}
(\secref{sec:dpd-current}).} Our power model evaluation
(\secref{sec:hbm2-power-model}) demonstrates the importance of capturing the
resulting data pattern dependence: existing power models that are unaware of the
data pattern dependence inaccurately predict HBM power \omcr{1}{consumption}.}
\omcr{2}{This work makes the following major contributions:}

\begin{itemize}
    \item We perform the first rigorous experimental characterization of High
    Bandwidth Memory (HBM) power consumption using a large-scale testing setup
    of \numStacks{} real HBM2 \omcr{1}{chip}s.
    \item We comprehensively analyze empirical HBM2 current measurements to
    demonstrate how power consumption is fundamentally sensitive to temperature,
    data rate, accessed HBM2 structures, and transferred data patterns. This
    analysis reveals that uniform power models fail to \omcr{1}{accurately}
    capture the \omcr{1}{real} power \omcr{1}{consumption} distribution across
    HBM2 bank groups and banks.
    \item We demonstrate that standard JEDEC-specified current measurement loops fail
    to capture worst-case HBM2 power consumption. By using our empirical
    insights into structural and data-pattern dependencies, we carefully
    engineer a new access pattern that consumes up to \param{35.4}\% more power than
    the most power-consuming standard current measurement loop (IDD4R).
    \item \microrevcommon{We develop and evaluate a new, accurate HBM power
    model (built on DRAMPower~\cite{steiner2025drampower5}) that captures both
    structural and data-pattern-dependent power variation and is substantially
    more accurate than state-of-the-art HBM power models. We validate our model
    against real HBM3E GPU memory power \omcr{1}{consumption measurements} and
    use it to project the read energy of future HBM4 \atbcr{1}{\omcr{1}{chip}s}.
    \omcr{1}{We open source our model as part of
    Ramulator2.1~\cite{luo2026ramulator21, luo2024ramulator, ramulator2github,
    kim2016ramulator, bostanci2026cleaning} and
    independently~\cite{hbmpower-github}.}}
\end{itemize}
\section{Background}
\label{sec:background}

We provide HBM2 background \omcr{3}{to ease understanding}.

\subsection{HBM2 Organization}

\begin{figure}[!ht]
\centering
\includegraphics[width=1.0\linewidth]{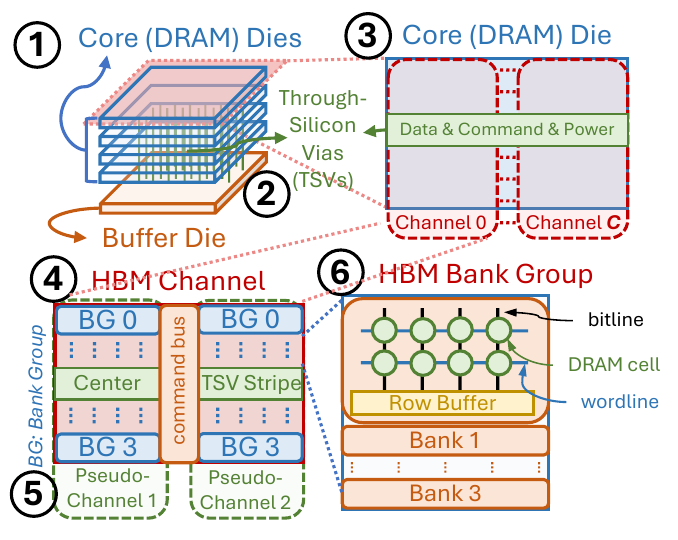}
\caption{HBM2 \atbcr{1}{chip}, die, channel, pseudo-channel, bank group, and
bank organization.}
\label{fig:hbm_stack_die_channel}
\end{figure}

\figref{fig:hbm_stack_die_channel} shows the organization of an HBM
\atbcr{1}{chip}~\cite{oconnor2021thesis,kim2022low,oh2020hbm2e,lee2020hbm2e,
lee2014high,cho2018hbm2,ryu2023hbm3,chun2020hbm2e,adhinarayanan2025folded,
lee2016simultaneous}.
\copied{The memory controller (not shown) communicates with \atbcr{1}{a chip}
(\dingOne{}) using the HBM2 interface via the \microlabel{E2}\emph{buffer
die}.}\footnote{\microrevc{Each HBM2 channel operates independently.
\atbcr{2}{An HBM-based system could implement} one memory controller per HBM2
channel \atbcr{2}{or \omcr{3}{one memory controller} per HBM2 pseudo-channel}.}}
\copied{An HBM2 \atbcr{1}{chip} contains \atbcr{2}{one or} multiple \emph{core
(DRAM) dies} stacked on top of the buffer die and connected and powered using
\emph{through-silicon vias (TSVs)} in a \emph{\omcr{2}{center} TSV stripe}
(\dingTwo{}). Each core die comprises \atbcr{2}{$1\leq{}C\leq{}8$} \emph{HBM2
channels} that operate independently (\dingThree{}). An HBM2 channel
(\dingFour{}) contains \atbcr{2}{two} \emph{pseudo-channels} and each
pseudo-channel has \atbcr{2}{four} \emph{bank groups} (\dingFive{}). A DRAM bank
group (\dingSix{}) comprises \atbcr{2}{four} DRAM \emph{banks}, each of which
has millions of \emph{DRAM cells} that are laid \atbcr{3}{out} in a
two-dimensional array of \emph{wordlines} and \emph{bitlines}.}
\microlabel{B2}\microrevb{DRAM cells are typically partitioned into multiple
DRAM subarrays~\cite{kim2012case, seshadri2013rowclone, chang2014improving,
yaglikci2023hira} (not shown in the figure). The cells in a subarray share a row
of \emph{sense amplifiers}, commonly referred to as a \emph{row buffer}; a row
of sense amplifiers is typically shared between two adjacent subarrays.}

\omcrcomment{2}{Can we have a die photo or real picture? give ranges for number of each structure (why bgs hardcoded???).}
\atbcrcomment{2}{hbm2 has 4 bgs and 4 banks in each bg}

\subsection{HBM2 Operation}

\microrevb{To access data in a DRAM chip, the memory controller issues the
following sequence of commands. First, the controller issues an activate (ACT)
command targeting a DRAM row. The row decoder asserts the row's
\microlabel{B2}wordline, which connects every cell in the row to its bitline;
the sense amplifiers then sense and amplify the resulting bitline perturbations,
latching the \emph{entire row} into the row buffer. Second, to read from or
write to a particular column of the activated row, the controller issues a
\omcr{2}{read (RD)} or \omcr{2}{a write (WR)} command. A RD command transfers
the addressed column's data from the sense amplifiers, over the bank's internal
data bus and the HBM2 \atbcr{1}{chip}'s I/O circuitry, to the memory controller.
A WR command drives data from the memory controller into the sense amplifiers
for the addressed column. \omcr{2}{When} accesses to the open row are complete,
the memory controller issues a precharge (PRE) command. PRE deasserts the
wordline and precharges the bitlines and sense amplifiers back to the reference
voltage.}

\noindent
\textbf{DRAM Periodic Refresh.}
\copied{Since a DRAM cell stores data as charge in its capacitor and the
capacitor naturally loses charge over time~\cite{liu2013experimental}, the
capacitor must be periodically refreshed to prevent data
corruption~\cite{jedechbm2,liu2012raidr,liu2013experimental}. Consequently, the
memory controller should issue a refresh (REF) command on average every
\SI{3.9}{\micro\second}~\cite{jedechbm2} ($t_{REFI}$), such that every DRAM cell
is refreshed once \atbcr{3}{within} a fixed refresh window (e.g.,
\SI{32}{\milli\second}).}
\section{Motivation}

Literature lacks empirical, fine-grained data on HBM power consumption. This
causes two important problems in HBM-based system design, research, and
evaluation. First, researchers {cannot} validate, tune, or improve the accuracy
of architectural power models. Second, HBM current values provided to system
designers by HBM vendors {cannot} be validated.\footnote{We expect that system
vendors \omcr{2}{could} receive IDD values from HBM vendors and that the \omcr{2}{receipt and} use
of these current values \nb{are} 
likely governed by strict non-disclosure agreements. However, we {cannot}
confirm that system designers receive accurate IDD values, and there are
\textit{no} publicly-available vendor-provided HBM IDD values today.} We
elaborate on these two problems.

\noindent
\textbf{\omcr{2}{Accurate} HBM Power Model\omcr{2}{ing}.} System designers and
researchers use architectural power models to estimate and budget for HBM power,
evaluate thermal performance, and evaluate system
throughput~\cite{steiner2025drampower5, vogelsang2010understanding,
ghose2018what, li2009mcpat, fan2001memory, micron2007power,stow2019power,
chen2012cacti,poddar2017mesap,park20183ddate,
weis2016dramspec,thoziyoor2008comprehensive,wang2005dramsim,rosenfeld2011dramsim2}.
Since DRAM technology \omcr{2}{(and especially HBM technology)} is highly
proprietary and considered a trade
secret~\cite{hassan2024self,patel2024rethinking}, low-level details about DRAM
architecture and technology parameters are \emph{not} publicly available. Thus,
architectural power models rely on reverse\nb{-}engineered or estimated DRAM
technology parameters. Therefore, architectural power models and technology
parameters must be continuously updated\nbcomment{or these models should be
configured using this data? I am not convinced about continuously update part}
using empirical data to yield accurate power estimations.
\omcr{2}{Unfortunately,} literature lacks fine-grained data on HBM power
consumption. The lack of data prevents \omcr{2}{researchers and designers} from
\omcr{2}{understanding} and improving the fidelity of architectural HBM power
models. HBM vendors may provide IDD values to a system
designer~\cite{micron2007power}. However, these values are provided under strict
non-disclosure agreements\nbcomment{do we need this? isnt it enough that these
are not publicly available? seems like we are guessing this happens but we dont
know for sure} and are \emph{not} publicly
available~\cite{xilinxForumPost}.\omcrcomment{2}{are these correct
references?}\atbcrcomment{2}{yes. 91 is micron technical note for power
calculation. it includes idd values} Open\omcr{2}{ly available} empirical HBM
power measurements can significantly improve the fidelity of architectural HBM
power models, enabling more accurate system power estimations. \nbcomment{I buy
this argument but isnt this paper provide first the analysis and maybe the data
as an artifact later? dont we need to argue for the analysis instead of the
data?}

\noindent
\textbf{Validating Vendor-Provided Current Measurements.}
Inaccurate HBM power estimations and inadequate HBM power budgeting could lead
to severe system robustness and performance issues. System designers rely on
vendor-provided IDD values in early prototyping stages to comprehensively
evaluate power characteristics, e.g., design the power delivery network, and
make design decisions trading off power for other important performance metrics.
However, a prior work\omcr{2}{, the VAMPIRE paper~\cite{ghose2018what},} has
shown that relying solely on vendor datasheets for DDR3L DRAM leads to
significant discrepancy in predicted vs measured DRAM power
\omcr{2}{consumption}. 
Inaccurate IDD values could lead to misguided power budgeting in the design of
an HBM-based computing system and incur important robustness and performance
implications, such as job failures~\cite{kokolis2025revisiting},
\atbcr{2}{thermal throttling}~\atbcr{2}{\cite{lin2007thermal, liu2011hardware,
eckert2014thermal, son2023thermal, paul2015harmonia, li2020dramsim3}}, or power
delivery failures~\atbcr{2}{\cite{shevgoor2013quantifying, fan2007power,
wu2016dynamo, kokolis2025revisiting}} when real workloads \omcr{1}{(such as
memory-intensive AI/ML training or serving)} trigger high-power
states~\atbcr{1}{\cite{ganesan2010sympo, fan2007power, wu2016dynamo,
hajyahya2022agilewatts, hajyahya2021ichannels, antoniou2022agilepkgc}}.
\omcr{2}{Openly available empirical} HBM power measurements would provide a
strong foundation for system designers to \atbcr{2}{build accurate HBM power
models}, optimize thermal management strategies, and make design
trade-offs much earlier in the system development lifecycle.

\omcr{2}{\textbf{Our goal}} is to \omcr{3}{1)} \omcr{2}{provide accurate and
openly available HBM power consumption measurements \omcr{3}{from real HBM chips
and 2) build a new HBM power model based on such real chip measurements.} To
\omcr{4}{this} end, we conduct \atbcr{3}{the} first detailed experimental characterization of power
consumption in real HBM2 chips and} demonstrate how HBM2 power
\atbcr{2}{consumption} varies 1)~across standard IDD current measurement loops,
2)~across 3D-stacked HBM2 components, and 3)~with carefully constructed access
and data patterns. \omcr{2}{These measurements provide} the foundational basis
for \atbcr{2}{constructing accurate} HBM power models, \omcr{3}{including the
new one we introduce,} and for system designers to independently
\omcr{1}{validate} vendor-provided HBM current values and accelerate HBM-based
system development lifecycle. 
\section{Experimental Methodology}
\label{sec:methodology}

We describe our HBM2 DRAM testing infrastructure and the HBM2 DRAM
\atbcr{1}{chip}s tested. \atbcr{1}{We publicly release our testing
infrastructure \omcr{2}{as part of DRAM
Bender~\cite{safari-drambender,olgun2023drambender} and
independently~\cite{hbmpower-github}}.}

\begin{figure}[!ht]
  \centering
  \begin{subfigure}{0.9\linewidth}
    \centering
    \includegraphics[width=\linewidth]{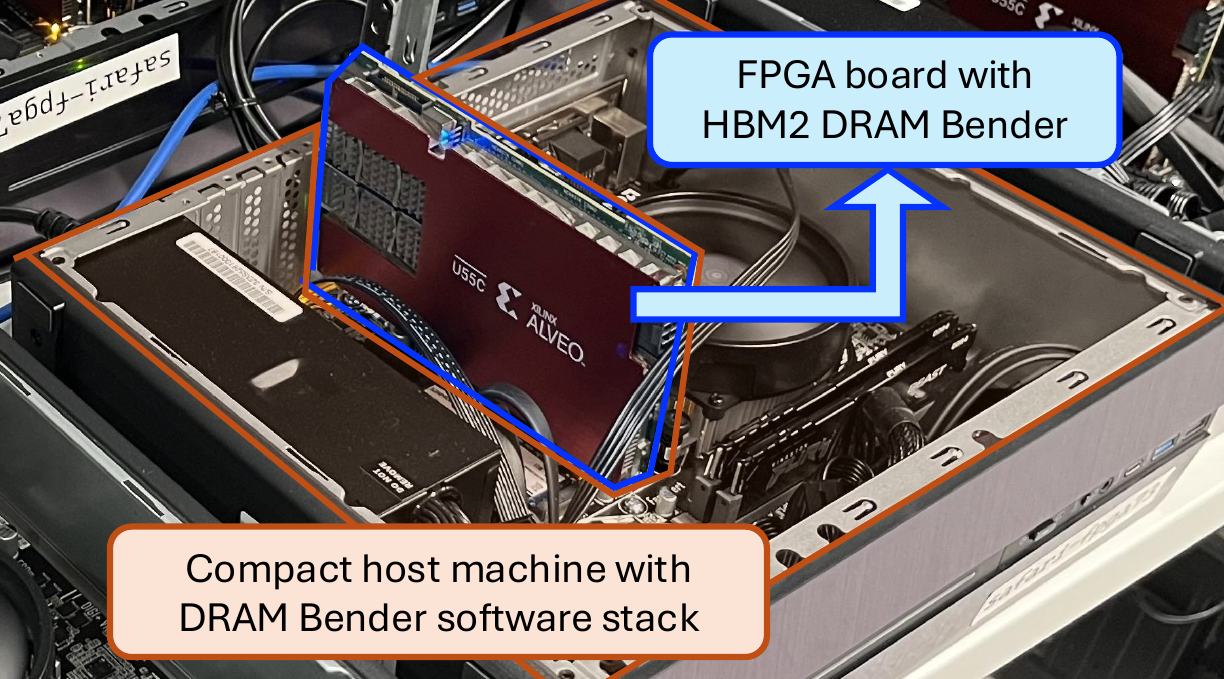}
    \caption{\omcr{2}{Our} HBM2 testing and power measurement setup. Adapted
  from~\cite{olgun2026laboratory}.}
    \label{fig:testing_setup_board}
  \end{subfigure}\\[4pt]
  \begin{subfigure}{0.9\linewidth}
    \centering
    \includegraphics[width=\linewidth]{figures/dram_bender_laboratory.pdf}
    \caption{Our laboratory for understanding DRAM~\cite{olgun2026laboratory}.}
    \label{fig:testing_setup_lab}
  \end{subfigure}
  \caption{\atbcr{1}{Our HBM2 testing infrastructure: (a)~HBM2 testing and power
  measurement setup and (b)~our laboratory for understanding DRAM \omcr{3}{with
  120+ setups for testing HBM2 and DDR4 chips}.}}
  \label{fig:testing_setup}
\end{figure}

\noindent
\textbf{HBM2 DRAM Testing Setup.} We build our testing setup on DRAM
Bender~\cite{olgun2023drambender, safari-drambender, olgun2026laboratory,
olgun2023hbm, olgun2024read}, an open source FPGA-based DRAM testing
infrastructure (which builds on SoftMC~\cite{hassan2017softmc,
softmc-safarigithub}).\footnote{We augment the open source version
of DRAM Bender~\cite{safari-drambender} with two features to enable our
experiments. First, DRAM Bender's open source version can \emph{only} issue DRAM
commands to one HBM2 channel at a time, and switching between HBM2 channels
incurs a non-negligible latency cost. We add the ability to use an arbitrary
combination of HBM2 channels at a time. Second, the open source version does not
support the Alveo U55C board. We extend DRAM Bender to i)~support the Alveo U55C
board and ii)~read HBM2 current sensors from the \microrevb{U55C board's}
satellite controller~\cite{alveo-u55c-manual} \omcr{2}{and open source the
extended version in the DRAM Bender repository~\cite{safari-drambender} and
independently~\cite{hbmpower-github}}.} \atbcr{1}{\figref{fig:testing_setup} shows a picture of
our testing setup \atbcr{1}{(\figref{fig:testing_setup}a)} and our laboratory
for understanding DRAM \omcr{2}{where we scaled our experiments across 18
machines} \atbcr{1}{(\figref{fig:testing_setup}b)}.} \omcr{2}{Our testing} setup
\atbcr{1}{(\figref{fig:testing_setup}a)} consists of two main components: 1)~a
host machine that generates the test program and collects experimental results,
and 2)~an FPGA development board Alveo U55C~\cite{alveo-u55-c}, programmed with
our extended version of DRAM Bender, \atbcr{3}{which we use} to execute test
programs and analyze experimental data.

\atbcrcomment{2}{Top picture adapted from ics paper}
\noindent
\textbf{Tested HBM2 \atbcr{1}{Chip}s.} We test \numStacks{} HBM2
\atbcr{1}{chip}s in \numFPGAs{} FPGA boards (each of which has two HBM2
\microlabel{C2}\atbcr{1}{chip}s).\footnote{\microrevc{We do not have conclusive
insight into tested HBM2 vendor information and part numbers. This information
is typically available to the FPGA system designer via the HBM2
\atbcr{1}{chip}'s IEEE1500 interface~\cite{jedechbm2}. Unfortunately, the FPGA
system designer does not provide user logic (our testing setup design) access to
the IEEE1500 interface. Moreover, the HBM2 vendor information is kept secret by
the FPGA board manufacturer due to vendor non-disclosure
agreements~\cite{xilinxForumPost}. These chips are \atbcr{2}{likely}
manufactured by Samsung because a whitepaper published by the FPGA
vendor~\cite{xilinx-samsung-hbm} indicates that their HBM-based FPGA boards use
Samsung HBM2 \atbcr{1}{chip}s.}} Tested HBM2 \atbcr{1}{chip}s all have a density
of \SI{8}{\giga\byte}, 8 channels, 2 pseudo-channels, 4 bank groups, 4 banks,
and 8K DRAM rows. The HBM2 \atbcr{1}{chip} is composed of two ranks of
4-high\atbcrcomment{2}{not defined in background} DRAM dies (i.e., the
\atbcr{1}{chip} is 8 \microrevcommon{DRAM dies} high)~\cite{axi-hbm-ip}. A
single \atbcr{2}{stack} ID (SID) bit is used to select a rank of 4 HBM2 dies. In
our experiments, we set SID = 0 and test the first rank of 4 DRAM
dies\atbcr{2}{, and the tested HBM2 chips are clocked at 600 MHz (1200 MT/s data
rate)} \microlabel{B3}\microrevb{unless stated
otherwise}.\atbcr{3}{\footnote{\atbcr{3}{The U55C FPGA board supports HBM2 clock
frequencies between 225~MHz (450~MT/s)~--~900~MHz (1800~MT/s) (inclusive
range)~\cite{axi-hbm-ip}. We encounter FPGA design timing closure problems in
our testing setup for clock frequencies higher than 600 MHz. Thus, the highest
clock frequency we use is 600 MHz (1200 MT/s).}}}

\subsection{Measuring HBM2 Power} 

To enable fine-grained power measurements, we integrate the AMD Alveo Card
Management Solution (CMS) subsystem~\cite{amd-alveo-cms-manual} into DRAM
Bender. The CMS subsystem is a specialized management logic that communicates
with a \microrevb{satellite controller} to periodically collect telemetry data
from various sensors, including HBM2 current. We extend DRAM Bender to allow the
host machine to access CMS registers that store telemetry data. Using the
extended version, we retrieve average \atbcr{2}{HBM2} current consumption
values.

\noindent
\textbf{Core (VDDC), I/O (VDDQ), and Wordline (VPP) Current.}
The HBM2 standard~\cite{jedechbm2} specifies three \atbcr{2}{power} supply
rails, VDDC, VDDQ, and VPP.
\microlabel{A2}\microreva{Table~\ref{tab:nomenclature} summarizes the current
(IDD) and voltage (supply rail) nomenclature we use throughout this paper.}
\atbcr{2}{T}he CMS subsystem on our FPGA platform provides \atbcr{2}{an
aggregated} current value for the 1.2 V \atbcr{2}{VDDC and VDDQ rails
\atbcr{2}{through a single CMS register}~\cite{amd-ultrascale-plus-datasheet}.}

\begin{table}[!ht]
\centering
\caption{\microreva{Summary of the HBM2 current (IDD) and voltage (supply rail)
nomenclature used throughout this paper.}}
\label{tab:nomenclature}
\resizebox{\linewidth}{!}{%
\microreva{
\begin{tabular}{ll}
\hline
Symbol & Description \\
\hline
VDDC  & Core supply rail (powers the DRAM core and peripheral logic) \\
VDDQ  & I/O supply rail (powers the data I/O interface) \\
VPP   & Pumped supply rail (high voltage used for row/wordline activation) \\
\hline
IDD0  & Operating current: repeated activate--precharge to one bank \\
IDD2N & Precharge standby (idle) current: all banks precharged \\
IDD3N & Active standby (idle) current: banks left active \\
IDD4R & Burst read current \\
IDD4W & Burst write current \\
IDD5B & Burst refresh current \\
\hline
\end{tabular}%
}
}
\end{table}

\subsection{HBM2 Current Measurement Loops}
\label{sec:hbm2-current-measurement-loops}

We develop several HBM2 command sequences, each of which we call a current
measurement loop, to measure current (IDD values) on \microreva{VDDC and VDDQ
rails}. There are four types of IDD values that we measure: (1) \emph{idle}:
IDD2N and IDD3N, (2) \emph{activate and precharge}: IDD0, (3) \emph{read and
write}: IDD4R and IDD4W, and (4) \emph{refresh}: IDD5B. The HBM2 standard
specifies the IDD0, IDD4R, and IDD4W current measurement loops in detail; we use
these loops from the HBM2 standard~\cite{jedechbm2} \atbcr{3}{with minor
modifications} 
\atbcr{2}{to independently measure power consumed by
each pseudo-channel}. We \atbcr{2}{use} other relevant DRAM
standards~\cite{jedecddr5c,jedechbm3} \omcr{3}{and VAMPIRE~\cite{ghose2018what}}
to construct the IDD2N, IDD3N, and IDD5B current measurement loops.
\microlabel{B6}\microrevb{Each loop isolates a single operation by repeatedly
issuing only that operation's command(s). Table~\ref{tab:idd-loops} summarizes
the command(s) issued and the DRAM bank state during each loop.}

\begin{table}[!ht]
\centering
\caption{\microrevb{Command(s) issued and DRAM bank state during each HBM2
current measurement loop.}}
\label{tab:idd-loops}
\resizebox{\linewidth}{!}{%
\microrevb{
\begin{tabular}{lll}
\hline
Loop & DRAM bank state during loop & Command(s) issued \\
\hline
IDD2N   & All banks precharged (closed)        & NOP only \\
IDD3N1  & \omcr{2}{One} bank open, all others precharged   & NOP only \\
IDD3N16 & All 16 banks open                    & NOP only \\
IDD0    & One bank, repeatedly opened/closed   & ACT, PRE \\
IDD4R   & Rows activated beforehand, kept open & Back-to-back RD \\
IDD4W   & Rows activated beforehand, kept open & Back-to-back WR \\
IDD5B   & All banks precharged (closed)        & Back-to-back REF \\
\hline
\end{tabular}%
}
}
\end{table}

\subsection{HBM2 Power Testing Methodology}
\label{sec:hbm2-power-testing-methodology}

\begin{figure}[!ht]
  \centering
  \includegraphics[width=\linewidth]{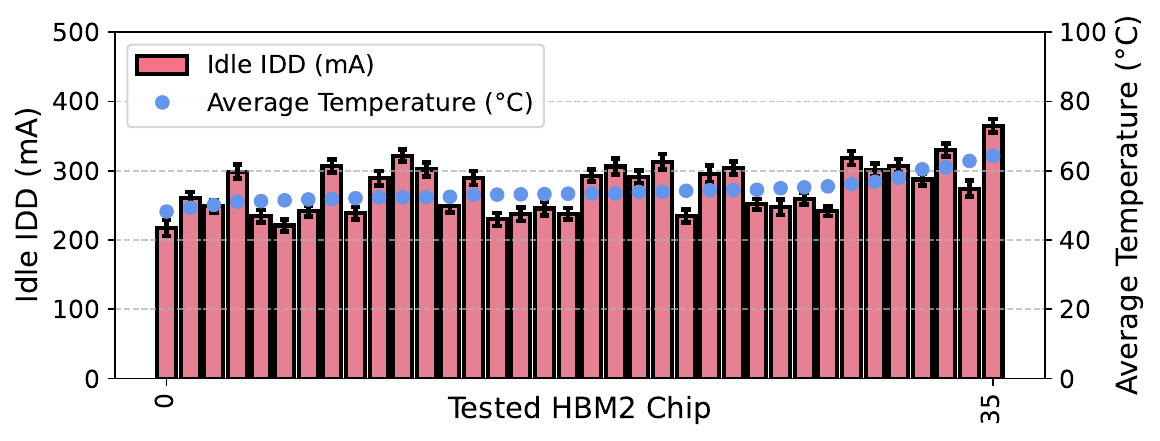}
  \caption{Idle HBM2 VDDC \atbcr{3}{and VDDQ} current and temperature across
  \omcr{2}{36} tested HBM2 \atbcr{1}{chip}s in increasing average temperature
  order. The large bars show average and the smaller error bars show standard
  deviation across 3600 measurements.} 
  \label{fig:idle-current-noise}  
\end{figure}

To calibrate the current sensor in each FPGA board, we \atbcr{2}{compute}
the average IDD across 100 seconds of \atbcr{2}{repeated} current sensor
measurements while all HBM2 \atbcr{1}{chip}s are powered off
\atbcr{2}{(\emph{power-off current})}. We execute each current measurement loop
(Table~\ref{tab:idd-loops}) repeatedly and record average current as described
in the HBM2 standard specification~\cite{jedechbm2}. We \atbcr{4}{subtract} the power-off
current from each average current value. \figref{fig:idle-current-noise} shows
how measured VDDC \atbcr{3}{and VDDQ} \emph{current} (IDD) varies with time
(3600 measurements across one hour of testing) while HBM2 is idle (i.e., DRAM
Bender is executing the IDD2N current measurement loop) and HBM2 temperature is
stable (within $\pm\SI{1}{\celsius}$ of the depicted average temperature). We
plot the mean and standard deviation across 3600 IDD measurements.

\observation{Idle current and temperature vary across the tested HBM2 \atbcr{1}{chip}s.}

\begin{figure}[!ht]
  \centering
  \includegraphics[width=\linewidth]{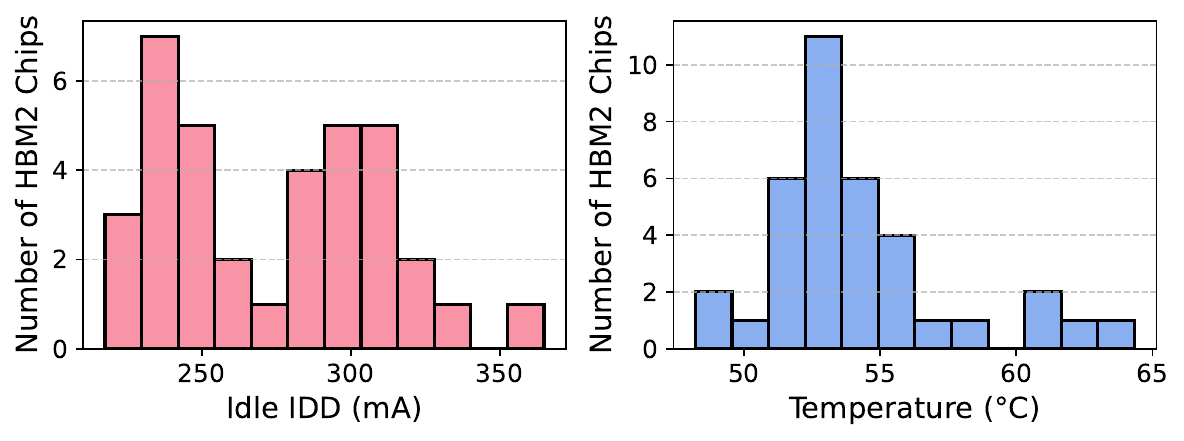}
  \caption{\microrevb{Distribution of mean idle HBM2 VDDC \atbcr{3}{and VDDQ}
  current (left) and mean idle temperature (right) across the \omcr{2}{36}
  tested HBM2 \atbcr{1}{chip}s. Each \atbcr{1}{chip} contributes its mean across
  3600 measurements.}}
  \label{fig:idle-current-histogram}
\end{figure}

First, we observe that the minimum and maximum mean current across all tested
HBM2 \atbcr{1}{chip}s are \param{\SI{217.5}{\milli\ampere}} and
\param{\SI{364.8}{\milli\ampere}}, respectively.
\microlabel{B4}\microrevb{\figref{fig:idle-current-histogram} shows the
distribution of mean idle current and mean idle temperature across the tested
HBM2 \atbcr{1}{chip}s as histograms.} Second, the average idle temperature
ranges between \param{\SI{48}{\celsius}} and \param{\SI{64}{\celsius}} across
tested HBM2 \atbcr{1}{chip}s.\footnote{The Alveo U55C board that we use in our
testing platform is passively cooled in a temperature-controlled room at room
temperature. Since the board does not have a dynamic cooling setup to target a
specific HBM2 temperature, tested HBM2 \atbcr{1}{chip}s operate at varying idle
temperatures.} Third, temperature variation solely does not explain idle current
variation across tested HBM2 \atbcr{1}{chip}s: we observe that current does not
consistently increase with temperature. We study the effect of temperature on
current in detail in~\secref{sec:temperature-sensitivity}. We hypothesize that
the idle current variation across tested HBM2 \atbcr{1}{chip}s is caused by
\atbcr{2}{manufacturing-induced} process variation~\cite{mehrotra2001modeling,
lee2017design} in HBM2 dies\atbcr{2}{, packaging, FPGAs and current sensor
integration.} \atbcr{2}{In particular, different HBM2 \atbcr{1}{chip}s may
\atbcr{2}{exhibit} manufacturing-induced \atbcr{3}{variation across} transistor
characteristics that leads to systematically high-clock-rate or low-clock-rate
and higher- or lower-current DRAM
dies}~\cite{al2003analyzing}.\atbcrcomment{3}{I don't think design-induced
variation would explain this current draw difference across chips.}
Unfortunately, we \emph{cannot} \atbcr{3}{definitively} confirm these hypotheses
because \omcr{2}{FPGA platform information is insufficient and
proprietary~\cite{xilinxForumPost}}.

\observation{Idle current for a \omcr{2}{given} HBM2 \atbcr{1}{chip} does not
significantly vary \atbcr{3}{across repeated} current measurements.}

For an HBM2 \atbcr{1}{chip}, current varies by a small amount
\atbcr{2}{(depicted by the small error bars in~\figref{fig:idle-current-noise})}
across measurements: standard deviation in current measurements is
\param{\SI{9.6}{\milli\ampere}} on average across all tested HBM2
\atbcr{1}{chip}s. 
\section{Characterization Results}
\label{sec:characterization-all}
\atbcrcomment{2}{These are all characterization results. Should we say so?}

We perform the first large-scale characterization of HBM2 power
\omcr{2}{consumption} using fine-grained current measurement loops in
\numStacks{} HBM2 \atbcr{1}{chip}s. Our goal is \param{twofold}. First, we would
like to present our empirical HBM2 current measurements for the broader research
community to build on and use. \param{Second, we would like to form the
empirical basis for \omcr{2}{a new and more accurate} power model we develop and
evaluate in \secref{sec:hbm2-power-model}.}

\subsection{Standard IDD Current Measurements}
\label{sec:standard-idd}

We measure IDD current using the \param{six} current measurement loops described
in \secref{sec:hbm2-current-measurement-loops}. We execute each loop for 30
minutes. \atbcr{2}{We observe that each test loop induces a different
temperature profile in an HBM2 \atbcr{1}{chip}. We see that temperature
stabilizes after $\approx{}\!$20 minutes of execution of each test. In the last
minute of testing, temperature varies by \omcr{3}{at most} \SI{1}{\celsius}. We} report the
average current measured by the current sensor \atbcr{2}{every second} in the
last minute of the loop \atbcr{2}{(i.e., the last 60 measurements)}.

\begin{figure}[!ht]
  \centering
  \includegraphics[width=1\linewidth]{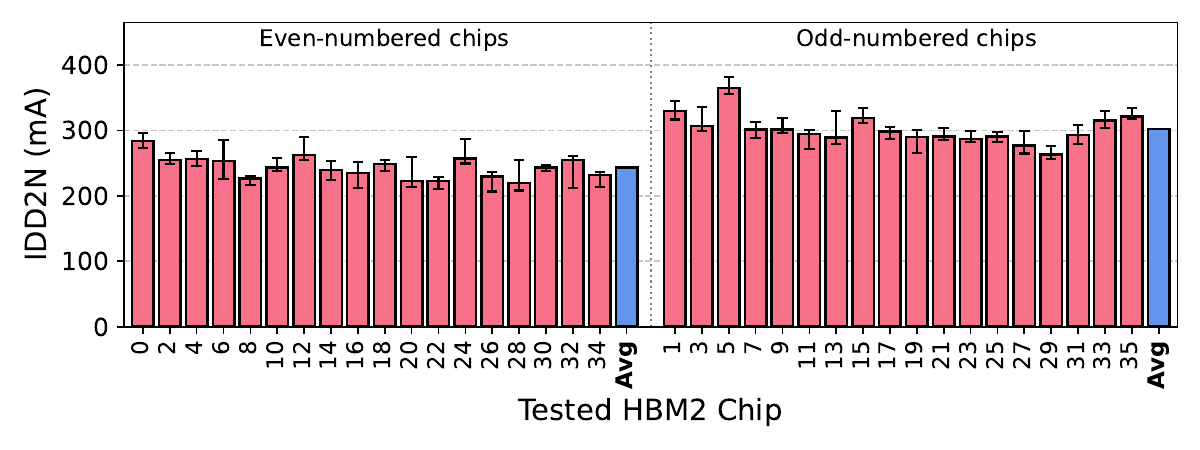}
  \caption{\microrevb{\atbcr{2}{Measured} IDD2N current across tested HBM2
  \atbcr{1}{chip}s, \atbcr{2}{grouped} by \atbcr{1}{the chip's position
  \omcr{2}{in} the package} \atbcr{2}{(even vs odd)}. \atbcr{2}{Bars show the
  average and error bars \omcr{3}{show} the full range of the last 60
  measurements.}}}
  \label{fig:idd2n_current_across_stacks}
\end{figure}

\noindent
\textbf{IDD2N.} We start by measuring the precharge standby (i.e., idle)
current.\footnote{We show IDD2N results in
\secref{sec:hbm2-power-testing-methodology} to demonstrate the variation in
current across tested HBM2 \atbcr{1}{chip}s and within each HBM2
\atbcr{1}{chip}. We present IDD2N results again, using \atbcr{2}{a current
measurement methodology that filters out temperature-induced current variation
(described earlier in this section)}, for completeness \atbcr{2}{and
consistency}.} For IDD2N, all DRAM banks remain closed and the HBM2 clock is
enabled throughout the measurement loop. The measurement loop for IDD2N solely
issues NOP commands to the HBM2 \atbcr{1}{chip}.
\microlabel{B4}\microrevb{\figref{fig:idd2n_current_across_stacks} shows the
distribution of measured IDD2N current across all tested HBM2 \atbcr{1}{chip}s
\atbcr{2}{as a bar plot, where each bar shows the average and the error bar the
full range of the last 60 current measurements}, with \atbcr{1}{chip}s
\atbcr{2}{grouped} by \atbcr{2}{their position in the
FPGA package (each tested FPGA has two HBM2 chips) as even-numbered and
odd-numbered chips, and the last bar in each group showing the group average}.}

First, we observe that IDD2N varies across tested HBM2 \atbcr{1}{chip}s. The average and
standard deviation across all tested HBM2 \atbcr{1}{chip}s for IDD2N are
\atbcr{1}{\param{\SI{272.7}{\milli\ampere}} and
\param{\SI{35.0}{\milli\ampere}}}, respectively. Second, odd-numbered HBM2
\atbcr{1}{chip}s tend to have higher current than even-numbered HBM2
\atbcr{3}{chips}. The average and standard deviation for IDD2N across
\emph{odd-numbered} HBM2 \atbcr{1}{chip}s are \param{\SI{301.9}{\milli\ampere}}
and \param{\SI{22.5}{\milli\ampere}}, respectively. Meanwhile, the average and
standard deviation for IDD2N across \emph{even-numbered} HBM2 \atbcr{1}{chip}s
are \atbcr{1}{\param{\SI{243.6}{\milli\ampere}} and
\param{\SI{17.0}{\milli\ampere}}},
respectively.

\microrevc{Whether a \atbcr{1}{chip} sits at an odd- or even-numbered
position\microlabel{C1} is determined by the FPGA board layout. Therefore, the
odd/even current difference cannot originate from the HBM2 \atbcr{1}{chip}s. We
attribute the odd/even current difference to systematic,
\emph{position-dependent} variation introduced by the FPGA package and board
integration process~\cite{mehrotra2001modeling}. Odd- and even-numbered
\atbcr{1}{chip}s occupy different physical positions on the
FPGA~\cite{alveo-u55-c}, and these positions can differ in trace lengths, power
delivery network electrical characteristics, silicon interposer routing between
the FPGA die and the HBM2 \atbcr{1}{chip}, and current-sensor integration. We
observe the same odd/even pattern across \atbcr{3}{other} subsequent IDD
measurements (\omcr{2}{\secref{sec:characterization-all}}).}

\omcrcomment{3}{can we validate this and control it somehow?}
\atbcrcomment{3}{we need to carefully redo this analysis and see. the difference
 seems to vanish for activate power and reappears with idd4r/w} 
 
\take{IDD2N varies relatively widely between odd- and even-numbered HBM2
\atbcr{1}{chip}s. \microrevc{Because the odd/even grouping reflects board
position rather than any property of the \atbcr{1}{chip}s, this variation is not
caused by the HBM2 \atbcr{1}{chip}s themselves, but by systematic,
position-dependent process variation in the FPGA package and board
integration.}}

\begin{figure}[!ht]
  \centering
  \includegraphics[width=1\linewidth]{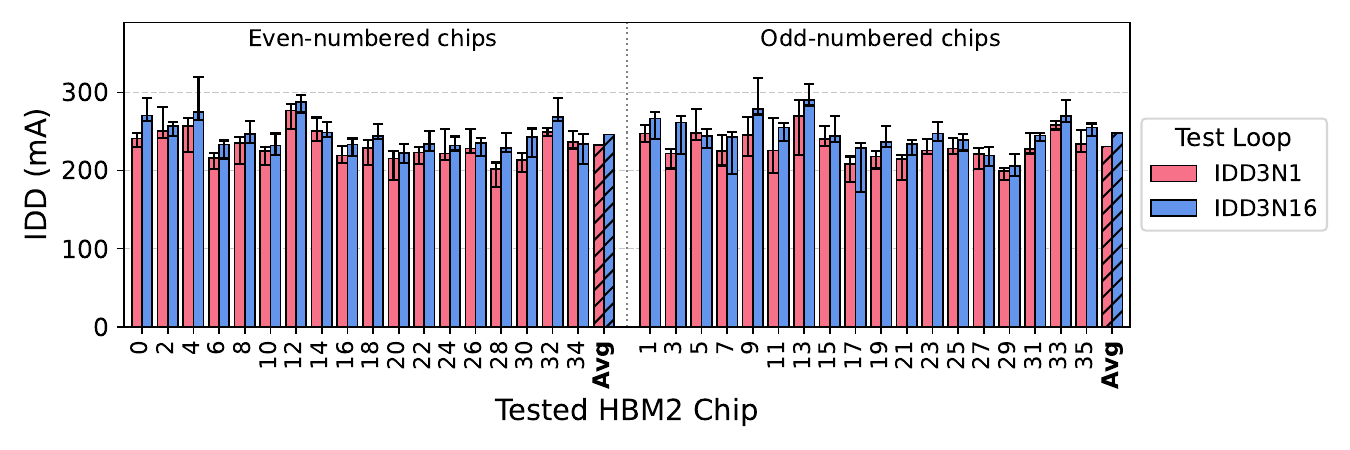}
  \caption{\microrevb{\atbcr{2}{Measured} IDD3N1 and IDD3N16 currents across
  tested HBM2 \atbcr{1}{chip}s, \atbcr{2}{grouped} by \atbcr{1}{the chip's
  position \omcr{2}{in} the package} \atbcr{2}{(even vs odd)}. \atbcr{2}{Bars
  show the average and error bars \omcr{3}{show} the full range of the last 60 measurements.}}}
  \label{fig:idd3n_current_across_stacks} 
\end{figure}

\noindent
\textbf{IDD3N.} Next, we measure active standby current. For IDD3N1 (defined as
IDD3N in~\cite{jedechbm2}), \emph{only} one DRAM bank has an active row (i.e.,
one bank is open) and all other DRAM banks remain precharged throughout the
measurement loop. For IDD3N16, all 16 DRAM banks have an active DRAM row (i.e.,
16 banks are open). The measurement loop for IDD3N solely issues NOP commands to
the HBM2 \atbcr{1}{chip}. \figref{fig:idd3n_current_across_stacks} shows IDD3N1 and
IDD3N16 currents over tested HBM2 \atbcr{1}{chip}s.

\observation{Active standby current increases with the number of open DRAM banks.}

We observe that as the number of open DRAM banks increases from 1 to 16, the
average IDD3N current increases by \param{6.5}\%. The average IDD3N1 and IDD3N16
currents across all tested \atbcr{1}{chip}s are \param{\SI{231.8}{\milli\ampere}}
and \param{\SI{246.9}{\milli\ampere}}, and the standard deviations for IDD3N1 and
IDD3N16 are \param{\SI{17.7}{\milli\ampere}} and
\param{\SI{19.1}{\milli\ampere}}, respectively. Opening more banks keeps more
sense amplifiers \omcr{2}{and} peripheral control \omcr{2}{\&} data circuitry
\omcr{2}{concurrently} active, which increases standby current. This result is
consistent with by prior work~\cite{jung2016new} on DDR3
DRAM.\atbcrcomment{2}{VAMPIRE does not study power vs \# of banks (it studies
bank power normalized to bank 0). This is data from one of the DRAMPower
papers.}

\begin{figure}[!ht]
  \centering
  \includegraphics[width=1\linewidth]{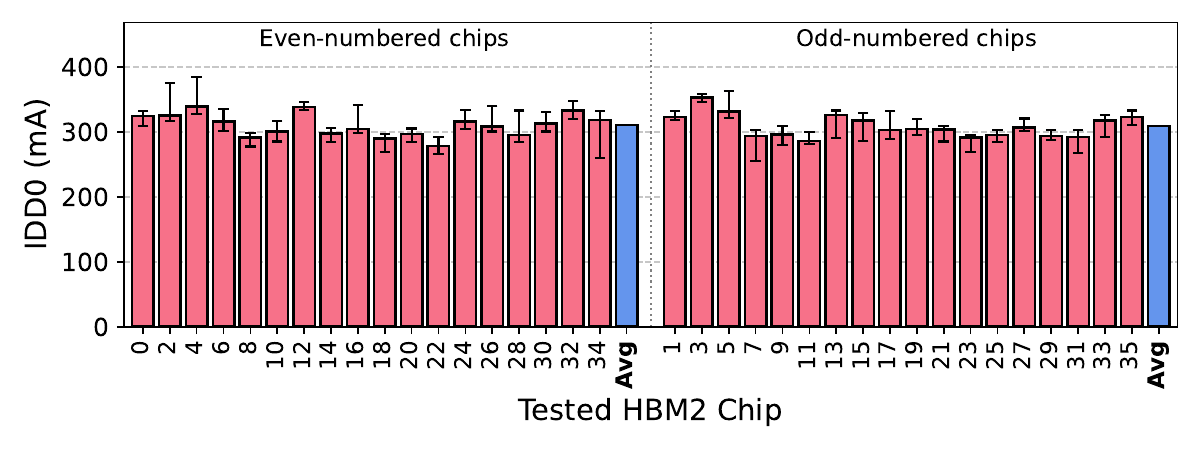}
  \caption{\microrevb{\atbcr{2}{Measured} IDD0 current across tested HBM2
  \atbcr{1}{chip}s, \atbcr{2}{grouped} by \atbcr{1}{the chip's position
  \omcr{2}{in} the package} \atbcr{2}{(even vs odd)}. \atbcr{2}{Bars show the
  average and error bars \omcr{3}{show} the full range of the last 60 measurements.}}}
  \label{fig:idd0_current_across_stacks}
\end{figure}

\noindent
\textbf{IDD0.} We measure the one bank activate-precharge current. For IDD0, we
repeatedly issue activate (ACT) and precharge (PRE) commands at the highest
allowed rate to \atbcr{2}{each DRAM bank in a round-robin
manner}.\atbcr{2}{\footnote{\atbcr{2}{We do not observe significant variation in
IDD0 across different tested banks and bank groups. However, read current
(IDD4R) significantly varies across tested banks and bank groups
(\secref{sec:structural-variation}).}}}
\figref{fig:idd0_current_across_stacks} shows IDD0
current across all tested HBM2 \atbcr{1}{chip}s.

\atbcr{2}{We observe that} the average and standard deviation for IDD0 current
across tested HBM2 \atbcr{1}{chip}s are {{\SI{309.7}{\milli\ampere}}} and
{{\SI{17.1}{\milli\ampere}}}, respectively. \atbcr{2}{We conclude that IDD0
current is slightly higher than IDD2N/IDD3N. This conclusion is consistent with
the results of a prior work~\cite{ghose2018what} on DDR3L DRAM.}

\begin{figure}[!ht]
  \centering
  \includegraphics[width=1\linewidth]{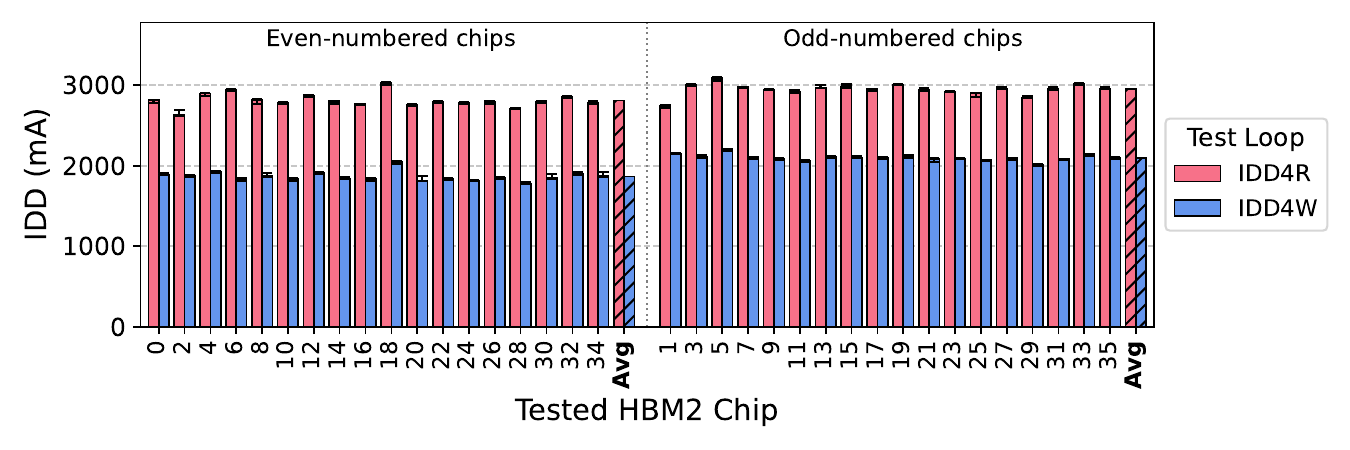}
  \caption{\microrevb{\atbcr{2}{Measured} IDD4R and IDD4W current across tested
  HBM2 \atbcr{1}{chip}s, \atbcr{2}{grouped} by \atbcr{1}{the chip's position
  \omcr{2}{in} the package} \atbcr{2}{(even vs odd)}. \atbcr{2}{Bars show the
  average and error bars \omcr{3}{show} the full range of the last 60 measurements.}}}
  \label{fig:idd4_current_across_stacks}
\end{figure}

\noindent
\textbf{IDD4R and IDD4W.} We measure read and write burst current. For IDD4R, we
repeatedly issue read (RD) commands to open rows across all 16 DRAM banks at the
highest allowed rate. For IDD4W, we repeatedly issue write (WR) commands to open
rows across all \omcr{2}{16} banks. The measurement loop, for both IDD4R and IDD4W, solely
issues back-to-back RD or WR commands to the HBM2 \atbcr{1}{chip}.
\figref{fig:idd4_current_across_stacks} shows IDD4R and IDD4W current across
all tested HBM2 \atbcr{1}{chip}s.

\observation{HBM2 read current is significantly higher than HBM2 write current.}

First, we observe that IDD4R is \atbcr{1}{\param{45.4}\%} higher than IDD4W.
\microlabel{B5}\microrevb{We attribute this difference to the fact that, on a
read, the HBM2 \atbcr{1}{chip}'s peripheral circuitry must drive the data across
the HBM2 channel to the memory controller, whereas on a write the memory
controller drives the channel and the DRAM \atbcr{1}{chip} only receives the
data. Thus, the read data and the write data travel the same distance, but are
driven by opposite ends of the channel, and the channel-driving current is drawn
from the HBM2 \atbcr{1}{chip}'s supply only for reads.} This observation is in
line with that of prior work on DDR3L DRAM chips~\cite{ghose2018what}. Second,
the average IDD4R and IDD4W across all tested \atbcr{1}{chip}s are
\param{\SI{2881.0}{\milli\ampere}} and \param{\SI{1981.9}{\milli\ampere}}, and
the standard deviations for IDD4R and IDD4W are \param{\SI{103.5}{\milli\ampere}}
and \param{\SI{126.6}{\milli\ampere}}, respectively. Third, we observe that
\omcr{2}{both} IDD4R and IDD4W vary across tested HBM2 \atbcr{1}{chip}s, with
odd-numbered \atbcr{1}{chip}s again exhibiting higher current than even-numbered
\atbcr{1}{chip}s.

The average IDD4R across all tested chips is \param{9.6}$\times{}$,
\param{11.4}$\times{}$, \param{10.7}$\times{}$, \param{8.3}$\times{}$,
\atbcr{1}{\param{45.4}}\%, and \param{3.1}$\times{}$ higher than the average
IDD2N, IDD3N1, IDD3N16, IDD0, IDD4W, and IDD5B across all tested chips,
respectively. This observation differs from a prior work's observation on planar
DDR3L DRAM chips, where the authors observe DRAM refresh operations (IDD5B) to
consume more current than any other DRAM operation~\cite{ghose2018what}. High
Bandwidth Memory typically offers very high bandwidth compared to planar DRAM
via many 3D-stacked channels. Thus, read current at high memory throughput is
higher in High Bandwidth Memory than in planar DRAM. \atbcr{2}{For example, the
VAMPIRE paper~\cite{ghose2018what} reports an average read current of
\SI{343.5}{\milli\ampere} for tested DDR3L modules from one vendor at a
\atbcr{3}{data rate} of 800 MT/s (approximately \SI{515}{\milli\ampere} at 1200
MT/s scaled linearly according to~\secref{sec:data-rate-sensitivity}), much
lower than the read current we observe (\SI{2881.0}{\milli\ampere}) on average
across all tested HBM2 chips at 1200 MT/s.} However, the tested HBM2
\atbcr{1}{chip}s are not much denser than a planar DRAM module. Since refresh
current scales with the number of DRAM cells (i.e., refresh current is a
function of DRAM density)~\cite{liu2012raidr}, the HBM2 refresh current observed
in this work, described in the following paragraph, is approximately the same as
planar DRAM refresh current observed in~\cite{ghose2018what}.

\take{HBM2 read and write operations consume significantly higher current (IDD)
than any other tested DRAM operation.}

\begin{figure}[!ht]
  \centering
  \includegraphics[width=1\linewidth]{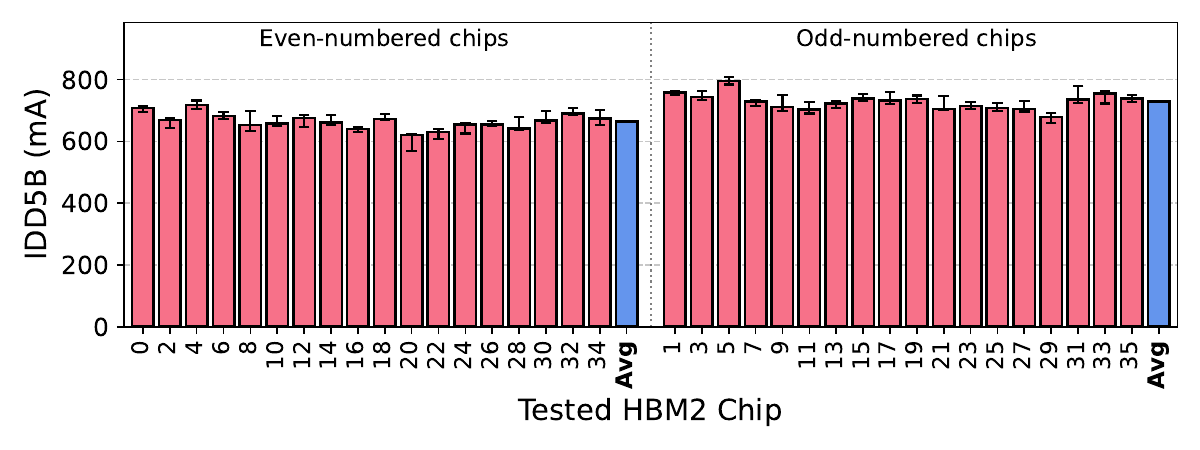}
  \caption{\microrevb{\atbcr{2}{Measured} IDD5B current across tested HBM2
  \atbcr{1}{chip}s, \atbcr{2}{grouped} by \atbcr{1}{the chip's position
  \omcr{2}{in} the package} \atbcr{2}{(even vs odd)}. \atbcr{2}{Bars show the
  average and error bars \omcr{3}{show} the full range of the last 60
  measurements.}}}
  \label{fig:idd5_current_across_stacks}
\end{figure}

\noindent
\textbf{IDD5B.} We measure all-bank refresh burst current. For IDD5B, we
issue refresh (REF) commands to all DRAM banks at the highest allowed rate
throughout the measurement loop. The measurement loop for IDD5B solely issues
back-to-back REF commands to the HBM2 \atbcr{1}{chip}.
\figref{fig:idd5_current_across_stacks} shows IDD5B current across all tested
HBM2 \atbcr{1}{chip}s.

We make two observations. First, we again observe that IDD5B current varies
across tested HBM2 \atbcr{1}{chip}s. Odd-numbered \atbcr{1}{chip}s tend to have higher current than
even-numbered \atbcr{1}{chip}s. Second, the average and standard deviation for IDD5B
across tested HBM2 \atbcr{1}{chip}s are \param{\SI{696.8}{\milli\ampere}} and
\param{\SI{40.4}{\milli\ampere}}, respectively.

\subsubsection{Temperature Sensitivity}
\label{sec:temperature-sensitivity}

We study HBM2 current sensitivity to temperature because HBM2 current depends on
operating temperature. Understanding how current changes with temperature is
important for both accurate HBM2 power modeling and system design. We
\emph{cannot} directly set HBM2 \atbcr{1}{chip}s to a desired temperature
because our testing setups are passively cooled. Instead, we control temperature
indirectly by selecting and repeatedly executing a current measurement loop for
30 minutes while periodically recording power and temperature every second. Each
loop drives the HBM2 \atbcr{1}{chip} toward a loop-dependent steady-state
temperature, which depends on both the current drawn by the loop and the initial
temperature of the HBM2 \atbcr{1}{chip}.

\begin{figure}[!th]
  \centering
  \includegraphics[width=1\linewidth]{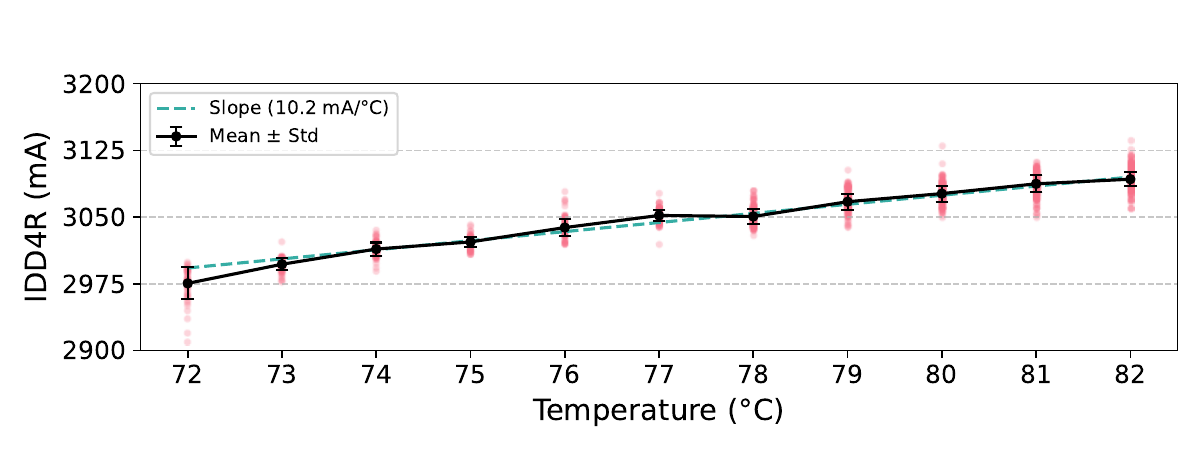}
  \caption{\atbcr{2}{HBM2 current (y-axis) sensitivity to temperature (x-axis).
  We plot IDD4R from one representative HBM2 \omcr{3}{chip} as a pictorial
  example.}}
  \label{fig:temperature_sensitivity}
\end{figure}

\atbcr{2}{\figref{fig:temperature_sensitivity} shows measured IDD4R current
\omcr{3}{for} one representative HBM2 \omcr{3}{chip} as a pictorial example for
temperature dependen\omcr{3}{ce}. The dashed line depicts a linear fit (obtained
using regression with linear least squares~\cite{gauss1809theoria}) over the
mean current across all tested temperature values. \atbcr{3}{We define the slope
as the change in mean current per \SI{1}{\celsius} increase in temperature.} We
note that other tested HBM2 \omcr{3}{chips} and other current measurement loops
exhibit similar current vs temperature behavior.}

\observation{\atbcr{2}{A test loop consumes higher current at higher temperatures}.}

\atbcr{2}{We observe that for the representative HBM2 \omcr{3}{chip}, IDD4R
draws \param{3.9}\% more current at the highest evaluated temperature (rightmost
x-axis value) than at the lowest evaluated temperature (leftmost x-axis value).
We make similar observations across all \omcr{3}{\extended{36}} tested \omcr{3}{chips} and
all evaluated current measurement loops. The linear fit over the mean current
values ha\atbcr{3}{s root} mean squared error of \param{\SI{9.4}{\milli\ampere}}.
We conclude that current and temperature likely have a linear
relationship across the entire evaluated range of temperature values
(\param{\SI{48}{\celsius}} -- \param{\SI{86}{\celsius}}).}

\begin{figure}[!ht]
  \centering
  \includegraphics[width=1\linewidth]{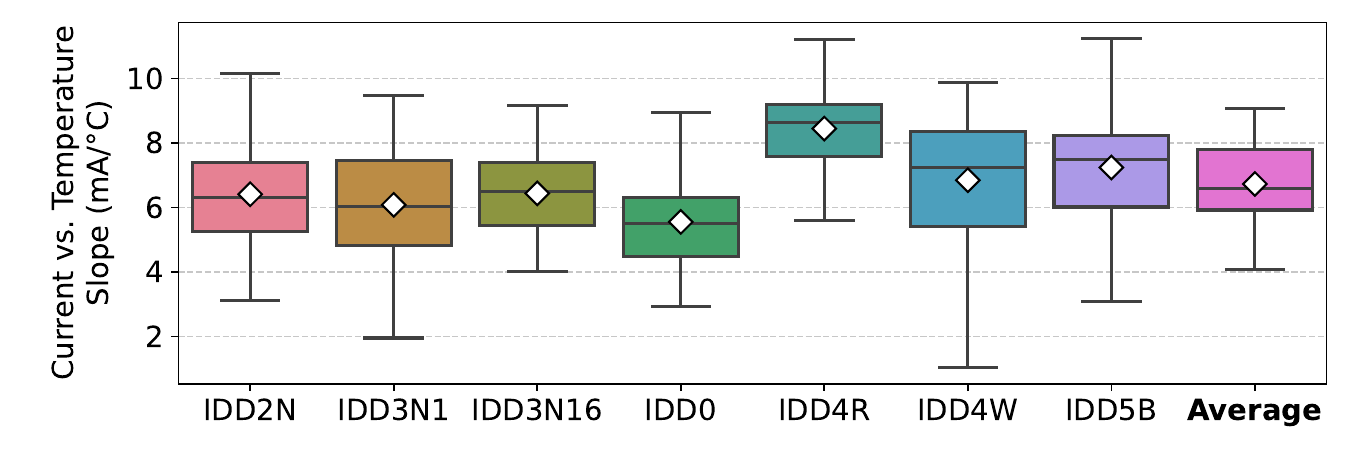}
  \caption{HBM2 current vs temperature linear fit slopes over current
  measurement loops. A higher value indicates that current increases more
  rapidly with temperature.}
  \label{fig:temperature_sensitivity_all}
\end{figure}

\figref{fig:temperature_sensitivity_all} depicts the distribution of the slopes
for the linear fit \atbcr{2}{(e.g., the \atbcr{3}{green} dashed line
in~\figref{fig:temperature_sensitivity}) of IDD vs temperature} across each HBM2
\atbcr{1}{chip}. 
We plot one box for each tested current measurement loop (x-axis). 
\atbcr{4}{For each HBM2 chip, we average the slopes across all tested
measurement loops. The Average box shows the distribution of these per-chip
averages.} \atbcr{2}{A higher slope value indicates that current increases more
rapidly with temperature.}

We observe that on average across all measurement loops and tested HBM2
\atbcr{1}{chip}s, current increases by
\atbcr{1}{\param{\SI{6.7}{\milli\ampere}}} every \param{\SI{1}{\celsius}}. We
note that the slope of current vs temperature varies significantly across test
loops (e.g., IDD4R slope distribution is higher than IDD0 slope distribution in
the plot). We hypothesize that the slope varies across current measurement loops
because each loop stresses a different mix of temperature-sensitive circuit
components.

\take{HBM2 current linearly increases with temperature across all tested \atbcr{1}{chip}s
and IDD measurement loops within the tested temperature range.}

\subsubsection{Data Rate Sensitivity}
\label{sec:data-rate-sensitivity}

We study current sensitivity to data rate because data rate directly determines
the memory bandwidth in an HBM-based system. Quantifying how current changes
with data rate is important both for building accurate bandwidth-dependent HBM
power models and for reasoning about the power cost of modern and future
higher-data-rate HBM generations. 
We use HBM2 clock frequencies of \SI{300}{\mega\hertz} (600 MT/s),
\SI{450}{\mega\hertz} (900 MT/s) and \SI{600}{\mega\hertz} (1200 MT/s) and
report the corresponding double data rate.

\begin{figure}[!ht]
  \centering
  \includegraphics[width=1\linewidth]{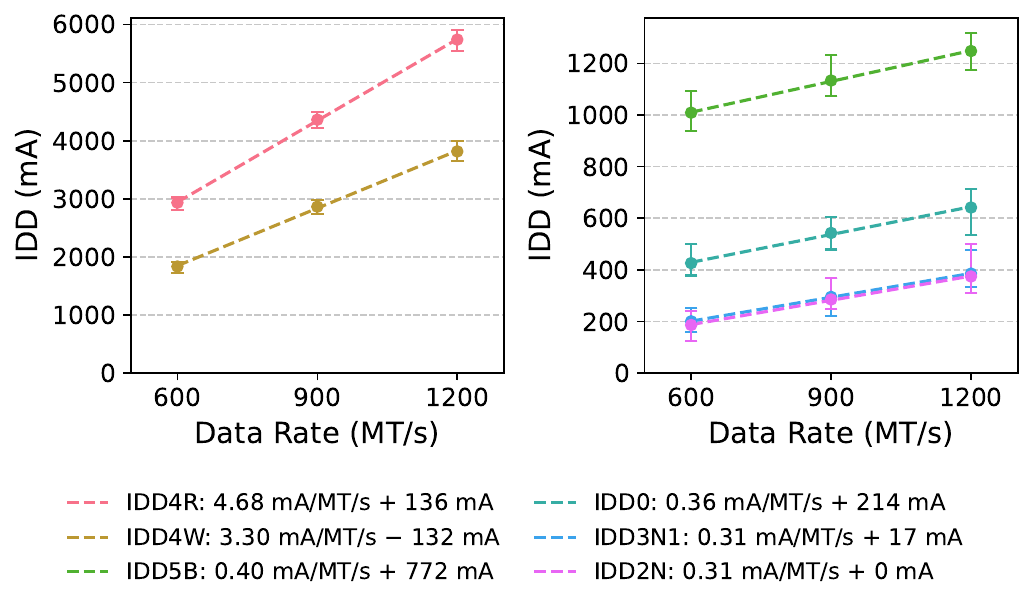}
  \caption{{\atbcr{2}{Current sensitivity to data rate for six standard IDD
  measurement loops.
  Points show the average and error bars \omcr{3}{show} the full range of the
  measurements across tested FPGAs.}}}
  \label{fig:data_rate_sensitivity}
\end{figure}

\atbcr{2}{\figref{fig:data_rate_sensitivity} shows current over data rate for
six standard current measurement loops (IDD4R and IDD4W on the left, the
remaining loops on the right).
For each loop and data rate, we measure current every second in the last minute
of the loop. Each point is the average of these measurements and the error bar
shows the full range across tested FPGAs (each of which has 2 HBM chips). The
dashed lines represent linear fits over the average current values across
varying data rates whose slope and intercept are shown in the legend.}

\observation{HBM2 current linearly increases with data rate.}

We make \atbcr{2}{three} observations. First, \omcr{2}{all tested command}
current\atbcr{2}{s} consistently increase with data rate for all tested HBM2
\atbcr{1}{chip}s \atbcr{2}{(as shown in~\figref{fig:data_rate_sensitivity}).}
\atbcr{2}{Second, read (IDD4R) and write (IDD4W) currents increase more rapidly
with data rate than other evaluated current values. This is because read and
write operations 1)~induce switching activity across in-HBM-chip signals with
larger total capacitance (e.g., bank group buses, TSVs, DQs) than other
operations, and 2)~induce higher switching frequency on the data, command, and
address bus signals with increasing data rate.} \modelcr{2}{In contrast, IDD0
and IDD5B increase with data rate at a similar rate as the standby currents
(IDD2N and IDD3N1). Because these loops issue a fixed number of activate and
refresh commands per second at every data rate, \omcr{3}{we attribute} the
data-rate-dependence of IDD0 and IDD5B largely to the standby current, i.e., the
power consumed by a single activate or refresh operation \atbcr{3}{\emph{only}
slightly increases} with data rate.} \atbcr{2}{Third}, the measured current vs
data rate relationship is approximately linear over the evaluated range of data
rate values. \atbcr{2}{For example,} across all tested HBM2 \atbcr{1}{chip}s,
the average IDD4R current at \param{600}~MT/s, \param{900}~MT/s, and
\param{1200}~MT/s is \param{\SI{1467.1}{\milli\ampere}},
\param{\SI{2181.0}{\milli\ampere}}, and \param{\SI{2870.1}{\milli\ampere}}, per
HBM2 \atbcr{1}{chip}, respectively. Given 1)~that the widely-known
\omcr{2}{dynamic} power \omcr{2}{consumption} equation~\cite{weste2011cmos}
describes that current is directly proportional to switching
frequency\footnote{$P=I*V\approx{}V^{2}f$ where P is power, I is current, V is
the voltage, and f is the \omcr{2}{switching} frequency.} and 2)~our empirical
observations, we consider a first-order linear model a reasonable fit for
current vs data rate.

\subsection{Structural Variation of HBM2 Current}
\label{sec:structural-variation}

We next study structural variation of HBM2 current because memory accesses can
target different internal structures, and current may vary across these
structures in currently unknown ways. Quantifying structural variation is
important for building accurate HBM power models and for understanding whether
certain memory access patterns can induce higher power than others by targeting
HBM2 structures that draw higher current than other structures.

\noindent
\textbf{Evaluation Methodology.} We use the standard IDD4R \atbcr{2}{and IDD4W}
current measurement loop\atbcr{2}{s} as the base patterns because
\atbcr{2}{they} draw the highest current among the standard current measurement
loops and therefore provide a strong baseline for studying structural current
variation. \atbcr{2}{We investigate structural current variation in four steps.}
1)~To evaluate \atbcr{2}{current variation across HBM2 \emph{stack IDs (SID)},
\emph{channels}, and \emph{pseudo-channels},} we execute \atbcr{2}{IDD4R/W} as
is and measure how current varies when memory accesses target different
\microrevb{stack IDs}, channels or pseudo-channels. 2)~To evaluate
\emph{bank-group-level} current variation, we \emph{cannot} use
\atbcr{2}{IDD4R/W} as is because they access all banks in all bank groups;
instead, we modify the current measurement loop to repeatedly access a selected
combination of two bank groups so that we preserve bank-group-level
parallelism~\cite{ghose2019demystifying, mutlu2008parallelism} and understand
structural variation of current across bank groups. 3)~To evaluate
\emph{bank-level} current variation, we modify \atbcr{2}{IDD4R/W to target}
\atbcr{3}{one bank in two bank groups}. The modified loop targets only one
selected bank in bank groups 0 and 2 (we later\omcr{2}{, in this section,} show
that bank groups 0 and 2 consume the most current out of all tested bank group
combinations). 4)~To evaluate \emph{row-level} current variation, we use a
separate access pattern that sequentially reads all HBM2 columns within a
selected DRAM row.

\noindent
\textbf{Power \omcr{2}{variation} across HBM2 \microrevb{stack IDs}, channels,
pseudo-channels, and rows in a bank.} We do not identify any significant
\atbcr{2}{(i.e., >1\%)} power variation across 3D-stacked \microrevb{stack IDs},
channels, pseudo-channels, and DRAM rows in a bank in the tested HBM2
\atbcr{1}{chip}s. In the interest of space, we omit these results. We
identify significant variation in current across bank groups and banks,
\omcr{2}{which we report next}.

\begin{figure}[!ht]
  \centering
  \includegraphics[width=1\linewidth]{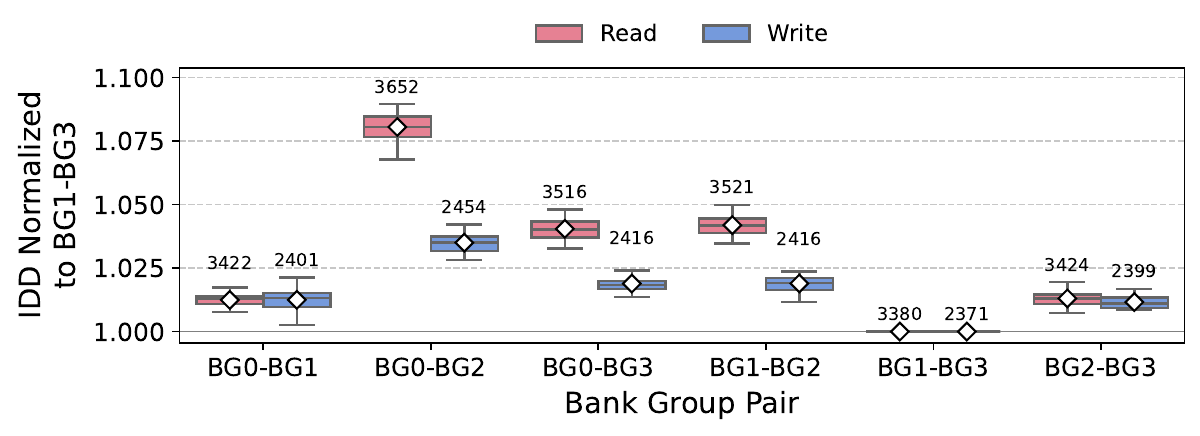}
  \caption{\atbcr{2}{Read and write current drawn by bank group pairs\modelcr{3}{,
  normalized to each chip's current for bank groups 1 and 3. Numbers above the
  boxes show the average current in mA across tested chips}.}}
  \label{fig:bank_group_variation}
\end{figure}

\noindent
\textbf{Power \omcr{3}{variation} across bank groups.}
\figref{fig:bank_group_variation} shows \atbcr{2}{the distribution of} HBM2
\atbcr{2}{read and write} current \atbcr{2}{for the access pattern described
earlier in this section} across \extended{35}
tested HBM2 chips for each of the evaluated bank group pairs.\atbcrcomment{2}{I
have the absolute numbers added. I do not like this figure because it does not
show that every single tested chip has this property where BG1-BG3 is the
smallest power consuming bg pair. Now that information is lost to the
distribution. We could do a (very) big barplot.}

\observation{HBM2 read \atbcr{2}{and write} currents significantly vary across
bank groups.}

We observe that repeatedly \atbcr{2}{reading from (writing to)} bank groups
\atbcr{3}{0 and 2} consumes \param{8.1}\% \atbcr{2}{(\param{3.5}\%)} higher
current than repeatedly \atbcr{2}{reading from (writing to)} bank groups 1 and
3, on average across all tested HBM2 \atbcr{1}{chip}s. \atbcr{2}{Importantly,
bank groups 1 and 3 consistently consume the smallest and bank groups 0 and 2
consume the largest power in all tested chips across all tested bank group
combinations.} This observation is important because bank group selection can
significantly affect HBM2 read and \atbcr{2}{write} power \atbcr{2}{consumption}
even if the memory throughput remains the same across varying burst
\atbcr{2}{read and write} access patterns.

\noindent
\textbf{\omcr{2}{Explanatory hypotheses}.} We hypothesize that the main reason
for power variation across bank group pairs is data movement distance. Several
prior works~\cite{lee2016hbm, lee2020hbm2e, cho2018hbm2, ryu2023hbm3,
park2022hbm3}\atbcrcomment{2}{SMLA does not talk about this} depict HBM designs
across multiple generations where two bank groups (e.g., bank groups 0 and 2
depicted in~\cite{ryu2023hbm3} as \param{``BG\_A'' and ``BG\_C''}) receive and
transmit data to the TSV stripe via a longer, more power-hungry bus than the
other two bank groups (e.g., bank groups 1 and 3). Therefore, reading data from
bank groups 0 and 2 draws more current than reading data from bank groups 1 and
3. This hypothetical explanation also consistently describes the current trend
across evaluated bank group combinations: any combination of bank groups that
includes either bank group 0 or 2 (the most power-hungry bank groups) consumes
more current than the combination of bank groups \microlabel{A1}1 and
3.\footnote{\microreva{These explanations are backed by insight into HBM
architecture disclosed in ISSCC and JSSC publications
(e.g.,~\cite{lee2016hbm,lee2020hbm2e,cho2018hbm2,ryu2023hbm3,park2022hbm3}),
usually authored by researchers and engineers that design and manufacture
advanced HBM \atbcr{1}{chip}s.}}

\begin{figure}[!ht]
  \centering
  \includegraphics[width=1\linewidth]{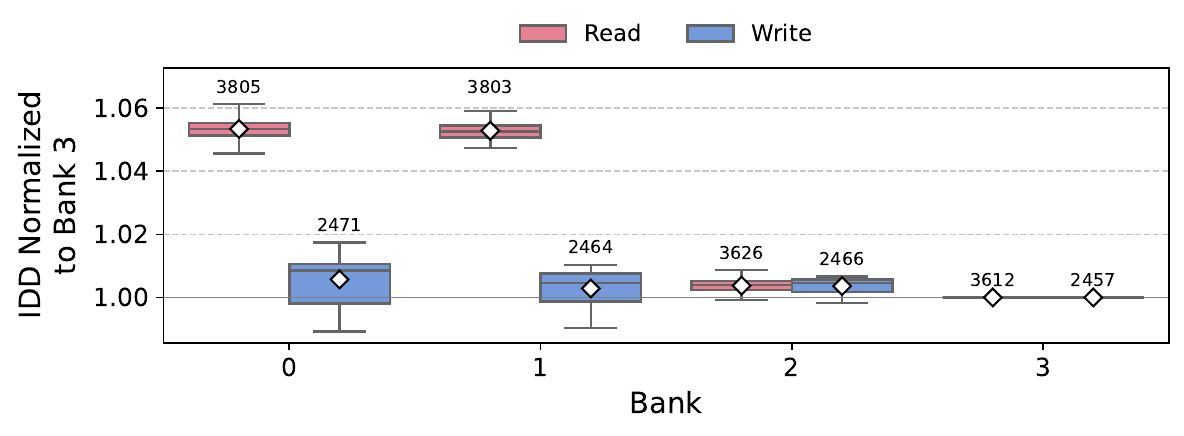}
  \caption{\atbcr{2}{Variation of read and write current across different
  banks\modelcr{3}{, normalized to each chip's current for bank 3. Numbers above
  the boxes show the average current in mA across tested chips}.}}
  \label{fig:bank_variation}
\end{figure}

\noindent
\textbf{Power \omcr{3}{variation} across banks.} \figref{fig:bank_variation}
shows HBM2 current across the evaluated banks.

\observation{HBM2 read current varies across banks.}

\atbcr{2}{We make two key observations. First,} we observe that repeatedly
\atbcr{2}{reading from} \atbcr{2}{bank 0 or 1} draws \param{5.3}\% and
\param{5.3}\% higher current than bank 3, respectively, on average across all
tested HBM2 \atbcr{1}{chip}s, whereas repeatedly \atbcr{2}{reading from} bank 2
consumes only \param{0.4}\% higher current than bank 3. This result indicates
that bank-level current variation is also systematic: some banks consistently
consume more current than others all else being equal. \atbcr{2}{Second, we do
\emph{not} observe significant write current variation across banks. Repeatedly
writing data to banks 0 and 1 draws \modelcr{2}{less than 1\%} higher current
than writing data to either bank 2 or bank 3.}

\noindent
\textbf{\atbcr{2}{Explanatory hypotheses}.} We hypothesize that the observed
spatial variation is caused by two fundamental architectural properties of
3D-stacked hierarchical DRAM organization: i)~power delivery network (PDN)
asymmetry and ii)~physical bus distance. First, because power is largely
delivered vertically via TSVs in the center TSV stripe
(see~\secref{sec:background}), the outer edges of an HBM2 DRAM die are more
vulnerable to voltage drop (IR drop)~\cite{oh2020hbm2e}. If banks 0 and 1 are
physically further away from the primary power \omcr{2}{rails} than banks 2 and
3, the resulting voltage drop can cause the peripheral circuitry to
\atbcr{3}{consume} additional energy to compensate and maintain signal
integrity.\omcrcomment{2}{double check and add more references} Second, reading
from or writing to banks located further from the central TSV stripe requires
driving data across longer internal global buses, which consume more dynamic
power to drive signals during read \atbcr{2}{and write} operations.

\take{HBM2 read \atbcr{2}{and write} currents vary across bank groups and banks.
We find no evidence that suggests HBM2 read current varies across 3D-stacked
HBM2 channels, across pseudo-channels, or across rows in a
bank.\label{take:structural-power}}

\subsection{Data Pattern Dependence of Current}
\label{sec:dpd-current}

We next study HBM2 current dependence on data patterns \atbcr{3}{for} two
reasons. First, the standard JEDEC current measurement loops use
\omcr{2}{\emph{fixed}} data patterns that might \emph{not} capture the
highest-power-consuming ways of accessing an HBM2 \atbcr{1}{chip}. Second, prior
work on planar DDR3L DRAM~\cite{ghose2018what} already shows that DRAM power
\omcr{2}{consumption heavily} depends on the data patterns transferred across
various data bus components in a DRAM chip.

\begin{figure}[!ht]
  \centering
  \includegraphics[width=1\linewidth]{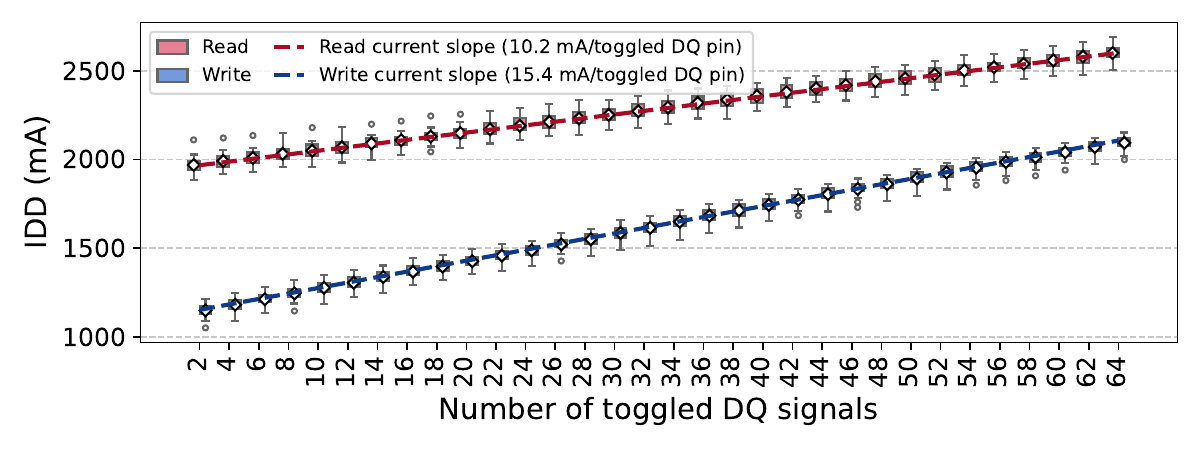}
  \caption{\atbcr{2}{Read and write} current vs number of toggled DQ \atbcr{2}{pins}
  across beats of a data transfer burst.}
  \label{fig:current_vs_number_of_toggles}
\end{figure}

\noindent
\textbf{Effect of \omcr{2}{data pin (DQ)} toggling.} We investigate how toggling
or switching activity in the DQ pins \atbcr{2}{(depicted as data I/O pins
in~\figref{fig:intro_figure})} affects HBM2 current. We execute an IDD4R-like
\omcr{2}{read} \atbcr{2}{and an IDD4W-like write} access pattern using 32
distinct 256-bit data patterns whereby we keep the number of \emph{logic-1}s
equal across beats of the data transfer burst (to isolate the effect of DQ
toggling on current). These data patterns vary \emph{only} in the number of DQs
they toggle every beat of the burst. \figref{fig:current_vs_number_of_toggles}
shows how current changes with the number of toggled DQs. The value on the
x-axis shows how many DQs toggle with every clock cycle (beat) of the data
transfer burst. Because there are only 64 DQs \atbcr{2}{per tested
pseudo-channel}, the x-axis ends at 64.

\observation{\atbcr{2}{HBM2 read and write current draw increases with the
number of toggled DQ signals}.} 

We observe that the average current \atbcr{2}{drawn by read and write
operations} across all tested HBM2 \atbcr{1}{chip}s increase\atbcr{3}{s} by
\param{32.0}\% and \atbcr{2}{\param{\modelcr{2}{82.0\%}}, respectively,} when
the number of toggled DQ signals increases from \param{2} to \param{64}. The
current vs number-of-toggles trend is approximately linear, with a\omcr{2}{n
increase of \SI{10.2}{\milli\ampere} \atbcr{2}{and
\modelcr{2}{\SI{15.4}{\milli\ampere}}} per toggled signal \atbcr{2}{for read and
write operations, respectively.}} 
We attribute this trend to higher switching activity in the HBM2 I/O circuitry:
toggling more DQ signals requires charging and discharging more signal
capacitance every beat of the data transfer burst, which increases dynamic
current. We note that data bus inversion (DBI)~\cite{stan2002bus,jedechbm2} or
other data encodings that aim to reduce data bus power
(e.g.,~\cite{lee2018reducing,childers2000reordering,
chatterjee2017architecting,stan1995coding})\omcrcomment{2}{cite more} could
reduce current for highly toggle-heavy data patterns by lowering the number of
DQ signal state transitions. We do not study DBI in this work because our goal
is to characterize baseline HBM2 current sensitivity to data patterns rather
than the effect of an optional controller-side data encoding mechanism.

\begin{figure}[!ht]
  \centering
  \includegraphics[width=1\linewidth]{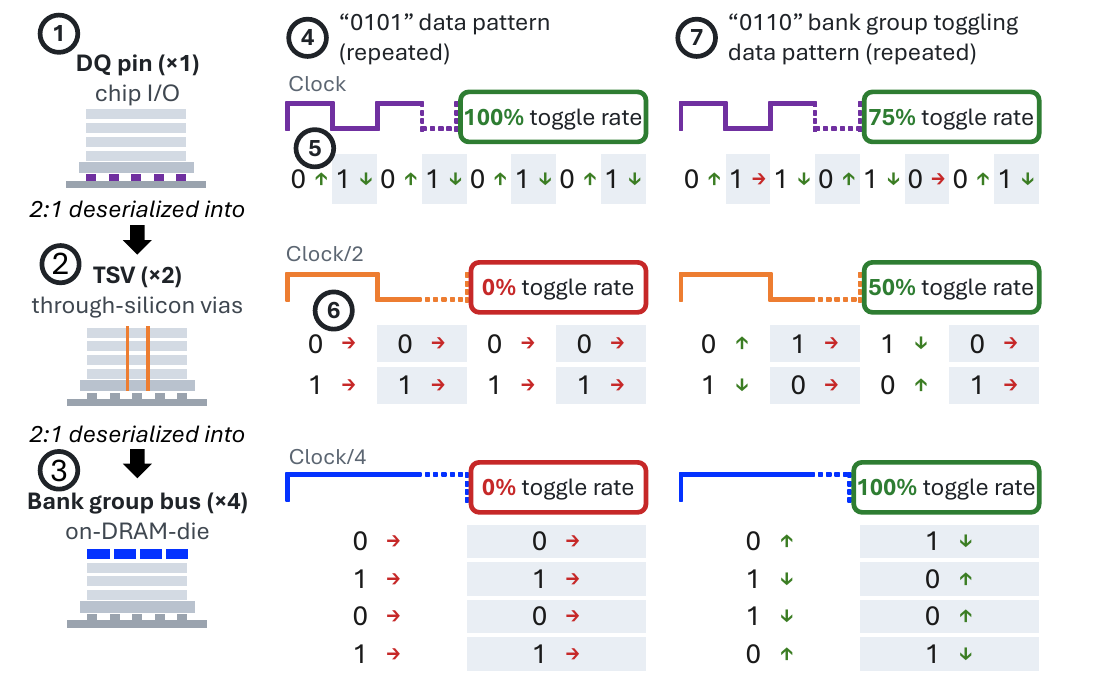}
  \caption{\atbcr{2}{HBM2 DQ (chip I/O; \dingOne{}), through-silicon via (TSV;
  \dingTwo{}), and bank group (\dingThree{}) buses in one pseudo-channel. Data
  transmission across the three depicted buses for two subsequent data transfer
  bursts (\dingFour{}, \dingSeven{}).}}
  \label{fig:toggling-depiction}
\end{figure}

\noindent
\textbf{Effect of TSV and bank-group bus toggling.} Next, we would like to
understand if and how toggling HBM2-\atbcr{1}{chip}-internal TSV
\atbcr{2}{(\textcolor{orange}{\ding{174}} in~\figref{fig:intro_figure})} and
bank-group bus
\atbcr{2}{(\textcolor{olive}{\ding{175}}--\textcolor{red}{\ding{176}}
in~\figref{fig:intro_figure})} signals affect current. To do so, we execute an
IDD4R-like \omcr{2}{read} access pattern using carefully constructed data
patterns that aim to toggle HBM2-\atbcr{1}{chip}-internal data.
\atbcr{2}{\figref{fig:toggling-depiction} illustrates the HBM2 data bus
hierarchy (left) and two example data transmissions over this hierarchy for two
consecutive data transfer bursts across these buses (right), based on prior
depictions of HBM2 data bus organization and design
~\cite{kim2022low,oh2020hbm2e,lee2020hbm2e,
lee2014high,cho2018hbm2,ryu2023hbm3,chun2020hbm2e,oconnor2021thesis}.}
\atbcr{2}{In this illustration, a DQ pin data signal between the logic die and
the memory controller (\dingOne{}) operates at the HBM2 interface clock speed
(Clock). The DQ pin's data is deserialized into two through-silicon via (TSV)
signals (\dingTwo{}), each of which operates at half the HBM2 interface clock
speed (Clock/2). The two TSV signals are further deserialized into four bank
group (BG) bus signals on a DRAM die (\dingThree{}). The bank group bus clock
speed (Clock/4) is four times slower than the HBM2 interface clock speed.}
\atbcr{2}{The right side of~\figref{fig:toggling-depiction} (\dingFour{},
\dingSeven{}) illustrates how an 8-bit data transfer over a single DQ pin is
laid out onto two TSV signals and four bank group bus signals. We display a
\emph{toggle rate} value per DQ pin, TSV, and bank group bus signal. We define
this value as the fraction of clock cycles where the bus value transitions from
a logic-0 to logic-1 (upward green arrow) or a logic-1 to logic-0 (downward
green arrow).}

\atbcr{2}{We make two key observations from~\figref{fig:toggling-depiction}.
First, the ``0101'' data pattern (\dingFour{}) causes the DQ pin signal to
toggle (toggle direction indicated with green arrows \dingFive{}) every clock
cycle (100\% toggle rate). However, this data pattern causes \emph{no toggling}
in TSV and bank group bus signals (lack of toggling indicated with red arrows
\dingSix{}). Second, the ``0110'' bank group toggling data pattern induces a
75\% toggle rate on the DQ pin. \atbcr{3}{Importantly}, this data pattern also
induces a \atbcr{3}{50\% toggle rate on TSV} and \atbcr{3}{a 100\% toggle rate
on} bank group bus signals. We conclude that a data pattern can induce a variety
of toggle rates across the HBM2 data bus hierarchy.}

\begin{figure}[!ht]
  \centering
  \includegraphics[width=1\linewidth]{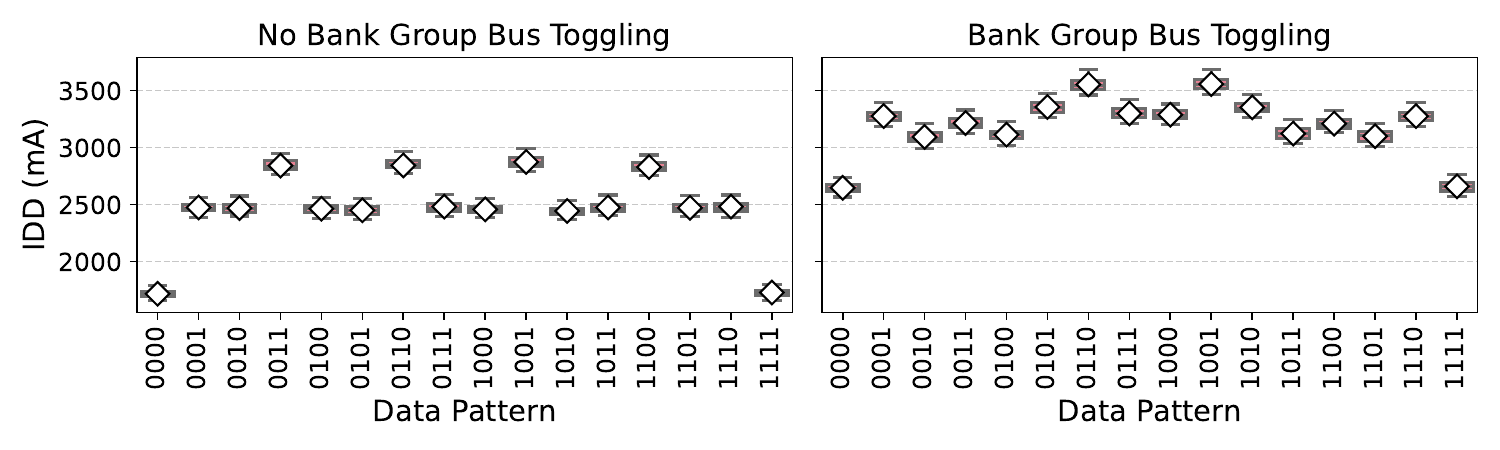}
  \caption{Current vs data transmitted by one HBM2 DQ signal over data transfer
  burst clock cycles. Each bit in an x-axis value depicts the data transmitted
  by a single DQ signal across 4 beats of the burst.}
  \label{fig:current_vs_bus_flip}
\end{figure}

{\figref{fig:current_vs_bus_flip} shows \atbcr{2}{distributions of HBM2
read} current \atbcr{2}{across \extended{35} tested HBM2 chips} over varying data patterns
transmitted by one DQ signal \atbcr{2}{in four consecutive data bus clock
cycles}. Each x-axis label is a 4-bit sequence, where each bit indicates the
value carried by the same DQ signal in one beat of the \atbcr{2}{data transfer}
burst. We group the patterns according to whether they induce toggling only in
the DQ bus or additionally in HBM2-\atbcr{1}{chip}-internal
\microrevcommon{bank group (BG)}
bus\microrevcommon{~\cite{kim2022low,oh2020hbm2e,lee2020hbm2e,
lee2014high,cho2018hbm2,ryu2023hbm3,chun2020hbm2e,oconnor2021thesis}.} For
example, ``1001'' in the left subplot and in the right subplot means that a DQ
bus signal transmits ``10011001'' and ``10010110'' over every 8 beats (clock
cycles) of two data transfer bursts, respectively.}

\vspace{2mm}
\observation{Current significantly varies with data pattern transferred over a
DQ bus signal across 8 data transfer clock cycles.}
\vspace{3mm}

The highest (lowest) average current (across tested HBM2 \atbcr{1}{chip}s) is
\SI{2873.6}{\milli\ampere} (\SI{1717.9}{\milli\ampere}) for no BG bus
toggl\omcr{2}{ing} and \SI{3556.3}{\milli\ampere} (\SI{2646.8}{\milli\ampere})
for BG bus toggl\omcr{2}{ing}, respectively.

\observation{BG bus toggling significantly increases HBM2 current compared to no
BG bus toggling. \omcr{2}{Thus, HBM2-\atbcr{1}{chip}-internal data movement
contributes significantly to \omcr{3}{power} consumption.}}

The average current across all \atbcr{2}{data} patterns and tested HBM2
\atbcr{1}{chip}s for BG bus toggl\omcr{2}{ing} is \param{29.4}\% higher than the
average current for no BG bus toggl\omcr{2}{ing}. {This observation suggests
that HBM2-\atbcr{1}{chip}-internal data movement contributes significantly to
\omcr{3}{power} consumption. Even when the DQ bus activity is similar, more
frequent toggling in the internal TSV and BG buses can increase read current. In
fact, the BG bus toggl\omcr{2}{ing} data patterns \atbcr{2}{(e.g., right
subplot, ``0110'' bank group bus toggling)} \emph{cannot} maximize the toggling
frequency in the DQ bus \atbcr{2}{(see~\figref{fig:toggling-depiction})}, yet
they \omcr{3}{lead to} higher current than a data pattern that maximizes the
toggling frequency in the DQ bus (e.g., left subplot, ``0101'').}

\observation{{Two \atbcr{2}{data} patterns \omcr{3}{induce} the highest current among all
evaluated patterns.}}

{We observe from the right side of~\figref{fig:current_vs_bus_flip} that two
particular \atbcr{2}{data} patterns ``0110'' and ``1001'' draw the highest
current across all tested HBM2 \atbcr{1}{chip}s.\omcrcomment{2}{can we try to
Explain this} This indicates that \atbcr{2}{identifying} the
highest-current data pattern \atbcr{2}{requires reasoning about the switching
activity that data pattern induces across the HBM data bus hierarchy.}

\take{HBM2 current is highly sensitive to data patterns transferred over burst
read operations. \omcr{2}{Thus, HBM2-\atbcr{1}{chip}-internal data movement
contributes significantly to \omcr{3}{power} consumption.} The data pattern that
\atbcr{1}{induces} the highest current induces toggling in the bank group, TSV,
and the DQ buses.\label{take:data-pattern-power}}

\subsection{Highly Power-Intensive \omcr{2}{HBM} Access Patterns}

Our results so far suggest that read current varies significantly across bank
groups, banks, and data patterns transferred across the various buses inside the
HBM2 \atbcr{1}{chip}. \omcr{2}{Unfortunately}, the standard IDD4R current
measurement loop's access pattern~\cite{jedechbm2} ignores \omcr{2}{such}
structural and data pattern dependence of current. \omcr{2}{As such, it could
easily} fall short of \atbcr{1}{inducing} highest-power-consuming conditions in
an HBM2 \atbcr{1}{chip}. \omcr{2}{To demonstrate this, we construct} carefully
engineered access and data patterns \omcr{2}{that} increase HBM2 current beyond
that of the standard measurement loops. 

\noindent
\textbf{Experimental Methodology \& Results.} We make two modifications to the
baseline IDD4R current measurement loop. First, we make it access \atbcr{3}{bank
0} in bank groups 0 and 2 repeatedly such that we exercise the highest power
consuming banks and bank groups (see Takeaway~\ref{take:structural-power}).
Second, we use the ``0110'' with BG bus toggl\omcr{2}{ing} data pattern to cause
toggling in and increase \omcr{3}{power} consumed by the HBM2-internal data
buses \atbcr{2}{as well as the DQs} (see
Takeaway~\ref{take:data-pattern-power}).
\atbcr{2}{\figref{fig:max_power_pattern}
shows the current drawn by the standard IDD4R loop and by our access pattern for
each tested HBM2 chip.}

\begin{figure}[!ht]
  \centering
  \includegraphics[width=1\linewidth]{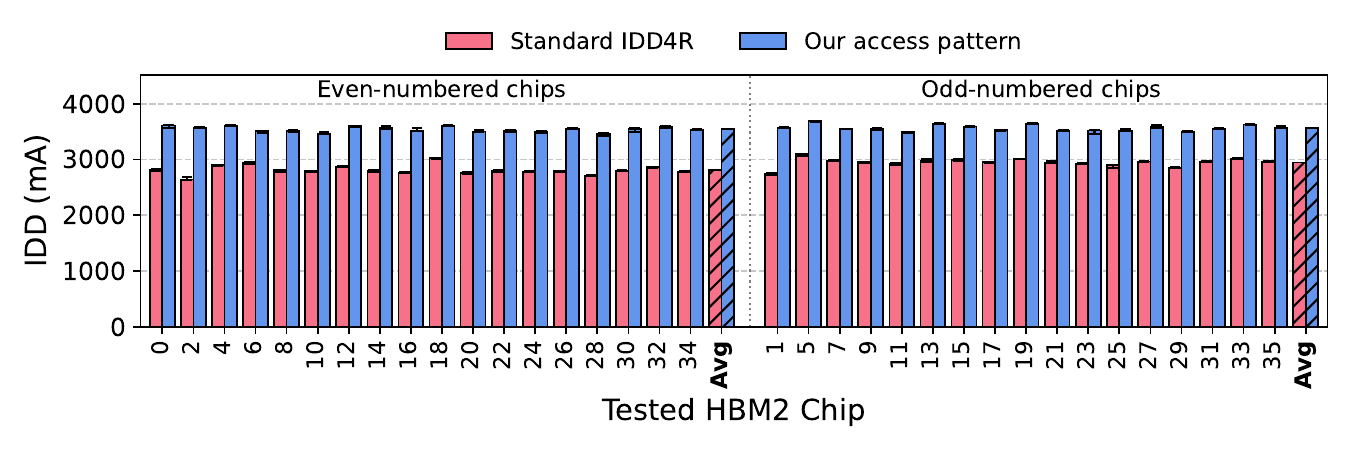}
  \caption{\atbcr{2}{Current drawn by the standard IDD4R loop and by our access
  pattern across tested HBM2 chips, grouped by the chip's position in the
  package (even vs odd).}}
  \label{fig:max_power_pattern}
\end{figure}

We find that by \omcr{2}{modifying} the standard IDD4R access pattern to
exercise higher-power structures and to stress data-dependent toggling, we
increase its power consumption. \atbcr{3}{Our access pattern consumes at least
19.1\%, on average 23.6\%, and up to 35.4\% more power than IDD4R across all
tested chips.} \atbcr{3}{This} indicates that the standard JEDEC loops do
\emph{not} necessarily capture worst-case HBM2 power. In other words, leveraging
our empirical observations, we provide a new access pattern that consumes
significantly higher current than the JEDEC standard IDD4R current measurement
loop.

\section{\atbcr{2}{Ayna: A More Accurate HBM Power Model}}
\label{sec:hbm2-power-model}

To improve 3D-stacked DRAM power modeling fidelity in open source architectural
simulation infrastructure\omcr{2}{s}, we develop
\atbcr{2}{Ayna,\footnote{\atbcr{2}{Ayna, the Turkish word for ``mirror'',
\omcr{3}{aims to accurately mirror} the power consumption behavior of real HBM chips across various memory
access \& data patterns and operating conditions.}}} a new power model guided by
our experimental characterization results, building on
DRAMPower~\cite{steiner2025drampower5, drampowergithub}. We make \atbcr{3}{four}
key modifications to DRAMPower: 1)~we extend \microrevcommon{it with an HBM2
organization and command timing specification}, 2)~we integrate the IDD values
obtained from our characterization study (\secref{sec:standard-idd}) into
DRAMPower, 3)~we extend the model to capture structural power variation across
bank groups and banks in the HBM2 \atbcr{1}{chip}
\microrevcommon{(\secref{sec:structural-variation})}, and \microrevcommon{4)~we
extend the model to capture data pattern dependence of power
(\secref{sec:dpd-current}).}\microlabel{CC4} \omcr{2}{We also include our new
power model in Ramulator2.1~\cite{luo2024ramulator, ramulator2github,
luo2026ramulator21, bostanci2026cleaning, kim2016ramulator} and release it
independently~\cite{hbmpower-github}.}

\noindent
\textbf{\microrevcommon{Tested Workloads.}} We evaluate \omcr{3}{our new} power
model using \param{\omcr{2}{six}} \microrevcommon{microbenchmarks} and
\microrevcommon{one large language model (LLM) \microlabel{CC3}decode workload
with varying batch sizes} to exercise unique mixes of power-consuming HBM2
components. \microrevcommon{These workloads heavily diverge from standard
current measurement loops, and they each use a different combination of DRAM
commands to access different parts of the HBM2 \atbcr{1}{chip}.
\atbcr{2}{\emph{Hammer} repeatedly activates and precharges rows in eight banks
(across two bank groups). \emph{Interleaved} activates a row, reads one column,
and precharges the bank, iterating over eight banks in two bank groups.
\emph{Streaming All Banks} opens a row in every one of the 16 banks and issues
eight consecutive column reads to each open row. This microbenchmark mimics a
high row-buffer-locality read access pattern that exercises all four bank
groups. \emph{Streaming One Bank Group} issues eight consecutive column reads
each to two open rows in a single bank group, alternating between two rows.
\emph{Read/Write} alternates bursts of back-to-back writes and back-to-back
reads to open rows in the four banks of a single bank group. \emph{Read BG0-BG2}
interleaves reads between the banks of bank groups 0 and 2 (the highest-power
bank group pair, \secref{sec:structural-variation}) and reads every column of
each activated row before moving to the next row.} For the LLM workload, we
generate DRAM command traces using
LLMSimulator~\cite{yun2024duplex,llm-simulator-github} integrated with
Ramulator2.0~\cite{ramulator2github, luo2024ramulator, kim2016ramulator} running
Llama3.1-8B~\cite{grattafiori2024llama} at FP16 precision. We (i) configure the
simulator to model an A100 GPU with the HBM2 configuration mimicking our
real-world system (see~\secref{sec:methodology}), and (ii) capture the DRAM
command trace of a single decode step from a single HBM2 channel at a context
length of 1024 tokens whereby we sweep the inference request batch size from 1
to 128 in powers of two.}

\microlabel{E1}
\noindent
\textbf{\microreve{Evaluated Power Models.}} \microreve{We compare three HBM2
power models to our empirical results: (1)~\emph{\atbcr{2}{Ayna}}, our own
model, built on DRAMPower~\cite{steiner2025drampower5, drampowergithub},
\atbcr{2}{(2)}~\emph{DRAMSim3}, DRAMSim3's~\cite{li2020dramsim3} publicly
available HBM2 IDD values~\cite{dramsim3github} (for a 128-bit HBM2 channel at
2000 MT/s data rate)\atbcrcomment{2}{we do not have empirical results for this
data rate and the best we could do is 1800 MT/s if we solve timing closure
issues. Our major characterization results are for 1200 MT/s.} that we integrate
into DRAMPower, \atbcr{3}{and} \atbcr{2}{(3)}~\emph{FGDRAM HBM2},
FGDRAM's~\cite{oconnor2017fine} energy-per-bit (pJ/bit) HBM2 model that we
integrate into DRAMPower.

We integrate DRAMSim3's~\cite{li2020dramsim3} IDD values~\cite{dramsim3github}
into DRAMPower by fairly accounting for the \atbcr{2}{three} differences
\atbcr{3}{between} our \atbcr{3}{HBM2} testing setup and DRAMSim3's simulated
HBM2 configuration. \atbcr{2}{First, w}e test eight 64-bit pseudo-channels (one
pseudo-channel in each HBM2 channel) whereas DRAMSim3 is configured with one
128-bit channel. We carefully scale background (e.g., IDD2N/3N) and dynamic (e.g.,
IDD4R/W) current components to match the HBM2 organization of our testing setup.
\atbcr{2}{Second,} our current measurements yield significant
(\param{\modelcr{2}{\SI{821}{\milli\ampere}}} on average across all tested
\atbcr{1}{chip}s) power-off current. We add this power-off current to all IDD
values reported in DRAMSim3. \atbcr{2}{Third,} our HBM2 \atbcr{1}{chip}s are
clocked at 1200 MT/s data rate whereas DRAMSim3 is configured at 2000 MT/s data
rate. \modelcr{2}{We scale down {every} DRAMSim3 IDD value to 1200 MT/s using
the data rate dependence we empirically measure for the corresponding current
measurement loop (\secref{sec:data-rate-sensitivity},
\figref{fig:data_rate_sensitivity}).}} \microreve{We integrate FGDRAM's HBM2
energy-per-bit values into DRAMPower fairly by solving each DRAMPower energy
calculation equation~\cite{dram-power-equation} for the IDD value that equation
is based on.} 

\begin{figure*}[!b]
  \centering
  \includegraphics[width=1\linewidth]{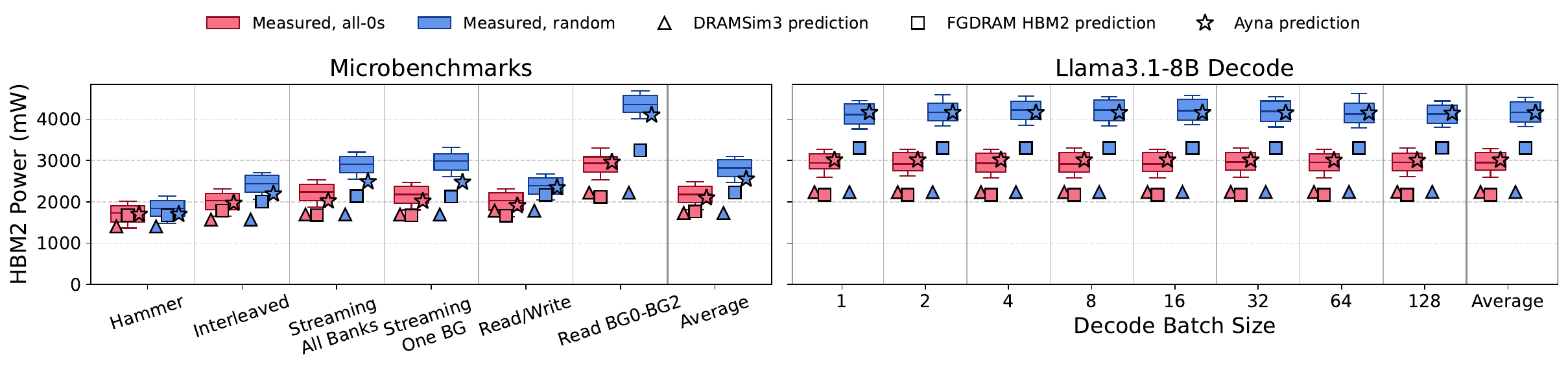}
  \caption{\modelcr{2}{Measured HBM2 power across \omcr{3}{all 36} tested chips
  (boxes) and the power predicted by DRAMSim3, FGDRAM HBM2, and Ayna (markers)
  for the evaluated microbenchmarks (left) and large language model decode
  workloads (right), with all-0s and random data patterns.} \modelcr{3}{The
  rightmost group in each plot (\emph{Average}) shows the distribution of each
  chip's power consumption averaged over all workloads \atbcr{3}{in that plot}.}}
  \label{fig:model_comparison_error}
\end{figure*}

\noindent
\atbcr{2}{\textbf{Evaluated Data Patterns.} We evaluate each workload using two
data patterns to study the effects of data patterns on power
\omcr{3}{consumption}.\footnote{\atbcr{2}{\atbcr{3}{In this section, we account
for data-pattern dependence using IDD values that we separately measure for
all-0s and random data. In~\secref{sec:data-dependent-power}, we evaluate Ayna's
explicit data-pattern model, which separately accounts for toggling on TSVs and
bank group buses to improve prediction accuracy across a wider range of data
patterns.}}} First, \emph{all-0s} where we initialize the HBM chip with logic-0
values and ensure that any write operation in the evaluated workload writes
logic-0 to its destination memory address. Second, \emph{random} where we
initialize the HBM chip with a pseudo-randomly-generated data pattern and ensure
that any write operation writes a new pseudo-randomly-generated value to its
destination memory address.} For \atbcr{2}{Ayna}, we experimentally characterize
the HBM2 \atbcr{1}{chip}s once using all-0s and once using random data patterns
and configure \atbcr{2}{Ayna} using real IDD measurements. DRAMSim3 does
\emph{not} have a data pattern dependent power model, \omcr{3}{so} we use the
same IDD values~\cite{dramsim3github} for both all-0s and random data patterns.
FGDRAM HBM2 has a tunable toggle rate parameter for data movement energy. We use
a 0\% toggle rate for the all-0s and a 50\% toggle rate for the random data
pattern.

For each \microrevcommon{tested workload}, we 1)~simulate the energy consumed by
\atbcr{2}{that workload} using \microrevcommon{each evaluated} power model, and
2)~execute the \microrevcommon{workload in one pseudo-channel in 8 HBM2 channels
across \numStacks{}} real HBM2 \atbcr{1}{chip}s and measure power.
\microrevcommon{The tested HBM2 \atbcr{1}{chip}s are clocked at 1200 MT/s data
rate.} \modelcr{2}{\figref{fig:model_comparison_error} shows, for every
evaluated workload (x-axis), the power measured across all tested HBM2 chips
(boxes) and the power predicted by each evaluated power model (markers), for
both the all-0s and the random data pattern. We quantify prediction accuracy
with the mean absolute percentage error (MAPE): for a workload, the absolute
difference between predicted and measured power, relative to the measured power,
averaged across all tested chips. A MAPE of \param{0}\% indicates a perfect
prediction.}

\microlabel{CC4}
We make three key observations from
\modelcr{2}{\figref{fig:model_comparison_error}}. First, we observe that
\atbcr{2}{Ayna} accurately predicts the power consumed by each evaluated
workload in each tested HBM2 \atbcr{1}{chip}. \atbcr{3}{For the all-0s data
pattern,} on average across all HBM2 \atbcr{1}{chip}s and evaluated workloads,
\atbcr{2}{Ayna} yields a \modelcr{2}{MAPE} of \param{\modelcr{2}{8.1\%}}. For
real Llama workloads, \atbcr{2}{Ayna} yields an average MAPE of
\param{\modelcr{2}{7.3\%}}.
\modelcr{2}{With the random data pattern, Ayna's average MAPE is \param{7.6\%}
across all workloads and \param{5.2\%} for the Llama workloads.}

\microreve{Second, \atbcr{2}{Ayna} is the most accurate power model across all
evaluated models for both all-0s and random data patterns. \modelcr{2}{For
all-0s, the average (minimum--maximum) MAPE across evaluated workloads is
\param{21.6\%} (\param{11.0}--\param{23.8\%}) for DRAMSim3, \param{22.6\%}
(\param{10.9}--\param{26.7\%}) for FGDRAM HBM2, and \param{8.1\%}
(\param{6.6}--\param{11.3\%}) for Ayna; for random, it is \param{41.6\%}
(\param{21.8}--\param{48.9\%}), \param{20.1\%} (\param{9.3}--\param{28.0\%}),
and \param{7.6\%} (\param{4.8}--\param{16.3\%}), respectively.} For all-0s,
\atbcr{2}{Ayna} yields a \modelcr{2}{13.5} and \modelcr{2}{14.5} percentage
points (pp) better prediction accuracy than DRAMSim3 and FGDRAM HBM2,
respectively. \atbcr{2}{We attribute this to DRAMSim3 and FGDRAM HBM2's
relatively low IDD4R (read current) values.} \modelcr{2}{Excluding the power-off
current of our testing platform, DRAMSim3 and FGDRAM HBM2 \atbcr{3}{use} an
all-0s IDD4R of \param{\SI{1079}{\milli\ampere}} and
\param{\SI{966}{\milli\ampere}}, respectively, whereas we measure
\param{\SI{1716}{\milli\ampere}} on real HBM2 chips.} For random,
\atbcr{2}{Ayna} yields a \modelcr{2}{34.0} and \modelcr{2}{12.5} pp better
prediction accuracy than DRAMSim3 and FGDRAM HBM2, respectively.
\modelcr{2}{Third}, a data-pattern-dependence oblivious power model (e.g.,
DRAMSim3) significantly underpredicts the power consumed by real HBM2
\atbcr{1}{chip}s. For example, DRAMSim3 predicts that the Llama3.1-8B decode
workload with batch size = 8 consumes \modelcr{2}{\SI{2235}{\milli\watt}}
whereas on average across all tested HBM2 \atbcr{1}{chip}s that same workload
consumes \modelcr{2}{\SI{4201}{\milli\watt}} \atbcr{3}{when we use the random
data pattern} (DRAMSim3 has \modelcr{2}{46.6\%} MAPE for this workload).
\omcrcomment{3}{UNFAIRNESS COMMENT (REMINDER): Give insight Why do we do. The
things we do. This is descriptive but not insightful. A reader will basically be
at a loss as to why we do. Not actually model values better and can easily
conclude that the modeling is wrong or unfair.} Thus, it is important to account
for data pattern dependence in modeling HBM2 power, as also demonstrated
empirically by our characterization study (see~\secref{sec:dpd-current}).}

\subsection{\atbcr{3}{Modeling Data Pattern Dependence in Ayna}}
\label{sec:data-dependent-power}

\microlabel{CC4}
\microrevcommon{Data pattern \omcr{2}{significantly} affects HBM2 read current
(see~\secref{sec:dpd-current}) and read current is significantly higher than all
other command currents evaluated in our work \omcr{2}{(as shown in
\atbcr{3}{\secref{sec:characterization-all}})}.\atbcrcomment{2}{A prominent
non-evaluated current loop is IDD7 but it is not described in HBM2 spec.}
Therefore, it is important to understand how well existing power models predict
HBM2 read current with varying data patterns and improve \omcr{2}{HBM power
modeling}. We use FGDRAM as a baseline data-pattern-dependent power model.
FGDRAM~\cite{oconnor2017fine} lays out a foundational basis for
data-pattern-dependent high bandwidth memory power modeling. At a high level,
this model offers two ``toggle rate'' parameters that specify 1)~the toggling
activity in HBM-die-internal data signals (e.g., TSVs) and 2)~the toggling
activity in I/O data signals (e.g., DQs).} \microrevcommon{Based on our
empirical power \omcr{2}{consumption} vs data pattern results \omcr{2}{in real
HBM2 DRAM chips} and our \omcr{2}{informed} architectural \omcr{2}{hypotheses}
that explain these results (\secref{sec:dpd-current}), we develop an improved
data-pattern-dependent power model that decomposes \atbcr{3}{FGDRAM}'s
in-HBM-die data signal toggle rate parameter into two toggle rate parameters and
two coefficients: 1)~\emph{TSV toggle rate}\omcr{2}{, which} quantifies the
switching activity caused by a data pattern in the TSVs in an HBM
\atbcr{1}{chip}, and 2)~\emph{bank group bus toggle rate}\omcr{2}{, which}
quantifies the switching in the bank group bus in an HBM die. \atbcr{3}{By
separately modeling TSV and bank group bus toggle rates, Ayna predicts read data
movement energy per bit with a MAPE of 2.3\%, compared to 11.4\% for the FGDRAM
baseline.}}

\noindent
\textbf{\microrevcommon{Tuning the \atbcr{3}{FGDRAM} Model.}}
\microrevcommon{FGDRAM's data movement energy model consists of three
components: 1)~a data-independent pre-global-sense-amplifier energy
($E_{Pre\text{-}GSA}$), 2)~a data-dependent post-global-sense-amplifier energy
($E_{Post\text{-}GSA}$), and 3)~a data-dependent I/O energy ($E_{IO}$).
\atbcr{2}{Adopting the model described in~\cite{oconnor2017fine},} we model
these components as:
\begin{align}
E_{Pre\text{-}GSA} &= Pre\text{-}GSA, \\
E_{Post\text{-}GSA} &= Post\text{-}GSA \cdot ToggleRate_{Post\text{-}GSA}, \\
E_{IO} &= IO \cdot ToggleRate_{IO}.
\end{align}
The total FGDRAM HBM2 data movement energy per bit is therefore:
\begin{equation}
E_{FGDRAM}^{HBM2}
=
E_{Pre\text{-}GSA}
+
E_{Post\text{-}GSA}
+
E_{IO}.
\end{equation}}

\noindent
\microrevcommon{We fit this model to our measurements (in
\figref{fig:current_vs_bus_flip}) by converting $E_{FGDRAM}^{HBM2}$ into a read
current via DRAMPower's read energy equation. We use
$ToggleRate_{Post\text{-}GSA}$ and $ToggleRate_{IO}$ to model the effect of
tested data patterns on power: as toggle rate increases from 0.0 to 1.0, power
increases linearly.\footnote{\microrevcommon{FGDRAM HBM2~\cite{oconnor2017fine}
builds on the Rambus power model~\cite{vogelsang2010understanding}, which scales
power linearly with toggle rate.}} FGDRAM's published coefficients (Table~3
in~\cite{oconnor2017fine}) reproduce neither end of our observed data-pattern vs
current range (\figref{fig:current_vs_bus_flip}): they \omcr{2}{underestimate}
both the minimum observed current (all-0s) and the maximum observed current
(``0110'' with BG Bus Toggle). Therefore, we recalibrate the model to
\omcr{2}{fit} our \omcr{2}{empirical} data \omcr{2}{from real chips}.
Specifically, we (i)~set the data-independent $Pre\text{-}GSA$
\atbcr{2}{component} to the measured all-0s (zero-toggle) current, (ii)~retain
FGDRAM's published $IO$ coefficient, and (iii)~fit the single $Post\text{-}GSA$
coefficient \omcr{2}{using} least squares \omcr{2}{fit} over all \param{32}
measured data patterns.}

\noindent
\textbf{\microrevcommon{Tuning \atbcr{3}{Ayna}.}}
\microrevcommon{\atbcr{2}{Ayna} replaces \atbcr{3}{FGDRAM's} single in-HBM-die
data-dependent term ($E_{Post\text{-}GSA}$) with two terms \atbcr{2}{that we
introduce} \atbcr{2}{to} capture the switching a data pattern induces on the
TSVs and on the bank group bus:
\begin{align}
E_{TSV} &= TSV \cdot ToggleRate_{TSV}, \\
E_{BG}  &= BG  \cdot ToggleRate_{BG},
\end{align}
thus the improved data movement energy per bit is
\begin{equation}
E_{Improved}^{HBM2}
= E_{Pre\text{-}GSA} + E_{TSV} + E_{BG} + E_{IO}.
\end{equation}} 
\microrevcommon{We compute $ToggleRate_{TSV}$ and $ToggleRate_{BG}$ per data
pattern from the switching each data pattern induces on the TSV and bank group
bus signals, and fit all coefficients \omcr{2}{using} least squares
\omcr{2}{fit} against the same per-pattern read-current measurements used for
\atbcr{3}{FGDRAM} (\figref{fig:current_vs_bus_flip}).}
\microrevcommon{\figref{fig:data_pattern_model} shows measured IDD4R (x-axis) vs
\atbcr{2}{IDD4R predicted using the two data-pattern-dependent power models
(y-axis)}: \modelcr{2}{the tuned} \emph{\modelcr{3}{FGDRAM}} using orange circles and
\modelcr{2}{the tuned} \emph{\modelcr{3}{Ayna}} using blue crosses.}

\begin{figure}[!ht]
  \centering
  \includegraphics[width=1\linewidth]{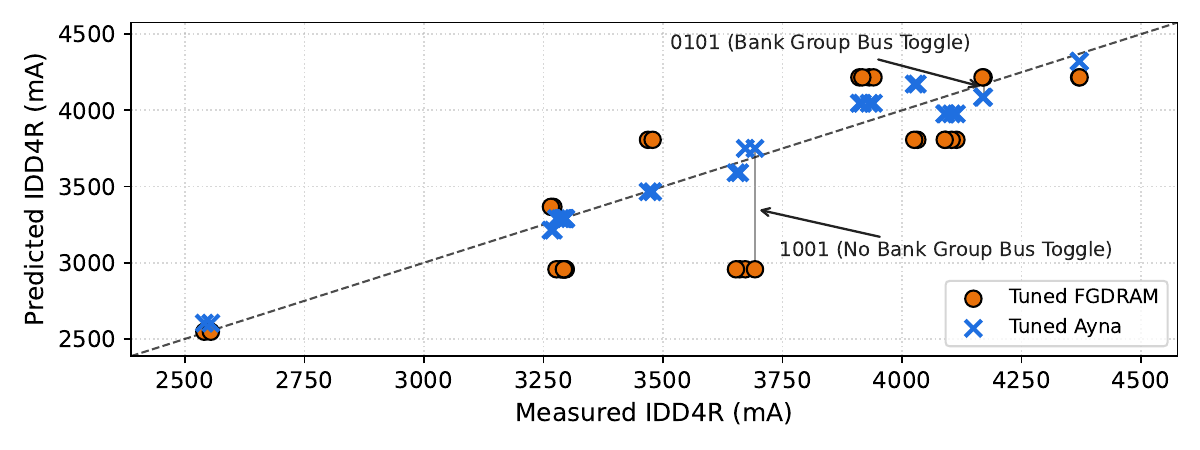}
  \caption{\microrevcommon{Empirically-observed IDD4R vs IDD4R predicted by
  \atbcr{2}{the tuned \modelcr{3}{FGDRAM} power model} and \modelcr{2}{the tuned}
  \modelcr{3}{Ayna}.}}
  \label{fig:data_pattern_model}
\end{figure}

\observation{\microrevcommon{Our improved data-pattern-dependent power model
predicts HBM2 read current substantially more accurately than the
\omcr{2}{state-of-the-art FGDRAM power model}.}}

\microrevcommon{We observe that across all \param{32} tested data patterns
\atbcr{2}{Ayna's} predictions are close to the $y\!=\!x$ line with coefficient
of \atbcr{3}{determination} ($R^2$) of $\param{0.97}$ whereas the
\omcr{2}{state-of-the-art} baseline's predictions are away from the $y\!=\!x$
line with $R^2\!=\!\param{0.41}$.}

\take{\microrevcommon{Accurately modeling data-pattern-dependent HBM2 read power
requires accounting for switching on the internal TSV and bank group buses.
\omcr{2}{We design a new, open source power model that takes this important
information into account \atbcr{3}{and thus provides much higher modeling
accuracy}.}}}

\subsection{\microrevcommon{HBM3E-Based Validation}}
\label{sec:hbm3e-validation}

\microlabel{CC5}
\microrevcommon{\omcr{2}{Ayna} can be used to accurately predict the power
consumption of more advanced high bandwidth memory systems. We augment
\atbcr{2}{Ayna} with HBM3E~\cite{jedechbm3} organization and command timing
parameters from
Ramulator2.1~\cite{ramulator2github,ramulator2hbm3timingparameters}. We evaluate
\atbcr{2}{Ayna's} accuracy in predicting HBM3E power. \modelcr{2}{We find that,
within HBM3E's specified supply voltage range (1.067 V -- 1.177
V)~\cite{jedechbm3}, Ayna predicts the power consumed by a
high-throughput\atbcrcomment{3}{this is with former Ayna-DPD} read-intensive
microbenchmark executed on three HBM3E-based~\cite{jedechbm3} NVIDIA H200
systems~\cite{nvidia-h200-datasheet} with as small as \param{2.4\%} error.}}

\noindent
\microrevcommon{\textbf{Evaluation.} We measure GPU memory power (reported by
the ``nvidia-smi'' tool~\cite{nvidia-smi}) in \omcr{2}{three} NVIDIA H200
systems with identical HBM3E configuration\omcr{2}{s}. We develop and execute a
microbenchmark that exerts high read throughput\footnote{\microrevcommon{We
focus on read power because read operations consume the most power across all
evaluated DRAM operations \omcr{2}{(see~\secref{sec:characterization-all})}. Our
H200 experiments confirm that read operations consume more GPU memory power than
write operations at equal memory throughput.}} to stress the available bandwidth
in the H200 system (4.8 TB/s)\atbcrcomment{2}{This is max HBM3E bw at h200 hbm
freq}. We develop a synthetic DRAM command trace that is designed to exert the
same sustained high read throughput pressure in the modeled HBM3E
system.\footnote{\modelcr{2}{The trace \modelcr{3}{activates a random row,
issues four reads to consecutive columns in that row, and precharges the bank,
cycling through all 16 banks. Consecutive reads alternate between two bank
groups, so that reads issue} back-to-back \modelcr{3}{and} fully utilize the
available memory bandwidth\modelcr{3}{, while the activates for the next banks
overlap with in-flight reads. The trace also issues} an all-bank refresh command
every standard refresh interval (tREFI, \SI{3.9}{\micro\second}). The resulting
read throughput (\SI{4.50}{TB/s}) matches the throughput we measure on the H200
systems (\SI{4.52}{TB/s}, 94\% of the peak 4.8 TB/s) to within 1\%.}}
\microrevcommon{Because no public HBM3E IDD values exist, we configure
\atbcr{2}{Ayna} by extrapolating our empirically-characterized HBM2 IDD values
to HBM3E's \param{6.4}~Gbps data rate and \param{32}-bit
pseudo-channel\modelcr{2}{.} \atbcr{3}{We sweep modeled HBM3E voltage from
1.067~V to 1.177~V (specified supply voltage range~\cite{jedechbm3}).}
\atbcr{3}{\figref{fig:empirical-hbm3e} (left) shows \modelcr{2}{predicted HBM3E
power consumption for the high read throughput trace (Ayna) and measured HBM3E
power for the microbenchmark (H200).\footnote{\atbcr{4}{The exact HBM3E supply voltage value
used in H200 is \emph{not} disclosed. The used voltage value is likely within the
supply voltage range (1.067~V to 1.177~V) specified in~\cite{jedechbm3}.}} \figref{fig:empirical-hbm3e} (right) shows
predicted HBM3E power consumption for a trace that \emph{only} issues refresh
commands (Ayna) and measured HBM3E idle power consumption (H200).}}}}

\begin{figure}[!ht]
  \centering
  \begin{subfigure}{0.49\linewidth}
    \centering
    \includegraphics[width=\linewidth]{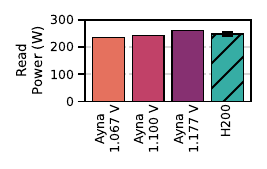}
  \end{subfigure}
  \hfill
  \begin{subfigure}{0.49\linewidth}
    \centering
    \includegraphics[width=\linewidth]{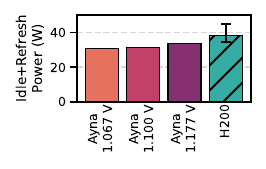}

  \end{subfigure}
  \caption{\microrevcommon{\atbcr{3}{Modeled vs measured HBM3E read (left) and
  idle + refresh (right) power consumption at HBM3E's minimum, nominal, and
  maximum supply voltage} \modelcr{3}{(VDD = 1.067 V, 1.100 V, and 1.177 V)}.
  \modelcr{2}{Error bars show the range across the three tested H200 systems.}}}
  \label{fig:empirical-hbm3e}
  \vspace{0.5em}
\end{figure}

\modelcr{2}{We make two observations. First,
\atbcr{3}{\figref{fig:empirical-hbm3e} (left) shows that} Ayna predicts the
measured power under the high throughput \omcr{3}{read} microbenchmark with
5.3\%, 2.4\%, and 4.5\% error at the minimum, nominal, and maximum VDD,
respectively (the measured power lies between the nominal- and maximum-VDD
predictions). Second, \atbcr{3}{\figref{fig:empirical-hbm3e} (right) shows that}
Ayna predicts the measured idle power with 20.3\%, 17.7\%, and 12.0\% error at
the minimum, nominal, and maximum VDD, respectively. We attribute the remaining
differences to process and circuit differences between HBM2 and HBM3E and to
power that the measured GPU memory power includes but Ayna does \emph{not}
model. We conclude that Ayna accurately predicts HBM3E power consumption under a
high read throughput microbenchmark and idle HBM3E power consumption across
HBM3E's specified supply voltage range.}

\subsection{\microrevcommon{Case Study: HBM4 Read Energy Estimation}}
\label{sec:hbm4-estimation}

\microlabel{CC5}
\microrevcommon{\omcr{2}{Ayna} enables exploration of future memory systems. As
a concrete case study, we show \atbcr{3}{what} HBM4 read energy could look like,
accounting for architectural and HBM core \atbcr{3}{voltage (VDDC)} and I/O
\atbcr{3}{voltage (VDDQ)} changes between HBM3E~\cite{jedechbm3} and
HBM4~\cite{jedechbm4}. We build on our HBM3E-validated power model and update it
with HBM4's doubled 2048-bit interface, faster speed bins, and split VDDC/VDDQ
power rails from the JEDEC HBM4 standard. Because no public HBM4 IDD values
exist, we 1)~carry over our HBM3E currents\modelcr{2}{, 2)~scale the
precharge/active standby (IDD2N/3N), read/write (IDD4R/W), activate--precharge
(IDD0), and refresh (IDD5B) currents \atbcr{3}{according to their measured
trends (as we measure on HBM2, \secref{sec:data-rate-sensitivity})}, and 3)~pair
the DQ read I/O energy with the lower VDDQ (0.7 -- 0.9 V) rail while pairing
core/TSV/bank-group read energy with VDDC (1.05 V).} \atbcr{3}{We use the
synthetic command trace with high read throughput described
in~\secref{sec:hbm3e-validation}.} \microrevcommon{\figref{fig:hbm4-energy}
\atbcr{3}{(left) shows modeled HBM3E read energy (pJ/bit) across the
4.8--6.4\,Gbps/pin speed bins for VDD = 1.10~V (red circles) and HBM4 read
energy across the 4.8--8.0\,Gbps/pin speed bins for three \omcr{2}{DRAM} core
(VDDC) \& I/O (VDDQ) voltage configurations.}} \modelcr{2}{The dashed line and
shaded band show the mean and range of the HBM3E read energy we measure on the
three H200 systems at 6.4\,Gbps/pin (\secref{sec:hbm3e-validation}).}}
\atbcr{3}{\figref{fig:hbm4-energy} (right) shows the modeled HBM4 read energy at
8 Gbps/pin over three VDDC and VDDQ configurations (x-axis), normalized to HBM3E
read energy at 6.4 Gbps/pin (dashed line).}

\begin{figure}[!ht]
  \centering
  \includegraphics[width=1\linewidth]{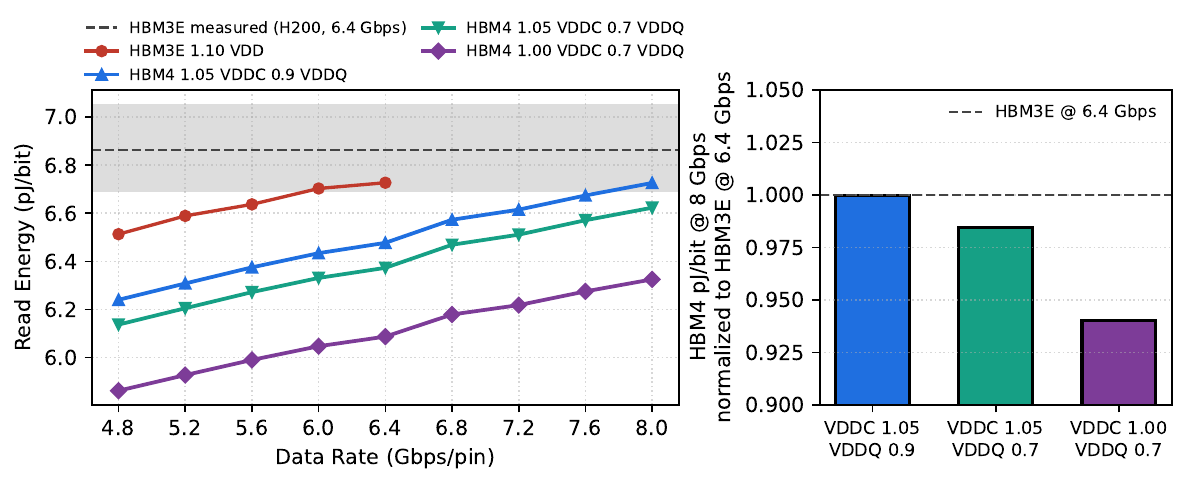}
  \caption{\microrevcommon{HBM4 and HBM3E modeled read energy with varying data
  rate (left). \omcr{2}{Modeled} HBM4 read energy \modelcr{2}{at 8.0\,Gbps/pin}
  \atbcr{3}{normalized} to HBM3\modelcr{2}{E} read energy \modelcr{2}{at
  6.4\,Gbps/pin} (right).}}
  \label{fig:hbm4-energy}
\end{figure}

\microrevcommon{\modelcr{2}{We make three major observations. First,
\atbcr{3}{\figref{fig:hbm4-energy} (left) shows that} at 6.4\,Gbps/pin data
rate, HBM4's lower supply voltages reduce read energy per bit by 3.7\%, 5.3\%,
and 9.5\% compared to HBM3E for (VDDC=1.05\,V, VDDQ=0.9\,V), (1.05\,V, 0.7\,V),
and (1.00\,V, 0.7\,V), respectively. Second, read energy per bit increases with
data rate for both HBM3E and HBM4 (e.g., from 6.24 to 6.73\,pJ/bit between 4.8
and 8.0\,Gbps/pin for HBM4 at VDDC=1.05\,V/VDDQ=0.9\,V). \atbcr{3}{There are two
reasons for this. (1) Standby (IDD2N/3N) and read (IDD4R) currents increase with
data rate and execution time decreases as data rate
increases (\secref{sec:data-rate-sensitivity}). These two effects largely offset each
other, keeping the read energy contributions of IDD2N/3N and IDD4R fixed across
the simulated data rates. (2)~The difference between the activate--precharge
current (IDD0) and the standby current (IDD2N/3N) grows with data rate (from
30.5 to 40.8 mA per pseudo-channel as data rate increases from 4.8 Gbps/pin to
8.0 Gbps/pin). Therefore, the energy contribution of an activate command
increases with data rate, causing the overall read energy per bit to increase.}
Third, \atbcr{3}{\figref{fig:hbm4-energy} (right) shows that} HBM4 at its
maximum 8.0\,Gbps/pin data rate consumes \atbcr{3}{similar} energy per bit (to
within 0.1\%) as HBM3E at its maximum 6.4\,Gbps/pin for (VDDC=1.05\,V,
VDDQ=0.9\,V). HBM4 at 8.0 Gbps/pin consumes 1.5\% and 6.0\% \atbcr{3}{lower}
energy per bit for (1.05\,V, 0.7\,V) and (VDDC=1.00\,V, VDDQ=0.7\,V),
respectively. We conclude that reduced core and I/O supply voltages could make
HBM4 reads more energy-efficient than HBM3E
reads.}}\footnote{\microrevcommon{Technology improvements could further improve
HBM4's read energy benefits over HBM3E. Our power model establishes the
scientific basis for exploring future high bandwidth memory systems. We believe
future work should experimentally characterize HBM4 \atbcr{1}{chip}s and improve
our power model.}}
\section{Related Work}

\omcr{2}{To our knowledge,} this is the first work to 1)~\atbcr{2}{rigorously
characterize High Bandwidth Memory (HBM) power consumption using 36 \emph{real
HBM2 \atbcr{1}{chip}s}} \microrevcommon{and 2)~develop \atbcr{3}{a new} \omcr{4}{and more}
accurate \atbcr{3}{HBM} power model \omcr{4}{(called Ayna)} \atbcr{3}{based on this rigorous
characterization}}. We discuss relevant prior work.

\noindent
\textbf{Experimental DRAM Power
Characterization~\cite{stow2019power,ghose2018what,jung2016new,
jung2017platform, mathew2018analysis}.} Stow et al.~\cite{stow2019power} study
power consumption in HBM2 \atbcr{1}{chip}s in a GPU. Their goal is to profile
and reason about die-stacked-memory power in a modern GPU system at a high
level. \omcr{2}{In contrast,} our goal is to characterize HBM2 power directly at
the DRAM-command level and expose fine-grained structural \omcr{2}{power
variation} and data-pattern-dependence effects. Our work
complements~\cite{stow2019power} with a lower-level characterization of HBM2.
Unlike~\cite{stow2019power}, we execute standard JEDEC IDD loops and carefully
designed fine-grained command sequences to reveal new insights into the spatial
variation and data pattern dependence of power in HBM2 \atbcr{1}{chip}s
\omcr{2}{and develop a new, accurate HBM power model.}

Several prior works~\cite{jung2016new, jung2017platform,
mathew2018analysis,ghose2018what} study real \atbcr{3}{planar} DDRx DRAM chips
to measure IDD values. VAMPIRE~\cite{ghose2018what} shows that vendor-specified
current values alone are insufficient for accurate DRAM power modeling. Our work
follows the same empirical philosophy for HBM2: we measure standard IDD values
on real HBM2 \atbcr{1}{chip}s, study data-pattern dependence, \atbcr{3}{and}
investigate structural variation. Several qualitative observations are
consistent across VAMPIRE and our work: we observe significant variation across
tested \atbcr{1}{chip}s and tested banks in a channel, higher read current than
write current, and sensitivity to the transferred data values and data bus
toggling activity. Our HBM2 characterization also uncovers new insights into
HBM2 power \omcr{3}{consumption} that these prior works~\cite{jung2016new,
jung2017platform, mathew2018analysis,ghose2018what} could \emph{not} study in planar
DRAM.

\noindent
\textbf{Reduced-Voltage DRAM Characterization~\cite{chang2017understanding,
yaglikci2022understanding, larimi2021understanding}.} Prior experimental studies
characterize DRAM behavior under reduced supply or wordline voltage, including
in HBM2 \atbcr{1}{chip}s~\cite{larimi2021understanding}, and show how
lower-voltage operation affects power consumption, reliability, latency, and
performance~\cite{chang2017understanding, yaglikci2022understanding,
larimi2021understanding}. These works are important because low-voltage
operation can reduce DRAM energy. Our \atbcr{4}{work} is orthogonal to these
characterization works. We characterize nominal-voltage HBM2 power consumption
across standard current measurement loops, temperature levels, data rates,
fine-grained structures, and transferred data patterns. Our results can help
future undervolting studies, e.g., \atbcr{3}{to} reason about which HBM2 access
patterns are likely to stress the power delivery network most.

\noindent
\textbf{DRAM Power Models~\powerModelsAllCitations{}.} Widely-used DRAM power
models and simulators~\powerModelsAllCitations{} are useful for
\omcr{2}{architectural innovation and} design space exploration. However, these
models rely on vendor-provided currents or highly-proprietary DRAM manufacturing
technology parameters, and thus do \emph{not} include or consider the
i)~empirically-obtained real HBM2 IDD values, ii)~HBM2-specific structural
variation, and iii)~HBM2-specific data-pattern sensitivity. Our characterization
study complements these tools by providing empirical measurements from real HBM2
devices\omcr{2}{, and, building on our real measurements, we \omcr{3}{develop
and open source} Ayna, the most accurate HBM power model to date.}
\section{Conclusion}

We introduced the first rigorous, comprehensive experimental characterization of
HBM2 power consumption using real HBM2 \atbcr{1}{chip}s. Our work, for the first
time, provides the empirical basis (IDD values \atbcr{3}{and data dependence
characteristics for internal HBM chip data buses}) for accurate HBM2 power
modeling. Our study makes \omcr{2}{three major} contributions: 1)~we
systematically characterize HBM2 current across standard IDD current measurement
loops, temperature values, data rates, HBM2 structures, and transferred data
patterns, 2)~we leverage \omcr{2}{our empirical} insights to engineer access and
data patterns that \atbcr{3}{induce higher power consumption} in real HBM2
\atbcr{1}{chip}s than standard JEDEC current measurement loops\omcr{2}{, and
3)~we build and open source Ayna, the most accurate HBM power model to date.}
\omcr{2}{We hope and believe that our findings and open source artifacts will
accelerate architectural innovation in HBM-based systems and enable \omcr{3}{the
design of better (i.e., both more efficient and higher performance)} \omcr{4}{computing systems that make use of HBM, including} AI/ML
training and serving systems.}

\section*{Acknowledgments}

\atbcr{1}{We thank the anonymous reviewers of MICRO 2026 for their feedback. We
thank the SAFARI Research Group members for their constructive feedback and for
providing a stimulating intellectual, scientific, scholarly environment. We
acknowledge the generous gift funding provided by our industrial partners
(especially Google, Huawei, Intel, Microsoft), which has been instrumental in
enabling the research we have been conducting on read disturbance in DRAM in
particular and memory systems in general~\cite{mutlu2013memory, mutlu2025memory,
mutlu2024memory, mutlu2023fundamentally, mutlu2025modern, mutlu2017rowhammer,
mutlu2019retrospective, mutlu2019processing, mutlu2015research,
mutlu2023retrospectiveretention, ghose2019processing,
mutlu2023retrospectiveflipping, mutlu2015main, luna2022benchmarking,
oliveira2022accelerating, singh2021fpga, oliveira2021damov,
kakolyris2026columnkeeper, cai2017flashtbd, mutlu2020intelligentdate,
mutlu2023tesseractretrospective, mutlu2023selfretrospective, luo2026ramulator21,
yuksel2026memory, bostanci2026cutmess, cai2017errors, olgun2026laboratory,
bera2026machine}. This work was in part supported by a Google Security and
Privacy Research Award and the Microsoft Swiss Joint Research Center.}

\balance
\bibliographystyle{IEEEtran}
\bibliography{refs}
\newpage
\nobalance  

\appendix
\section{Detailed Measurement Results}
\label{sec:appendix-measurements}

This appendix lists the measurements behind the characterization results of
\secref{sec:characterization-all}. Each table is computed from the same
measurements and with the same processing as the figure(s) it accompanies.
All currents are in mA, after subtracting the board's current with the HBM2 chip
powered off. In rows of individual chips, ``$\pm$'' shows the standard deviation
of that chip's measurements; in summary rows, it shows the standard deviation of
the per-chip averages across chips. As in \secref{sec:characterization-all}, we
group chips by their position in the FPGA package (even- vs odd-numbered). Some
experiments cover a subset of the tested chips; each table states the chips it
includes.

\subsection{Standard IDD Measurements}
\label{sec:appendix-idd}

Table~\ref{tab:idd-per-chip} lists the measured IDD values (in mA) of each tested HBM2
chip for all seven current measurement loops (Table~\ref{tab:idd-loops}). Each
value corresponds to one bar in
Figs.~\ref{fig:idd2n_current_across_stacks}--\ref{fig:idd5_current_across_stacks}.

\begin{table*}[!b]
\centering
\caption{Measured IDD values (in mA) of each tested HBM2 chip: average $\pm$
standard deviation of the last 60 measurements of each loop. The last three
rows show the average $\pm$ standard deviation of the per-chip averages across
chips.}
\label{tab:idd-per-chip}
\footnotesize
\renewcommand{\arraystretch}{0.93}
\begin{tabular}{rrrrrrrr}
\hline
Chip & IDD2N & IDD3N1 & IDD3N16 & IDD0 & IDD4R & IDD4W & IDD5B \\
\hline
0 & 283.8\,$\pm$\,5.4 & 240.9\,$\pm$\,3.3 & 270.2\,$\pm$\,6.2 & 324.3\,$\pm$\,3.0 & 2816.8\,$\pm$\,5.6 & 1891.4\,$\pm$\,4.7 & 707.8\,$\pm$\,6.0 \\
2 & 254.5\,$\pm$\,4.7 & 250.6\,$\pm$\,6.8 & 256.6\,$\pm$\,4.3 & 325.5\,$\pm$\,9.9 & 2633.3\,$\pm$\,12.7 & 1874.5\,$\pm$\,5.1 & 668.5\,$\pm$\,5.4 \\
4 & 256.4\,$\pm$\,4.4 & 257.1\,$\pm$\,7.9 & 274.9\,$\pm$\,7.4 & 339.5\,$\pm$\,8.5 & 2896.8\,$\pm$\,5.8 & 1923.2\,$\pm$\,5.3 & 717.7\,$\pm$\,4.1 \\
6 & 253.2\,$\pm$\,6.4 & 216.3\,$\pm$\,3.5 & 233.1\,$\pm$\,4.5 & 316.0\,$\pm$\,9.6 & 2939.7\,$\pm$\,4.2 & 1831.3\,$\pm$\,6.6 & 682.0\,$\pm$\,4.2 \\
8 & 226.4\,$\pm$\,3.4 & 235.0\,$\pm$\,5.4 & 246.4\,$\pm$\,5.1 & 291.4\,$\pm$\,4.8 & 2814.2\,$\pm$\,10.3 & 1880.8\,$\pm$\,5.9 & 653.4\,$\pm$\,15.7 \\
10 & 243.2\,$\pm$\,3.5 & 225.3\,$\pm$\,3.2 & 231.8\,$\pm$\,4.1 & 300.7\,$\pm$\,6.0 & 2786.2\,$\pm$\,3.8 & 1837.8\,$\pm$\,6.7 & 657.4\,$\pm$\,6.8 \\
12 & 262.5\,$\pm$\,6.2 & 276.9\,$\pm$\,5.9 & 287.8\,$\pm$\,3.2 & 338.9\,$\pm$\,3.1 & 2874.8\,$\pm$\,4.4 & 1908.2\,$\pm$\,3.0 & 676.3\,$\pm$\,6.9 \\
14 & 239.2\,$\pm$\,8.7 & 250.5\,$\pm$\,5.1 & 248.8\,$\pm$\,3.2 & 297.6\,$\pm$\,3.6 & 2801.5\,$\pm$\,4.6 & 1842.9\,$\pm$\,5.7 & 661.1\,$\pm$\,5.3 \\
16 & 234.9\,$\pm$\,8.0 & 219.2\,$\pm$\,4.2 & 233.2\,$\pm$\,3.7 & 304.4\,$\pm$\,6.3 & 2766.9\,$\pm$\,3.2 & 1831.5\,$\pm$\,7.1 & 639.5\,$\pm$\,4.5 \\
18 & 249.1\,$\pm$\,2.8 & 228.6\,$\pm$\,8.1 & 244.1\,$\pm$\,3.0 & 290.3\,$\pm$\,6.1 & 3025.7\,$\pm$\,6.5 & 2048.8\,$\pm$\,4.4 & 672.6\,$\pm$\,3.9 \\
20 & 222.6\,$\pm$\,7.3 & 215.2\,$\pm$\,4.6 & 222.6\,$\pm$\,3.8 & 296.8\,$\pm$\,6.8 & 2759.2\,$\pm$\,3.5 & 1814.2\,$\pm$\,8.0 & 619.9\,$\pm$\,7.5 \\
22 & 222.5\,$\pm$\,3.3 & 222.9\,$\pm$\,3.4 & 234.0\,$\pm$\,6.8 & 278.8\,$\pm$\,6.2 & 2792.7\,$\pm$\,6.7 & 1831.6\,$\pm$\,6.6 & 630.3\,$\pm$\,6.3 \\
24 & 257.1\,$\pm$\,6.9 & 221.7\,$\pm$\,6.5 & 232.1\,$\pm$\,3.4 & 316.3\,$\pm$\,8.8 & 2777.9\,$\pm$\,9.6 & 1813.2\,$\pm$\,3.8 & 655.2\,$\pm$\,4.7 \\
26 & 229.4\,$\pm$\,6.0 & 228.0\,$\pm$\,5.6 & 235.2\,$\pm$\,5.7 & 308.4\,$\pm$\,6.7 & 2793.6\,$\pm$\,5.7 & 1847.0\,$\pm$\,6.0 & 654.2\,$\pm$\,3.6 \\
28 & 219.3\,$\pm$\,9.8 & 202.2\,$\pm$\,7.1 & 228.7\,$\pm$\,4.9 & 295.7\,$\pm$\,9.8 & 2710.9\,$\pm$\,2.8 & 1774.9\,$\pm$\,5.8 & 642.1\,$\pm$\,7.5 \\
30 & 243.3\,$\pm$\,2.6 & 213.0\,$\pm$\,6.0 & 243.1\,$\pm$\,6.9 & 313.4\,$\pm$\,7.5 & 2799.9\,$\pm$\,4.6 & 1846.1\,$\pm$\,10.7 & 667.8\,$\pm$\,4.9 \\
32 & 254.9\,$\pm$\,7.9 & 249.2\,$\pm$\,3.2 & 268.3\,$\pm$\,3.9 & 333.2\,$\pm$\,8.7 & 2857.2\,$\pm$\,4.7 & 1896.1\,$\pm$\,5.4 & 690.8\,$\pm$\,4.3 \\
34 & 231.7\,$\pm$\,4.4 & 236.6\,$\pm$\,4.0 & 234.0\,$\pm$\,9.4 & 318.5\,$\pm$\,11.0 & 2777.3\,$\pm$\,4.5 & 1876.0\,$\pm$\,8.2 & 673.3\,$\pm$\,7.3 \\
\hline
1 & 329.4\,$\pm$\,4.8 & 247.3\,$\pm$\,4.5 & 266.3\,$\pm$\,5.6 & 323.1\,$\pm$\,3.0 & 2743.7\,$\pm$\,3.7 & 2148.2\,$\pm$\,4.2 & 758.0\,$\pm$\,2.9 \\
3 & 306.9\,$\pm$\,6.4 & 221.7\,$\pm$\,5.4 & 261.5\,$\pm$\,7.4 & 353.4\,$\pm$\,3.1 & 3006.0\,$\pm$\,4.4 & 2115.4\,$\pm$\,7.3 & 743.3\,$\pm$\,6.2 \\
5 & 365.0\,$\pm$\,7.3 & 247.7\,$\pm$\,8.3 & 244.4\,$\pm$\,4.1 & 331.2\,$\pm$\,6.5 & 3091.7\,$\pm$\,5.8 & 2197.3\,$\pm$\,6.2 & 796.2\,$\pm$\,6.0 \\
7 & 301.5\,$\pm$\,3.7 & 224.5\,$\pm$\,6.8 & 242.5\,$\pm$\,9.1 & 293.5\,$\pm$\,8.1 & 2976.6\,$\pm$\,2.6 & 2095.8\,$\pm$\,9.0 & 729.3\,$\pm$\,2.8 \\
9 & 301.1\,$\pm$\,5.2 & 245.7\,$\pm$\,6.6 & 278.4\,$\pm$\,9.0 & 296.1\,$\pm$\,4.9 & 2947.2\,$\pm$\,4.3 & 2081.6\,$\pm$\,5.6 & 711.4\,$\pm$\,7.5 \\
11 & 294.5\,$\pm$\,5.6 & 225.6\,$\pm$\,7.3 & 255.2\,$\pm$\,3.7 & 286.2\,$\pm$\,4.2 & 2915.7\,$\pm$\,4.0 & 2061.2\,$\pm$\,4.9 & 703.4\,$\pm$\,5.1 \\
13 & 289.5\,$\pm$\,7.5 & 269.5\,$\pm$\,9.4 & 290.4\,$\pm$\,5.7 & 325.9\,$\pm$\,7.9 & 2969.1\,$\pm$\,4.8 & 2113.2\,$\pm$\,3.2 & 721.9\,$\pm$\,5.2 \\
15 & 319.0\,$\pm$\,4.9 & 240.2\,$\pm$\,5.6 & 244.2\,$\pm$\,6.4 & 317.9\,$\pm$\,7.0 & 2980.5\,$\pm$\,12.6 & 2101.9\,$\pm$\,5.1 & 739.2\,$\pm$\,3.6 \\
17 & 298.1\,$\pm$\,3.9 & 208.0\,$\pm$\,6.1 & 229.1\,$\pm$\,9.2 & 302.9\,$\pm$\,4.8 & 2951.6\,$\pm$\,3.8 & 2099.5\,$\pm$\,5.7 & 732.2\,$\pm$\,7.5 \\
19 & 290.3\,$\pm$\,6.7 & 217.7\,$\pm$\,5.2 & 236.5\,$\pm$\,5.5 & 304.7\,$\pm$\,7.4 & 3003.6\,$\pm$\,2.5 & 2114.1\,$\pm$\,8.4 & 736.9\,$\pm$\,7.4 \\
21 & 290.7\,$\pm$\,4.1 & 214.7\,$\pm$\,6.4 & 234.2\,$\pm$\,3.7 & 303.6\,$\pm$\,5.3 & 2940.9\,$\pm$\,9.5 & 2085.3\,$\pm$\,8.0 & 705.1\,$\pm$\,6.5 \\
23 & 287.1\,$\pm$\,2.9 & 225.8\,$\pm$\,4.5 & 247.3\,$\pm$\,6.3 & 291.3\,$\pm$\,4.1 & 2921.1\,$\pm$\,6.1 & 2088.3\,$\pm$\,2.5 & 714.5\,$\pm$\,3.8 \\
25 & 291.1\,$\pm$\,3.8 & 228.0\,$\pm$\,5.3 & 238.7\,$\pm$\,3.8 & 295.5\,$\pm$\,4.3 & 2900.7\,$\pm$\,6.5 & 2067.6\,$\pm$\,4.0 & 708.2\,$\pm$\,3.6 \\
27 & 276.7\,$\pm$\,8.4 & 221.1\,$\pm$\,5.4 & 219.3\,$\pm$\,6.7 & 307.2\,$\pm$\,4.0 & 2961.7\,$\pm$\,3.8 & 2092.4\,$\pm$\,5.0 & 705.2\,$\pm$\,5.6 \\
29 & 263.5\,$\pm$\,4.1 & 199.3\,$\pm$\,2.6 & 205.4\,$\pm$\,10.5 & 294.1\,$\pm$\,3.4 & 2849.5\,$\pm$\,6.2 & 2008.9\,$\pm$\,4.2 & 679.7\,$\pm$\,5.0 \\
31 & 293.2\,$\pm$\,9.5 & 227.1\,$\pm$\,5.6 & 244.9\,$\pm$\,2.3 & 292.1\,$\pm$\,4.0 & 2954.7\,$\pm$\,4.0 & 2084.0\,$\pm$\,2.7 & 735.1\,$\pm$\,8.5 \\
33 & 315.6\,$\pm$\,7.4 & 258.0\,$\pm$\,3.2 & 269.5\,$\pm$\,6.6 & 317.8\,$\pm$\,4.5 & 3019.8\,$\pm$\,4.9 & 2126.8\,$\pm$\,4.1 & 755.5\,$\pm$\,6.4 \\
35 & 321.4\,$\pm$\,4.3 & 233.6\,$\pm$\,7.0 & 255.2\,$\pm$\,3.3 & 323.4\,$\pm$\,8.3 & 2956.9\,$\pm$\,5.8 & 2098.3\,$\pm$\,3.6 & 739.1\,$\pm$\,5.4 \\
\hline
\textbf{Even chips} & 243.6\,$\pm$\,16.5 & 232.7\,$\pm$\,18.2 & 245.8\,$\pm$\,17.9 & 310.5\,$\pm$\,17.0 & 2812.5\,$\pm$\,83.7 & 1865.0\,$\pm$\,57.7 & 665.0\,$\pm$\,24.3 \\
\textbf{Odd chips} & 301.9\,$\pm$\,21.9 & 230.9\,$\pm$\,17.2 & 247.9\,$\pm$\,20.1 & 308.9\,$\pm$\,17.2 & 2949.5\,$\pm$\,70.9 & 2098.9\,$\pm$\,37.3 & 728.6\,$\pm$\,25.6 \\
\textbf{All chips} & 272.7\,$\pm$\,35.0 & 231.8\,$\pm$\,17.7 & 246.9\,$\pm$\,19.1 & 309.7\,$\pm$\,17.1 & 2881.0\,$\pm$\,103.5 & 1981.9\,$\pm$\,126.6 & 696.8\,$\pm$\,40.4 \\
\hline
\end{tabular}

\end{table*}

\subsection{Temperature Sensitivity}
\label{sec:appendix-temperature}

Table~\ref{tab:temperature-per-chip} lists, for each tested chip, the chip's
temperature (in \textdegree C) during the IDD2N loop (\figref{fig:idle-current-noise}) and the
slope of a linear fit of current vs.\ temperature (in mA/\textdegree C) for each loop
(\secref{sec:temperature-sensitivity}, \figref{fig:temperature_sensitivity_all}).

\begin{table*}[!t]
\centering
\caption{Temperature during the IDD2N loop (average $\pm$ standard deviation,
in \textdegree C) and slope of current vs.\ temperature (in mA/\textdegree C) for
each loop and tested HBM2 chip. Average: the chip's average slope across loops.
The last three rows show the average $\pm$ standard deviation across chips.}
\label{tab:temperature-per-chip}
\footnotesize
\renewcommand{\arraystretch}{0.93}
\begin{tabular}{rrrrrrrrrr}
\hline
 & Idle temp. & \multicolumn{8}{c}{Slope of current vs.\ temperature (in mA/\textdegree C)} \\
Chip & (in \textdegree C) & IDD2N & IDD3N1 & IDD3N16 & IDD0 & IDD4R & IDD4W & IDD5B & Average \\
\hline
0 & 60.4\,$\pm$\,0.5 & 4.56 & 7.16 & 6.02 & 6.13 & 8.10 & 5.87 & 7.41 & 6.47 \\
2 & 53.2\,$\pm$\,0.4 & 6.55 & 8.15 & 8.34 & 6.93 & 11.68 & 7.08 & 7.96 & 8.10 \\
4 & 62.8\,$\pm$\,0.7 & 8.66 & 9.47 & 8.81 & 8.52 & 8.77 & 7.63 & 8.33 & 8.60 \\
6 & 54.5\,$\pm$\,0.5 & 3.12 & 4.79 & 5.21 & 3.02 & 5.59 & 7.92 & 5.71 & 5.05 \\
8 & 54.2\,$\pm$\,0.4 & 4.47 & 8.89 & 7.37 & 6.22 & 8.74 & 6.17 & 6.42 & 6.90 \\
10 & 55.5\,$\pm$\,0.5 & 6.29 & 6.81 & 6.92 & 5.63 & 9.28 & 6.19 & 7.00 & 6.88 \\
12 & 55.0\,$\pm$\,0.6 & 6.45 & 8.12 & 7.89 & 3.06 & 7.31 & 9.09 & 7.46 & 7.06 \\
14 & 53.4\,$\pm$\,0.5 & 4.86 & 4.82 & 6.80 & 6.42 & 6.17 & 8.97 & 3.96 & 6.00 \\
16 & 51.2\,$\pm$\,0.4 & 5.78 & 7.84 & 4.36 & 3.60 & 5.01 & 4.15 & 8.23 & 5.57 \\
18 & 52.5\,$\pm$\,0.5 & 5.77 & 5.42 & 5.97 & 5.46 & 6.36 & 2.80 & 7.82 & 5.66 \\
20 & 51.5\,$\pm$\,0.5 & 3.54 & 1.95 & 4.01 & 2.92 & 9.07 & 1.04 & 5.92 & 4.06 \\
22 & 53.1\,$\pm$\,0.4 & 4.33 & 5.53 & 7.02 & 3.51 & 7.64 & 2.67 & 3.20 & 4.84 \\
24 & 50.1\,$\pm$\,0.4 & 5.24 & 6.05 & 5.48 & 4.62 & 10.95 & 5.19 & 4.88 & 6.06 \\
26 & 53.2\,$\pm$\,0.4 & 8.45 & 4.21 & 8.26 & 6.52 & 6.25 & 7.00 & 7.53 & 6.89 \\
28 & 48.2\,$\pm$\,0.4 & 4.17 & 6.28 & 4.38 & 5.49 & 8.69 & 8.24 & 3.07 & 5.76 \\
30 & 51.7\,$\pm$\,0.7 & 6.28 & 5.52 & 6.21 & 7.14 & 8.27 & 6.69 & 5.92 & 6.58 \\
32 & 55.2\,$\pm$\,0.4 & 5.40 & 7.47 & 6.54 & 5.31 & 7.35 & 9.75 & 8.27 & 7.15 \\
34 & 52.1\,$\pm$\,0.4 & 6.32 & 4.66 & 6.53 & 4.01 & 7.35 & 5.05 & 6.36 & 5.75 \\
\hline
1 & 61.0\,$\pm$\,0.8 & 6.60 & 6.02 & 6.30 & 5.72 & 8.06 & 5.55 & 7.96 & 6.60 \\
3 & 54.4\,$\pm$\,0.5 & 8.80 & 9.21 & 7.48 & 5.22 & 8.16 & 9.80 & 11.02 & 8.53 \\
5 & 64.3\,$\pm$\,0.6 & 9.63 & 7.44 & 9.17 & 8.12 & 10.23 & 8.44 & 9.40 & 8.92 \\
7 & 58.0\,$\pm$\,0.7 & 5.92 & 4.41 & 4.35 & 5.21 & 8.38 & 6.14 & 7.40 & 5.97 \\
9 & 54.4\,$\pm$\,0.5 & 8.15 & 9.12 & 6.44 & 7.09 & 8.86 & 7.69 & 8.28 & 7.95 \\
11 & 56.8\,$\pm$\,0.8 & 8.72 & 7.32 & 6.64 & 6.14 & 10.08 & 9.86 & 9.65 & 8.34 \\
13 & 53.4\,$\pm$\,0.5 & 7.32 & 5.71 & 8.29 & 8.93 & 8.94 & 8.21 & 7.97 & 7.91 \\
15 & 53.9\,$\pm$\,0.7 & 7.12 & 6.02 & 5.18 & 5.19 & 8.59 & 8.16 & 7.83 & 6.87 \\
17 & 51.8\,$\pm$\,0.7 & 6.53 & 5.12 & 7.14 & 5.32 & 7.79 & 8.81 & 7.58 & 6.90 \\
19 & 53.4\,$\pm$\,0.5 & 5.25 & 5.46 & 6.02 & 4.36 & 8.28 & 8.32 & 7.98 & 6.53 \\
21 & 52.3\,$\pm$\,0.5 & 4.06 & 2.03 & 4.61 & 3.34 & 8.66 & 4.79 & 8.62 & 5.16 \\
23 & 53.9\,$\pm$\,0.7 & 5.96 & 3.66 & 5.71 & 3.79 & 6.99 & 5.42 & 6.37 & 5.41 \\
25 & 51.1\,$\pm$\,0.6 & 6.83 & 3.31 & 4.57 & 5.63 & 9.17 & 7.58 & 6.62 & 6.25 \\
27 & 53.0\,$\pm$\,0.7 & 10.16 & 6.74 & 6.72 & 5.44 & 11.20 & 5.35 & 9.80 & 7.91 \\
29 & 49.3\,$\pm$\,0.5 & 5.62 & 5.08 & 5.27 & 5.75 & 9.04 & 9.74 & 5.40 & 6.55 \\
31 & 52.4\,$\pm$\,0.5 & 7.61 & 7.09 & 7.53 & 6.21 & 10.47 & 9.31 & 6.03 & 7.75 \\
33 & 56.2\,$\pm$\,0.6 & 8.92 & 7.51 & 8.49 & 8.14 & 9.27 & 7.40 & 11.23 & 8.71 \\
35 & 52.3\,$\pm$\,0.5 & 7.22 & 4.52 & 5.69 & 5.35 & 9.22 & 4.30 & 5.96 & 6.04 \\
\hline
\textbf{Even chips} & 53.8\,$\pm$\,3.3 & 5.57\,$\pm$\,1.45 & 6.29\,$\pm$\,1.84 & 6.45\,$\pm$\,1.37 & 5.25\,$\pm$\,1.58 & 7.92\,$\pm$\,1.70 & 6.19\,$\pm$\,2.32 & 6.41\,$\pm$\,1.66 & 6.30\,$\pm$\,1.09 \\
\textbf{Odd chips} & 54.6\,$\pm$\,3.5 & 7.25\,$\pm$\,1.56 & 5.88\,$\pm$\,1.87 & 6.42\,$\pm$\,1.36 & 5.83\,$\pm$\,1.42 & 8.97\,$\pm$\,1.00 & 7.49\,$\pm$\,1.76 & 8.06\,$\pm$\,1.63 & 7.13\,$\pm$\,1.12 \\
\textbf{All chips} & 54.2\,$\pm$\,3.5 & 6.41\,$\pm$\,1.72 & 6.08\,$\pm$\,1.86 & 6.44\,$\pm$\,1.37 & 5.54\,$\pm$\,1.53 & 8.44\,$\pm$\,1.49 & 6.84\,$\pm$\,2.16 & 7.24\,$\pm$\,1.84 & 6.71\,$\pm$\,1.18 \\
\hline
\end{tabular}

\end{table*}

\subsection{Data Rate Sensitivity}
\label{sec:appendix-data-rate}

Table~\ref{tab:data-rate-per-board} lists the current (in mA) of each tested FPGA board
(i.e., both of its HBM2 chips) at each data rate for the six loops in
\figref{fig:data_rate_sensitivity} (\secref{sec:data-rate-sensitivity}). The
Mean row corresponds to the points and the Slope row to the dashed lines in
\figref{fig:data_rate_sensitivity}.

\begin{table*}[!t]
\centering
\caption{Current (in mA) of each tested FPGA board (two HBM2 chips) at 600, 900,
and 1200~MT/s. IDD2N, IDD3N1, IDD4R, and IDD4W: 11 boards; IDD0 and IDD5B
(activate and refresh commands issued at a fixed rate per second at every data
rate): 16 boards. --: not measured due to intermittent errors in testing
infrastructure. Mean: average across boards; Std.: standard deviation of the
boards' averages; Slope: linear fit of the mean current vs.\ data rate.}
\label{tab:data-rate-per-board}
\footnotesize
\renewcommand{\arraystretch}{0.93}
\setlength{\tabcolsep}{2.5pt}
\begin{tabular}{rrrrrrrrrrrrrrrrrrr}
\hline
Board & \multicolumn{3}{c}{IDD2N} & \multicolumn{3}{c}{IDD3N1} & \multicolumn{3}{c}{IDD0} & \multicolumn{3}{c}{IDD4R} & \multicolumn{3}{c}{IDD4W} & \multicolumn{3}{c}{IDD5B} \\
(chips) & 600 & 900 & 1200 & 600 & 900 & 1200 & 600 & 900 & 1200 & 600 & 900 & 1200 & 600 & 900 & 1200 & 600 & 900 & 1200 \\
\hline
0/1 & 183.4 & 275.0 & 384.2 & 187.6 & 278.0 & 383.2 & 428.7 & 547.0 & 660.8 & 2944.0 & 4370.0 & 5815.2 & 1844.5 & 2872.3 & 3822.4 & 993.9 & 1148.1 & 1226.0 \\
2/3 & 232.3 & 362.7 & 458.8 & 222.9 & 360.4 & 468.9 & 461.0 & 602.6 & 705.6 & 3011.7 & 4483.3 & 5894.3 & 1906.5 & 2964.2 & 3990.7 & 1065.1 & 1211.4 & 1311.5 \\
6/7 & -- & -- & -- & -- & -- & -- & 411.6 & 549.7 & 645.9 & -- & -- & -- & -- & -- & -- & 994.6 & 1134.3 & 1265.8 \\
8/9 & 183.1 & 276.4 & 401.7 & 198.2 & 297.5 & 418.1 & 457.4 & 564.2 & 684.5 & 2970.7 & 4389.7 & 5812.3 & 1872.4 & 2857.4 & 3786.7 & 1042.0 & 1156.8 & 1251.4 \\
10/11 & 183.8 & 263.0 & 358.1 & 193.3 & 266.4 & 356.1 & 401.1 & 527.5 & 634.9 & 2907.9 & 4300.6 & 5619.6 & 1799.5 & 2797.1 & 3718.0 & 1007.5 & 1085.3 & 1240.7 \\
12/13 & 192.3 & 297.0 & 396.4 & 233.4 & 325.3 & 414.0 & 457.6 & 570.7 & 678.1 & 2968.5 & 4423.3 & 5831.6 & 1860.5 & 2927.4 & 3899.1 & 1023.0 & 1167.7 & 1294.2 \\
14/15 & 213.7 & 280.3 & 363.9 & 215.3 & 302.7 & 388.8 & 432.0 & 552.4 & 636.3 & 2942.1 & 4395.0 & 5766.4 & 1845.9 & 2885.7 & 3838.6 & 1023.0 & 1131.4 & 1246.7 \\
16/17 & 174.2 & 299.3 & 357.9 & 190.3 & 293.9 & 371.2 & 416.2 & 524.7 & 617.5 & 2907.0 & 4293.7 & 5616.1 & 1815.5 & 2823.9 & 3776.5 & -- & 1123.8 & 1226.8 \\
18/19 & 194.4 & 285.7 & 373.8 & 227.6 & 292.5 & 395.3 & 441.4 & 560.9 & 660.0 & 3010.9 & 4434.6 & 5890.5 & 1893.5 & 2915.3 & 3905.9 & 1026.7 & 1182.4 & 1283.0 \\
20/21 & -- & -- & -- & -- & -- & -- & 399.9 & 491.3 & 586.4 & -- & -- & -- & -- & -- & -- & 975.1 & 1105.2 & 1183.8 \\
22/23 & 157.0 & 263.5 & 340.6 & 178.1 & 263.6 & 339.4 & 413.2 & 514.4 & 606.9 & 2880.8 & 4322.9 & 5652.6 & 1799.0 & 2828.5 & 3758.3 & 975.9 & 1093.8 & 1217.2 \\
26/27 & -- & -- & -- & -- & -- & -- & 437.7 & 559.9 & 636.1 & -- & -- & -- & -- & -- & -- & 1007.6 & -- & 1244.7 \\
28/29 & 162.5 & 257.8 & 329.6 & 177.7 & 269.0 & 344.4 & 387.1 & 494.5 & 581.7 & 2825.7 & 4235.6 & 5546.2 & 1746.5 & 2754.3 & 3662.8 & 968.6 & 1091.9 & 1189.8 \\
30/31 & 176.3 & 279.8 & 355.9 & 184.7 & 301.2 & 363.1 & 412.9 & 523.3 & 625.2 & 2907.2 & 4333.0 & 5698.1 & 1805.4 & 2893.0 & 3816.5 & -- & 1096.5 & -- \\
32/33 & -- & -- & -- & -- & -- & -- & 434.1 & 570.2 & 648.2 & -- & -- & -- & -- & -- & -- & 1019.1 & 1157.1 & 1297.3 \\
34/35 & -- & -- & -- & -- & -- & -- & 428.1 & 537.3 & 658.4 & -- & -- & -- & -- & -- & -- & 1004.4 & 1118.2 & 1237.1 \\
\hline
\textbf{Mean} & 186.6 & 285.5 & 374.6 & 200.8 & 295.5 & 385.7 & 426.2 & 543.2 & 641.7 & 2934.2 & 4362.0 & 5740.3 & 1835.4 & 2865.4 & 3816.0 & 1009.0 & 1133.6 & 1247.7 \\
\textbf{Std.} & 20.6 & 27.5 & 33.8 & 19.4 & 27.1 & 36.1 & 21.1 & 28.6 & 32.7 & 53.4 & 69.0 & 114.1 & 45.2 & 58.4 & 87.8 & 25.9 & 35.6 & 36.3 \\
\textbf{Slope} & \multicolumn{3}{c}{0.31 mA/(MT/s)} & \multicolumn{3}{c}{0.31 mA/(MT/s)} & \multicolumn{3}{c}{0.36 mA/(MT/s)} & \multicolumn{3}{c}{4.68 mA/(MT/s)} & \multicolumn{3}{c}{3.30 mA/(MT/s)} & \multicolumn{3}{c}{0.40 mA/(MT/s)} \\
\hline
\end{tabular}

\end{table*}

\subsection{Structural Variation}
\label{sec:appendix-structural}

Tables~\ref{tab:structural-read} and~\ref{tab:structural-write} list each
tested chip's read and write current (in mA) for each bank-group pair and bank
(\secref{sec:structural-variation}, Figs.~\ref{fig:bank_group_variation}
and~\ref{fig:bank_variation}). The figures normalize each chip's current to its
BG1--BG3 and bank-3 current, respectively; the tables show absolute currents.

\begin{table*}[!t]
\centering
\caption{Read current (in mA) of each tested HBM2 chip for each bank-group pair and
bank (36 chips). The last rows show the average and standard deviation across
chips.}
\label{tab:structural-read}
\footnotesize
\renewcommand{\arraystretch}{0.93}
\begin{tabular}{rrrrrrrrrrr}
\hline
 & \multicolumn{6}{c}{Bank-group pair} & \multicolumn{4}{c}{Bank} \\
Chip & BG0--BG1 & BG0--BG2 & BG0--BG3 & BG1--BG2 & BG1--BG3 & BG2--BG3 & 0 & 1 & 2 & 3 \\
\hline
0 & 3613.9 & 3836.8 & 3701.3 & 3700.6 & 3574.7 & 3613.2 & 3843.7 & 3832.3 & 3670.7 & 3656.1 \\
2 & 3629.6 & 3845.1 & 3709.9 & 3717.7 & 3581.7 & 3621.3 & 3840.7 & 3847.0 & 3672.1 & 3660.4 \\
4 & 3725.6 & 3933.7 & 3816.8 & 3815.4 & 3683.9 & 3723.9 & 3937.0 & 3945.2 & 3763.3 & 3765.1 \\
6 & 3365.9 & 3584.3 & 3471.1 & 3460.9 & 3319.4 & 3360.8 & 3582.5 & 3584.6 & 3405.5 & 3401.9 \\
8 & 3572.9 & 3789.8 & 3656.3 & 3663.4 & 3530.4 & 3564.9 & 3775.3 & 3768.6 & 3615.7 & 3597.9 \\
10 & 3554.6 & 3789.5 & 3645.1 & 3658.6 & 3514.5 & 3556.7 & 3779.2 & 3770.9 & 3595.7 & 3586.7 \\
12 & 3629.7 & 3866.1 & 3714.8 & 3727.3 & 3583.8 & 3621.1 & 3859.8 & 3857.9 & 3680.9 & 3649.3 \\
14 & 3638.0 & 3871.5 & 3733.6 & 3736.4 & 3590.4 & 3637.7 & 3875.1 & 3865.8 & 3696.2 & 3676.9 \\
16 & 3571.6 & 3783.5 & 3649.7 & 3670.1 & 3521.4 & 3559.5 & 3796.4 & 3792.0 & 3626.6 & 3609.3 \\
18 & 3645.1 & 3896.1 & 3742.7 & 3744.5 & 3609.5 & 3650.0 & 3886.5 & 3892.6 & 3697.1 & 3675.2 \\
20 & 3533.5 & 3772.0 & 3617.7 & 3630.6 & 3484.4 & 3528.8 & 3762.5 & 3748.2 & 3570.0 & 3547.6 \\
22 & 3580.6 & 3801.6 & 3658.1 & 3670.4 & 3526.5 & 3556.5 & 3800.6 & 3796.5 & 3610.6 & 3608.1 \\
24 & 3545.3 & 3774.2 & 3635.2 & 3625.4 & 3495.3 & 3533.4 & 3778.4 & 3772.0 & 3585.9 & 3578.7 \\
26 & 3602.0 & 3826.2 & 3694.0 & 3690.0 & 3555.1 & 3600.5 & 3844.1 & 3818.3 & 3645.9 & 3631.3 \\
28 & 3470.3 & 3687.9 & 3565.9 & 3564.9 & 3417.7 & 3467.1 & 3704.3 & 3685.8 & 3525.8 & 3512.6 \\
30 & 3379.6 & 3587.0 & 3474.9 & 3472.1 & 3341.8 & 3371.6 & 3585.6 & 3582.7 & 3394.9 & 3397.7 \\
32 & 3445.6 & 3660.6 & 3541.6 & 3535.3 & 3409.0 & 3437.1 & 3658.8 & 3650.7 & 3476.9 & 3474.4 \\
34 & 3384.8 & 3595.4 & 3463.5 & 3469.7 & 3353.7 & 3378.1 & 3593.7 & 3593.7 & 3408.0 & 3386.2 \\
\hline
1 & 3351.4 & 3599.4 & 3465.0 & 3448.5 & 3313.4 & 3357.4 & 3891.2 & 3876.3 & 3691.7 & 3679.0 \\
3 & 3314.8 & 3550.4 & 3412.2 & 3416.3 & 3271.9 & 3326.2 & 3881.2 & 3886.9 & 3704.7 & 3699.0 \\
5 & 3386.6 & 3651.5 & 3503.5 & 3512.8 & 3365.9 & 3412.1 & 4021.0 & 4017.6 & 3837.9 & 3819.7 \\
7 & 3289.8 & 3517.2 & 3394.3 & 3393.7 & 3244.7 & 3304.6 & 3823.6 & 3824.6 & 3647.0 & 3628.0 \\
9 & 3281.6 & 3502.8 & 3371.5 & 3367.4 & 3239.1 & 3284.0 & 3788.7 & 3785.6 & 3622.4 & 3613.3 \\
11 & 3224.9 & 3470.1 & 3323.7 & 3343.0 & 3195.9 & 3258.1 & 3797.4 & 3804.6 & 3603.4 & 3594.1 \\
13 & 3303.3 & 3571.7 & 3420.7 & 3441.4 & 3278.1 & 3321.0 & 3867.1 & 3863.4 & 3683.9 & 3664.2 \\
15 & 3312.0 & 3542.5 & 3413.7 & 3426.7 & 3266.3 & 3325.5 & 3858.4 & 3850.1 & 3676.3 & 3665.7 \\
17 & 3257.1 & 3473.7 & 3358.2 & 3355.6 & 3226.5 & 3256.7 & 3779.1 & 3783.8 & 3610.8 & 3609.0 \\
19 & 3341.8 & 3579.9 & 3443.3 & 3447.0 & 3286.8 & 3346.2 & 3911.1 & 3910.3 & 3732.7 & 3707.2 \\
21 & 3262.8 & 3501.4 & 3363.1 & 3369.7 & 3231.3 & 3274.9 & 3788.6 & 3797.7 & 3619.0 & 3607.7 \\
23 & 3260.1 & 3486.9 & 3348.8 & 3363.9 & 3217.0 & 3271.5 & 3790.1 & 3794.8 & 3615.1 & 3600.1 \\
25 & 3257.4 & 3488.0 & 3348.4 & 3351.8 & 3218.7 & 3260.6 & 3776.9 & 3779.5 & 3593.3 & 3581.7 \\
27 & 3301.2 & 3529.9 & 3391.8 & 3402.0 & 3245.2 & 3300.9 & 3830.7 & 3841.9 & 3655.5 & 3639.2 \\
29 & 3232.1 & 3451.0 & 3301.6 & 3308.9 & 3169.1 & 3217.1 & 3709.2 & 3711.8 & 3541.2 & 3515.0 \\
31 & 3310.6 & 3529.3 & 3400.1 & 3410.7 & 3258.3 & 3301.0 & 3830.9 & 3827.2 & 3658.4 & 3636.6 \\
33 & 3341.9 & 3585.0 & 3462.0 & 3457.1 & 3303.4 & 3356.0 & 3916.3 & 3921.5 & 3738.5 & 3734.5 \\
35 & 3275.9 & 3528.2 & 3376.1 & 3394.5 & 3256.6 & 3301.1 & 3820.9 & 3830.1 & 3658.8 & 3634.8 \\
\hline
\textbf{Even chips: mean} & 3549.4 & 3772.3 & 3638.5 & 3641.8 & 3505.2 & 3543.5 & 3772.5 & 3766.9 & 3591.2 & 3578.6 \\
\textbf{Even chips: std.} & 99.1 & 104.0 & 97.5 & 99.9 & 97.9 & 100.5 & 104.1 & 104.9 & 106.0 & 103.7 \\
\textbf{Odd chips: mean} & 3294.7 & 3531.1 & 3394.3 & 3400.6 & 3254.9 & 3304.2 & 3837.9 & 3839.3 & 3660.6 & 3646.0 \\
\textbf{Odd chips: std.} & 41.9 & 50.4 & 50.7 & 48.6 & 44.4 & 45.0 & 68.4 & 66.3 & 65.2 & 65.4 \\
\textbf{All chips: mean} & 3422.1 & 3651.7 & 3516.4 & 3521.2 & 3380.0 & 3423.8 & 3805.2 & 3803.1 & 3625.9 & 3612.3 \\
\textbf{All chips: std.} & 148.3 & 145.7 & 144.7 & 143.9 & 146.4 & 142.7 & 94.0 & 94.9 & 94.6 & 93.0 \\
\hline
\end{tabular}

\end{table*}

\begin{table*}[!t]
\centering
\caption{Write current (in mA) of each tested HBM2 chip for each bank-group pair
and bank (16 chips). The last rows show the average and standard deviation
across chips.}
\label{tab:structural-write}
\footnotesize
\renewcommand{\arraystretch}{0.93}
\begin{tabular}{rrrrrrrrrrr}
\hline
 & \multicolumn{6}{c}{Bank-group pair} & \multicolumn{4}{c}{Bank} \\
Chip & BG0--BG1 & BG0--BG2 & BG0--BG3 & BG1--BG2 & BG1--BG3 & BG2--BG3 & 0 & 1 & 2 & 3 \\
\hline
0 & 2422.5 & 2486.5 & 2440.5 & 2438.7 & 2393.5 & 2424.8 & 2505.1 & 2504.4 & 2498.6 & 2483.9 \\
2 & 2486.0 & 2538.8 & 2496.8 & 2502.0 & 2449.7 & 2479.3 & 2549.5 & 2541.2 & 2543.0 & 2529.7 \\
6 & 2331.1 & 2394.0 & 2355.9 & 2363.0 & 2308.4 & 2329.1 & 2407.9 & 2414.8 & 2429.2 & 2415.3 \\
8 & 2466.9 & 2509.2 & 2448.7 & 2444.2 & 2416.0 & 2450.9 & 2522.7 & 2515.6 & 2510.9 & 2496.6 \\
10 & 2381.0 & 2419.7 & 2388.2 & 2389.8 & 2347.1 & 2381.3 & 2447.2 & 2439.2 & 2425.4 & 2426.4 \\
12 & 2472.9 & 2498.9 & 2473.1 & 2477.7 & 2430.3 & 2457.0 & 2530.7 & 2522.0 & 2511.5 & 2496.3 \\
14 & 2449.5 & 2497.9 & 2459.1 & 2453.1 & 2416.3 & 2438.5 & 2504.7 & 2486.4 & 2486.6 & 2478.0 \\
16 & 2410.4 & 2459.8 & 2407.4 & 2410.0 & 2360.4 & 2398.2 & 2472.4 & 2457.7 & 2450.7 & 2439.0 \\
18 & 2485.9 & 2540.3 & 2505.3 & 2493.6 & 2464.1 & 2488.4 & 2552.3 & 2542.1 & 2525.6 & 2530.1 \\
20 & 2281.4 & 2347.9 & 2320.5 & 2319.7 & 2253.2 & 2291.0 & 2367.1 & 2368.2 & 2386.8 & 2379.2 \\
22 & 2393.0 & 2443.1 & 2412.4 & 2400.8 & 2355.5 & 2382.1 & 2449.0 & 2447.6 & 2440.1 & 2429.2 \\
26 & 2364.2 & 2428.5 & 2393.3 & 2393.9 & 2358.3 & 2380.4 & 2451.7 & 2441.3 & 2467.5 & 2455.7 \\
28 & 2337.1 & 2389.7 & 2351.1 & 2351.9 & 2305.3 & 2333.3 & 2411.6 & 2394.2 & 2385.9 & 2370.1 \\
30 & 2417.0 & 2464.8 & 2433.3 & 2433.7 & 2388.3 & 2408.7 & 2488.1 & 2471.3 & 2464.5 & 2462.2 \\
32 & 2372.6 & 2433.7 & 2396.9 & 2409.0 & 2357.9 & 2378.5 & 2450.5 & 2459.1 & 2482.5 & 2477.3 \\
34 & 2341.1 & 2411.1 & 2370.4 & 2372.7 & 2333.8 & 2356.5 & 2432.6 & 2425.4 & 2450.7 & 2449.0 \\
\hline
\textbf{All chips: mean} & 2400.8 & 2454.0 & 2415.8 & 2415.9 & 2371.1 & 2398.6 & 2471.5 & 2464.4 & 2466.2 & 2457.4 \\
\textbf{All chips: std.} & 59.5 & 53.6 & 51.4 & 49.9 & 54.8 & 54.1 & 52.0 & 49.8 & 44.8 & 45.3 \\
\hline
\end{tabular}

\end{table*}

\subsection{Data Pattern Dependence}
\label{sec:appendix-data-pattern}

Table~\ref{tab:toggled-dqs} lists the read and write current (in mA) for each number of
toggled DQ signals (\secref{sec:dpd-current},
\figref{fig:current_vs_number_of_toggles}), and
Table~\ref{tab:beat-patterns} lists the current (in mA) for each four-beat data pattern
with and without bank-group bus toggling (\figref{fig:current_vs_bus_flip}).

\begin{table}[!t]
\centering
\caption{Read and write current (in mA, average $\pm$ standard deviation across
chips) for each number of toggled DQ signals. Chips: number of chips measured.
Slope: linear fit of the average current vs.\ the number of toggled DQs.}
\label{tab:toggled-dqs}
\footnotesize
\renewcommand{\arraystretch}{0.93}
\begin{tabular}{rrrrr}
\hline
Toggled & \multicolumn{2}{c}{Read current} & \multicolumn{2}{c}{Write current} \\
DQs & (in mA) & Chips & (in mA) & Chips \\
\hline
2 & 1969.1\,$\pm$\,50.5 & 17 & 1150.1\,$\pm$\,39.9 & 16 \\
4 & 1991.3\,$\pm$\,49.9 & 17 & 1182.2\,$\pm$\,39.0 & 16 \\
6 & 2009.1\,$\pm$\,49.4 & 17 & 1212.4\,$\pm$\,37.4 & 16 \\
8 & 2031.2\,$\pm$\,48.5 & 17 & 1245.6\,$\pm$\,43.5 & 16 \\
10 & 2049.6\,$\pm$\,51.2 & 17 & 1278.0\,$\pm$\,41.5 & 16 \\
12 & 2068.5\,$\pm$\,49.4 & 17 & 1306.2\,$\pm$\,38.9 & 16 \\
14 & 2090.0\,$\pm$\,45.3 & 17 & 1338.3\,$\pm$\,39.6 & 16 \\
16 & 2108.3\,$\pm$\,43.4 & 17 & 1368.0\,$\pm$\,38.2 & 16 \\
18 & 2129.0\,$\pm$\,44.7 & 17 & 1396.7\,$\pm$\,39.0 & 16 \\
20 & 2148.8\,$\pm$\,45.8 & 17 & 1426.6\,$\pm$\,37.4 & 16 \\
22 & 2169.8\,$\pm$\,44.4 & 17 & 1458.7\,$\pm$\,39.2 & 16 \\
24 & 2190.1\,$\pm$\,47.0 & 17 & 1488.2\,$\pm$\,37.1 & 16 \\
26 & 2212.1\,$\pm$\,46.5 & 17 & 1521.2\,$\pm$\,38.4 & 16 \\
28 & 2230.4\,$\pm$\,49.0 & 17 & 1550.4\,$\pm$\,41.0 & 16 \\
30 & 2251.0\,$\pm$\,47.4 & 17 & 1585.8\,$\pm$\,42.1 & 16 \\
32 & 2271.2\,$\pm$\,49.4 & 17 & 1615.6\,$\pm$\,41.8 & 16 \\
34 & 2292.8\,$\pm$\,51.0 & 17 & 1649.8\,$\pm$\,41.1 & 16 \\
36 & 2318.4\,$\pm$\,49.2 & 17 & 1683.1\,$\pm$\,41.0 & 16 \\
38 & 2335.5\,$\pm$\,49.8 & 17 & 1713.3\,$\pm$\,43.3 & 16 \\
40 & 2357.6\,$\pm$\,48.6 & 17 & 1743.6\,$\pm$\,42.7 & 16 \\
42 & 2379.2\,$\pm$\,48.0 & 17 & 1773.5\,$\pm$\,40.5 & 16 \\
44 & 2400.2\,$\pm$\,44.2 & 17 & 1804.5\,$\pm$\,41.8 & 16 \\
46 & 2418.4\,$\pm$\,46.1 & 17 & 1833.8\,$\pm$\,42.9 & 16 \\
48 & 2439.4\,$\pm$\,48.5 & 17 & 1860.8\,$\pm$\,40.4 & 16 \\
50 & 2456.6\,$\pm$\,47.6 & 17 & 1892.9\,$\pm$\,41.4 & 16 \\
52 & 2477.7\,$\pm$\,47.8 & 17 & 1926.6\,$\pm$\,43.3 & 16 \\
54 & 2501.0\,$\pm$\,47.5 & 17 & 1953.9\,$\pm$\,41.3 & 16 \\
56 & 2519.0\,$\pm$\,44.8 & 17 & 1984.8\,$\pm$\,43.0 & 16 \\
58 & 2537.9\,$\pm$\,42.9 & 17 & 2012.5\,$\pm$\,42.7 & 16 \\
60 & 2557.8\,$\pm$\,43.4 & 17 & 2039.6\,$\pm$\,40.9 & 16 \\
62 & 2579.6\,$\pm$\,49.4 & 17 & 2070.0\,$\pm$\,40.7 & 16 \\
64 & 2599.3\,$\pm$\,49.3 & 17 & 2093.6\,$\pm$\,41.6 & 16 \\
\hline
\textbf{Slope} & \multicolumn{2}{c}{10.21 mA/toggled DQ} & \multicolumn{2}{c}{15.38 mA/toggled DQ} \\
\hline
\end{tabular}

\end{table}

\begin{table}[!t]
\centering
\caption{Current (in mA, average $\pm$ standard deviation across chips) for each
four-beat data pattern, without and with bank-group (BG) bus toggling. Chips:
number of chips measured.}
\label{tab:beat-patterns}
\footnotesize
\renewcommand{\arraystretch}{0.93}
\begin{tabular}{rrrrr}
\hline
Data & \multicolumn{2}{c}{No BG-bus toggling} & \multicolumn{2}{c}{BG-bus toggling} \\
pattern & (in mA) & Chips & (in mA) & Chips \\
\hline
0000 & 1717.9\,$\pm$\,35.2 & 35 & 2646.8\,$\pm$\,43.5 & 35 \\
0001 & 2474.6\,$\pm$\,46.1 & 35 & 3275.5\,$\pm$\,50.2 & 35 \\
0010 & 2468.6\,$\pm$\,43.5 & 35 & 3093.3\,$\pm$\,49.6 & 35 \\
0011 & 2841.5\,$\pm$\,48.8 & 35 & 3214.8\,$\pm$\,51.1 & 35 \\
0100 & 2463.9\,$\pm$\,42.7 & 35 & 3112.3\,$\pm$\,49.3 & 35 \\
0101 & 2449.6\,$\pm$\,46.9 & 35 & 3355.3\,$\pm$\,53.4 & 35 \\
0110 & 2842.8\,$\pm$\,96.7 & 35 & 3553.0\,$\pm$\,52.9 & 35 \\
0111 & 2481.6\,$\pm$\,75.0 & 35 & 3298.1\,$\pm$\,51.7 & 35 \\
1000 & 2457.2\,$\pm$\,40.2 & 35 & 3287.9\,$\pm$\,46.0 & 35 \\
1001 & 2873.6\,$\pm$\,50.7 & 35 & 3556.3\,$\pm$\,53.0 & 35 \\
1010 & 2444.4\,$\pm$\,43.3 & 35 & 3351.9\,$\pm$\,51.5 & 35 \\
1011 & 2475.0\,$\pm$\,46.4 & 35 & 3122.1\,$\pm$\,51.2 & 35 \\
1100 & 2828.6\,$\pm$\,59.3 & 35 & 3209.4\,$\pm$\,47.3 & 35 \\
1101 & 2471.8\,$\pm$\,46.2 & 35 & 3099.4\,$\pm$\,51.0 & 35 \\
1110 & 2481.9\,$\pm$\,77.1 & 35 & 3275.0\,$\pm$\,49.8 & 35 \\
1111 & 1728.7\,$\pm$\,40.1 & 35 & 2659.9\,$\pm$\,49.3 & 35 \\
\hline
\end{tabular}

\end{table}

\subsection{Highly Power-Intensive Access Patterns}
\label{sec:appendix-max-power}

Table~\ref{tab:max-power} compares each tested chip's IDD4R current with the
current of our highly power-intensive access pattern (both in mA)
(\figref{fig:max_power_pattern}).

\begin{table}[!t]
\centering
\caption{IDD4R current and current of our access pattern (in mA, average $\pm$
standard deviation) and the increase of our carefully-engineered highly
power-intensive access pattern's current over IDD4R for each tested HBM2 chip
(35 chips). The last three rows show the average $\pm$ standard deviation across
chips.}
\label{tab:max-power}
\footnotesize
\renewcommand{\arraystretch}{0.93}
\begin{tabular}{rrrr}
\hline
Chip & IDD4R (in mA) & Our pattern (in mA) & Increase (in \%) \\
\hline
0 & 2817.5\,$\pm$\,6.4 & 3606.3\,$\pm$\,8.6 & 28.0 \\
2 & 2639.2\,$\pm$\,11.8 & 3574.0\,$\pm$\,5.8 & 35.4 \\
4 & 2899.2\,$\pm$\,3.4 & 3611.5\,$\pm$\,5.3 & 24.6 \\
6 & 2940.5\,$\pm$\,4.1 & 3504.7\,$\pm$\,15.0 & 19.2 \\
8 & 2811.2\,$\pm$\,8.0 & 3508.5\,$\pm$\,9.0 & 24.8 \\
10 & 2786.9\,$\pm$\,4.4 & 3463.3\,$\pm$\,5.2 & 24.3 \\
12 & 2876.1\,$\pm$\,4.4 & 3602.0\,$\pm$\,4.0 & 25.2 \\
14 & 2802.2\,$\pm$\,5.4 & 3561.4\,$\pm$\,9.6 & 27.1 \\
16 & 2766.6\,$\pm$\,3.9 & 3521.4\,$\pm$\,10.4 & 27.3 \\
18 & 3022.2\,$\pm$\,5.1 & 3603.8\,$\pm$\,2.9 & 19.2 \\
20 & 2758.6\,$\pm$\,3.9 & 3500.3\,$\pm$\,13.7 & 26.9 \\
22 & 2794.7\,$\pm$\,7.5 & 3505.6\,$\pm$\,5.9 & 25.4 \\
24 & 2784.1\,$\pm$\,4.4 & 3492.1\,$\pm$\,9.5 & 25.4 \\
26 & 2792.7\,$\pm$\,6.8 & 3555.5\,$\pm$\,6.3 & 27.3 \\
28 & 2710.4\,$\pm$\,3.2 & 3461.0\,$\pm$\,7.4 & 27.7 \\
30 & 2798.4\,$\pm$\,4.9 & 3545.7\,$\pm$\,14.4 & 26.7 \\
32 & 2859.7\,$\pm$\,3.8 & 3588.2\,$\pm$\,6.1 & 25.5 \\
34 & 2776.0\,$\pm$\,5.0 & 3533.0\,$\pm$\,2.3 & 27.3 \\
\hline
1 & 2743.0\,$\pm$\,4.3 & 3565.4\,$\pm$\,5.9 & 30.0 \\
5 & 3092.0\,$\pm$\,7.0 & 3686.3\,$\pm$\,4.2 & 19.2 \\
7 & 2976.2\,$\pm$\,3.1 & 3545.3\,$\pm$\,2.9 & 19.1 \\
9 & 2949.7\,$\pm$\,3.0 & 3544.5\,$\pm$\,7.7 & 20.2 \\
11 & 2915.3\,$\pm$\,4.8 & 3492.2\,$\pm$\,8.3 & 19.8 \\
13 & 2968.8\,$\pm$\,5.7 & 3643.9\,$\pm$\,4.1 & 22.7 \\
15 & 2988.5\,$\pm$\,6.9 & 3592.3\,$\pm$\,6.6 & 20.2 \\
17 & 2950.4\,$\pm$\,4.0 & 3527.1\,$\pm$\,3.1 & 19.5 \\
19 & 3004.7\,$\pm$\,2.4 & 3640.2\,$\pm$\,3.5 & 21.1 \\
21 & 2946.0\,$\pm$\,7.4 & 3512.4\,$\pm$\,5.7 & 19.2 \\
23 & 2924.7\,$\pm$\,3.8 & 3520.8\,$\pm$\,15.8 & 20.4 \\
25 & 2900.1\,$\pm$\,7.9 & 3517.4\,$\pm$\,12.0 & 21.3 \\
27 & 2962.8\,$\pm$\,4.1 & 3589.7\,$\pm$\,7.4 & 21.2 \\
29 & 2846.4\,$\pm$\,5.4 & 3495.1\,$\pm$\,4.9 & 22.8 \\
31 & 2954.6\,$\pm$\,4.7 & 3551.7\,$\pm$\,4.8 & 20.2 \\
33 & 3017.3\,$\pm$\,4.1 & 3625.6\,$\pm$\,5.7 & 20.2 \\
35 & 2956.4\,$\pm$\,7.0 & 3567.0\,$\pm$\,5.3 & 20.7 \\
\hline
\textbf{Even chips} & 2813.1\,$\pm$\,82.8 & 3541.0\,$\pm$\,48.0 & 26.0\,$\pm$\,3.4 \\
\textbf{Odd chips} & 2946.9\,$\pm$\,72.0 & 3565.7\,$\pm$\,54.9 & 21.0\,$\pm$\,2.5 \\
\textbf{All chips} & 2878.1\,$\pm$\,102.6 & 3553.0\,$\pm$\,52.9 & 23.6\,$\pm$\,3.8 \\
\hline
\end{tabular}

\end{table}

\clearpage

\omcrcomment{2}{does this come after refs?}
\section{Artifact Appendix}

\subsection{Abstract}

Our artifact contains the data, source code, and scripts needed to reproduce our
results. We provide: 1)~the source code of our real HBM2 testing infrastructure
based on DRAM Bender, 2)~all power characterization results in our paper,
3)~\atbcr{3}{Ayna} building on our experimental characterization results, and
4)~all evaluated workloads' memory access traces. We provide Python scripts and
Jupyter Notebooks to analyze and plot the results.

Our artifact is designed to remotely perform the FPGA-based HBM2 and GPU-based
HBM3E power characterization experiments. We list the hardware and software
dependencies for remote execution. The artifact evaluators are given remote
access to our HBM2 testing infrastructure and to cloud-hosted HBM3E power
testing hardware (GPUs) via the HotCRP submission page.

\subsection{Artifact check-list (meta-information)}
\label{sec:artifact-checklist}

{\small
\begin{itemize}
  \item {\bf Run-time environment: } Docker (29.1.3+), docker-compose (v5.1.1+)
  \item {\bf Hardware: } An Ubuntu 22.04 (or newer) computer with approximately
  32 GiB of memory and 1 GiB of storage. 
  \item {\bf Metrics: } Electrical current, power consumption, temperature
  \item {\bf Output: } CSV, PDF 
  \item {\bf How much disk space required (approximately)?: } 1 GiB
  \item {\bf How much time is needed to prepare workflow (approximately)?: } 30
  minutes
  \item {\bf How much time is needed to complete experiments (approximately)?: }
  Approximately 2 hours of interactive work, and approximately 10 hours of
  background experiment run-time
  \item {\bf Publicly available?: } Publicly available at
  \url{https://github.com/CMU-SAFARI/HBM-Power/tree/artifact}
  \item {\bf Code licenses (if publicly available)?: } MIT
  \item {\bf Data licenses (if publicly available)?: } MIT
  \item {\bf Archived (provide DOI)?: }
  \url{https://doi.org/10.5281/zenodo.21531486}
\end{itemize}
}

\subsection{Description}

To facilitate the artifact evaluation process, we provide SSH access to our
internal FPGA-based testing infrastructure with all the software dependencies
already installed. We describe hardware/software dependencies as well as
installation instructions for two artifact evaluation scenarios: 1) remote
access, where the artifact evaluator remotely accesses our infrastructure to
evaluate the artifact, and 2) personal computer, where the artifact evaluator
evaluates the artifact using their own personal computer and hardware. 

We provide the instructions and credentials for remote access as part of the
artifact evaluation process via the HotCRP submission website.

\noindent
\textbf{How to access.} All experimental characterization data, results, source
code and analysis scripts are provided in our open source repository at
\url{https://github.com/CMU-SAFARI/HBM-Power/tree/artifact}. A copy of the submitted artifact
is also archived at \url{https://doi.org/10.5281/zenodo.21531486}.

\noindent
\textbf{Hardware dependencies (Remote).} One Ubuntu 22.04 (or newer) computer
with approximately 32 GiB of main memory and 1 GiB of storage.

\noindent
\textbf{Hardware dependencies (Personal Computer).} One or multiple AMD Alveo
U55C boards. One or multiple NVIDIA H200 GPUs. One Ubuntu 22.04 (or newer)
computer with approximately 32 GiB of main memory and 1 GiB of storage.

\noindent
\textbf{Software dependencies (Remote).} Docker (29.1.3+), docker-compose
(v5.1.1+). 

\noindent
\textbf{Software dependencies (Personal Computer).} Docker (29.1.3+),
docker-compose (v5.1.1+), CUDA, nvidia-smi, DRAM Bender's software dependencies.

\subsection{Installation}

\noindent
\textbf{Remote.} Download the artifact's GitHub repository at
\url{https://github.com/CMU-SAFARI/HBM-Power/tree/artifact}. 

\lstset{frame=single} 
\lstset{basicstyle=\ttfamily}
\begin{lstlisting}[breaklines]
git clone --branch artifact --single-branch https://github.com/CMU-SAFARI/HBM-Power.git
cd HBM-Power
\end{lstlisting}

Refer to the README.md in the project root directory for reproduction instructions.

\noindent
\textbf{Personal Computer.} Install the AMD Alveo U55C FPGA board in your
personal computer following the installation
guide~\cite{alveo-u55-c-installation}. Install DRAM Bender following the
instructions in its public GitHub repository~\cite{safari-drambender}. Install
CUDA and nvidia-smi for H200 GPU memory power measurement experiments. Download
the artifact's GitHub repository at
\url{https://github.com/CMU-SAFARI/HBM-Power/tree/artifact}.

\subsection{Experiment workflow}

At a high level, our artifact contains four major components: 1)~experimental
HBM2 power characterization, 2)~experimental H200 GPU memory (HBM3E) power
characterization, 3)~power model simulation, and 4)~figure and table generation.
Our artifact repository provides instructions for reproducing all four
components. We also provide a fast-forward path to power model simulation and
figure and table generation by using the experimental power characterization
data included in the artifact.

\subsection{Evaluation and expected results}

The artifact provides all necessary data and sources to comprehensively
reproduce all results in the paper. Please refer to our artifact repository
\url{https://github.com/CMU-SAFARI/HBM-Power/tree/artifact} for detailed reproduction
instructions.

\subsection{Methodology}

Submission, reviewing and badging methodology:

\begin{itemize}
  \item \url{https://www.acm.org/publications/policies/artifact-review-and-badging-current}
  \item \url{https://cTuning.org/ae}
\end{itemize}

\end{document}